\documentclass[reprint,amsmath,amssymb,aps,prb,floatfix]{revtex4-2}

\usepackage{amsmath,amsfonts,amssymb,graphicx,braket,qcircuit} 
\usepackage[]{units} 
\usepackage[T1]{fontenc} 
\usepackage{hyperref} \hypersetup{ colorlinks=true, linkcolor=blue, filecolor=blue, urlcolor=blue, citecolor=blue } 
\usepackage{xcolor}
\usepackage[normalem]{ulem}
\usepackage[export]{adjustbox}
\DeclareMathOperator{\Tr}{Tr}
\begin{document} 
\title{Phonon-Induced Exchange Gate Infidelities in Semiconducting Si-SiGe Spin Qubits} 
\author{Matthew Brooks$^1$} 
\email{matthew.brooks@lps.umd.edu} 
\author{Rex Lundgren$^1$} 
\author{Charles Tahan$^2$}

\affiliation{$^1$Laboratory for Physical Sciences, 8050 Greenmead Dr., College Park, Maryland 20740, USA}

\affiliation{$^2$Department of Physics, University of Maryland, College Park, Maryland 20740, USA}

\begin{abstract}
	Spin-spin exchange interactions between semiconductor spin qubits allow for fast single and two-qubit gates. During exchange, coupling of the qubits to a surrounding phonon bath may cause errors in the resulting gate. Here, the fidelities of exchange operations with semiconductor double quantum dot spin qubits in a Si-SiGe heterostructure coupled to a finite temperature phonon bath are considered. By employing a master equation approach, the isolated effect of each spin-phonon coupling term may be resolved, as well as leakage errors of encoded qubit operations. As the temperature is increased, a crossover is observed from where the primary source of error is due to a phonon induced perturbation of the two electron spin states, to one where the phonon induced coupling to an excited orbital state becomes the dominant error. Additionally, it is shown that a simple trade-off in pulse shape and length can improve robustness to spin-phonon induced errors during gate operations by up to an order of magnitude. Our results suggest that for elevated temperatures within $\unit[200-300]{mK}$, exchange gate operations are not currently limited by bulk phonons. This is consistent with recent experiments.
\end{abstract}
\maketitle

\section{Introduction} \label{sec:Intro}

Presently, spin qubit operational fidelity is predominately limited by charge noise and hyperfine interactions\cite{burkard2021semiconductor}. Other physical noise sources affecting operations include spin-phonon interactions. Such interactions could impact performance as spin qubit processors are scaled-up, due to an increase need for control electronics resulting in higher operational temperatures\cite{petit2020high}. Moreover, the operation of quantum systems at elevated temperatures offers greater cooling power compared to the standard regime of tens of milli-Kelvin\cite{vandersypen2017interfacing,yang2020operation,petit2022design}. As such, understanding the fundamental limits of spin qubit operations at higher temperatures could help guide spin qubit processor design\cite{vandersypen2017interfacing,yang2020operation,petit2022design,yang2020operation,petit2020high,huang2021high,undseth2023hotter}.

Operation of spin qubits at temperatures $\unit[200]{mK}-\unit[2.0]{K}$ has been primarily experimentally investigated in silicon SiMOS devices. A single spin-half qubit system was shown to achieve single qubit gate fidelities of $98\%$ at $\unit[1.5]{K}$ by electrically driven spin resonance of dots that are isolated from the electron reservoir\cite{yang2020operation}. Entangling gates with two single spin-half qubits have been shown with $86\%$ fidelity at $\unit[1.1]{K}$ with standard exchange pulses\cite{petit2020high}. This was improved to demonstrate entangling SWAP gates of fidelity above $99\%$ above $\unit[1.0]{K}$ with fast diabatic pulsing\cite{petit2022design,burkard1999physical}. Other advances in hot spin qubits include algorithmic spin state initialization\cite{philips2022universal} and thermal broadening resistant readout in SiMOS\cite{huang2021high} (up to $\unit[8]{K}$). In Si-SiGe heterostructure devices, recent results for a hot single spin-half qubit device have demonstrated that at temperatures around $\unit[200]{mK}$, the heating effect due to microwave control of individual is inhibited\cite{undseth2023hotter}. In turn, this limits crosstalk and improves spin initialization in microwave controlled devices without impacting coherence. Additionally, it has been demonstrated up to $\unit[300]{mK}$ that exchange strength is independent of temperature\cite{undseth2023hotter}.

Theoretical studies of spin qubits in bulk semiconductor crystals coupled to a phonon bath at finite temperature have previously focused primarily on extrapolating $T_1$ relaxation times and $T_2$ dephasing times\cite{kornich2018phonon,kornich2014phonon,golovach2004phonon,borhani2006spin} as well as spin-phonon dephasing processes\cite{he2023theory}, phonon assisted spin-flip and hopping processes\cite{kuroyama2023phonon} and electron-phonon interactions during transport\cite{krychowski2023electron}. In the case of Si-SiGe devices, at higher operating temperatures the primary channel of dephasing of double quantum dot (DQD) encoded spin qubits by phononic interaction in a simplified effective model was found to be dominated by two-phonon processes that scale quarticly ($\propto T^4$) with temperature in certain regimes\cite{kornich2018phonon}. These works set a useful foundation from which operational fidelities of high temperature spin qubits may be calculated, however, they do not consider the impact of phonons during gate operations. In spin-half qubits, barring unwanted electron charge transfer, errors occur within the qubit subspace. However, in two or more dot encoded qubits, errors also present as coupling of the encoded state information to spin states outside of the computation spin subspace. This is known as a leakage error\cite{wardrop2014exchange,andrews2019quantifying}, and is particularly problematic during two-qubit exchange gates\cite{wardrop2014exchange}, unless either long, complicated sequences of exchange pulses are employed\cite{mehl2015fault,fong2011universal}, or alternatively entangling gates are achieved by measurement-based approaches\cite{philips2022universal,huang2023high,brooks2021hybrid,brooks2023Meas}.

In this work, the fidelity and leakage of a DQD device coupled to a finite temperature ($>\unit[100]{mK}$) phonon bath undergoing an exchange gate will be theoretically investigated. The device architecture of focus will be Si-SiGe devices, which are not as well characterised experimentally at higher temperatures of operation compared to SiMOS devices. 

Throughout this work it is assumed that the electrons encoding information are localized to one of the valleys of a strained Si-SiGe heterostructure and that the orbital splitting is smaller than the valley splitting. For SiGe devices, the valley splitting tends to be $\mathcal{O}(\unit[100]{\mu eV})$\cite{friesen2007valley,shi2011tunable,zajac2015reconfigurable,philips2022universal,losert2023practical}, despite theoretical predictions\cite{boykin2004valley}. This can be improved by the addition of valley splitting engineering, up to $\sim\unit[0.5-1.5]{meV}$ \cite{goswami2007controllable,losert2023practical} making this regime potentially realizable. By focusing on a device parameter range where the orbital splitting is lower than the valley splitting, the spin-phonon interaction as a function of the controllable QD parameters, such as confinement length and inter-dot distance, may be investigated\cite{kornich2018phonon}. It is also expected that the decoherence effects observed in these calculations will be phenomenologically similar to a system where the orbital excited state is simply replaced with the first valley excitation state.

This paper is structured as follows, first the model for the DQD Si-SiGe qubits will be introduced in Sec.~\ref{sec:Model}, followed by an introduction to the master equation analysis employed throughout the paper in Sec.~\ref{sec:MasterEq}. Then an extensive analysis of the spin-phonon induced errors in exchange gate operations is given in Sec.~\ref{sec:Gate}, including perfect square wave, ramped square wave and Gaussian pulses in QD detuning. This is followed by analysis of the spin-phonon induced errors in exchange gate operations performed at zero detuning in Sec.~\ref{sec:SOP}. Isolation of the spin-phonon induced leakage errors for encoded spin qubits is given in Sec.~\ref{sec:Leakage} followed by a discussion of experimental signatures of the spin-phonon dependent phenomena in Sec.~\ref{sec:Experiment}. Finally, conclusions are given in Sec.~\ref{sec:Conclusion}.

\section{Model} \label{sec:Model}

The system considered here is that of an electrostatically defined DQD system charged with two electrons in a Si-SiGe planar heterostructure, coupled to a phonon bath at finite temperature. The Hamiltonian for this system can be written as 
\begin{equation}
	\mathcal{H}=H_q+H_{\text{int}}+H_{\text{bath}}, 
\label{eq:Htotal} \end{equation}

\noindent where
\begin{equation}
	H_q=\sum_{i=1}^2 H_0^i +H_{C}+H_{B}.
	\label{eq:HQubitRaw} 
\end{equation}

\noindent $H_q$ describes the DQD system, $H_0^i$ describes the kinematics of a single electron in the $i^{\text{th}}$ QD, $H_{C}$ describes the Coulomb interaction of the two electrons encoding our qubit, $H_{B}$ describes some applied external magnetic fields, $H_{\text{int}}$ describes the interaction between the DQD system and the phonon bath described by $H_{\text{bath}}$. 

\subsection{Qubit Hamiltonian} \label{sec:Hq}

It is assumed that the confinement potential given in $H_{q}$ is described in the $z$-direction with an infinite hard walled quantum well of depth $l_{z}$ such that
\begin{equation}
	V_{\perp}= 
	\begin{cases}
		0 & -\frac{l_{z}}{2}\leq z\leq \frac{l_{z}}{2},\\
		\infty & \text{otherwise}. 
	\end{cases}
\end{equation}

\noindent Confinement in the $x-y$ plane is assumed to be two harmonic potentials of equal confinement length $l_c$ and of inter-dot distance $L$ between the center of the two harmonic wells.

The two-body wavefunctions are given by solving the single particle wavefunctions $\phi_{L,R}$ of each dot and their leakage into the adjacent dot by the Wannier-overlap\cite{kornich2014phonon} given by
\begin{equation}
	\Phi_{L,R}(\mathbf{r})=\frac{\phi_{L,R}^{n_{L,R},l_{L,R}}(x,y)-g\phi_{R,L}^{n_{R,L},l_{R,L}}(x,y)}{\sqrt{1-2 s g +g^2}}\phi_{qw}(z), 
\label{eq:Wann} \end{equation}

\noindent where $n_{L,R}$ and $l_{L,R}$ are the radial and orbital quantum numbers of each harmonic well, $\phi_{qw}(z)$ is the ground state of the quantum well along the $z$-direction, $s$ is the overlap
\begin{equation}
	s=\braket{\phi_L^{n_L,l_L}|\phi_R^{n_R,l_R}}, 
\end{equation}

\noindent and $g$ is defined as
\begin{equation}
	g=\frac{1-\sqrt{1-s^2}}{s}. 
\end{equation}

\noindent From the definition in Eq. (\ref{eq:Wann}) the four necessary definitions for two-body wavefunctions for a DQD system are given as the bonding and antibonding states,
\begin{equation}
	\Psi_\pm(\mathbf{r}_1,\mathbf{r}_2)=\frac{\Phi_{L}(\mathbf{r}_1)\Phi_{R}(\mathbf{r}_2)\pm\Phi_{R}(\mathbf{r}_1)\Phi_{L}(\mathbf{r}_2)}{\sqrt{2}}, 
\end{equation}

\noindent and the doubly occupied states
\begin{equation}
	\Psi_{L,R}(\mathbf{r}_1,\mathbf{r}_2)=\Phi_{L,R}(\mathbf{r}_1)\Phi_{L,R}(\mathbf{r}_2). 
\end{equation}

\noindent The hybridized basis spin states of two electrons are the the singlet state and triplet states
\begin{subequations}
	\begin{equation}
		\ket{S}=\frac{\ket{\uparrow\downarrow}-\ket{\downarrow\uparrow}}{\sqrt{2}}, 
	\end{equation}
	\begin{equation}
		\ket{T_0}=\frac{\ket{\uparrow\downarrow}+\ket{\downarrow\uparrow}}{\sqrt{2}}, 
	\end{equation}
	\begin{equation}
		\ket{T_\pm}=\ket{\uparrow\uparrow}(\ket{\downarrow\downarrow}), 
	\end{equation}
\end{subequations}

\noindent which when combined with the orbital states gives the complete set of DQD basis state wavefunctions
\begin{subequations}
	\begin{equation}
		\ket{\Psi_S}=\Psi_+(\mathbf{r}_1,\mathbf{r}_2)\ket{S}, 
	\end{equation}
	\begin{equation}
		\ket{\Psi_{T_{0,\pm}}}=\Psi_-(\mathbf{r}_1,\mathbf{r}_2)\ket{T_{0,\pm}}, 
	\end{equation}
	\begin{equation}
		\ket{\Psi_{S_{L,R}}}=\Psi_{L,R}(\mathbf{r}_1,\mathbf{r}_2)\ket{S}. 
	\end{equation}
	\label{eq:basisStates} \end{subequations}

As discussed in Sec.~\ref{sec:Intro}, it is assumed that the electrons encoding the information in the DQD system are localized to one of the valleys of a strained Si-SiGe heterostructure and that the orbital splitting is smaller than the valley splitting. This regime is potentially realizable with valley splitting engineering \cite{goswami2007controllable,losert2023practical} and allows for one to investigate the spin-phonon interaction as a function of controllable QD parameters \cite{kornich2018phonon}. Additionally, we expect that the decoherence effects observed in our calculations will be phenomenologically similar to a system where the orbital excited state is replaced with the first valley excitation state. As such, in this work, only the ground states of each of the basis state ($n=0,l=0$) and the case when one of the electrons is in the first excited state ($n=1,l=0$) are considered. The difference in energy between a basis state in its ground state and its first excited state is given by 
\begin{equation}
	\Delta E_{\text{orb}}=\frac{\hbar^2}{m_{\text{eff}}l_c^2},
    \label{eq:E_orb}
\end{equation}

\noindent where $m_{\text{eff}}$ is the effective electron mass in silicon ($\unit[1.73 \times 10^{-31}]{kg}$). It is however assumed that the charging energy $U_c$ of the doubly occupied $\ket{\Psi_{S_{L,R}}}$ states is significantly larger than that of $\Delta E_{\text{orb}}$ for a reasonable range of $l_c$. Thus, doubly occupied states with orbital excitations are ignored.

The Coulomb interaction term has the standard form
\begin{equation}
	H_c(\mathbf{r_1},\mathbf{r_2})=\frac{e^2}{4 \pi \varepsilon_0 \varepsilon_r |\mathbf{r_1}-\mathbf{r_2}|}, 
\end{equation}

\noindent where $e$ is the elementary charge, $\varepsilon_0$ is the permittivity of free space and $\varepsilon_r$ is the relative permittivity of silicon. Here due to the strong confinement in the $z$-direction it is assumed that the electrons are confined in a 2D plane. This means the Coulomb integrals can be evaluated analytically for 3 of the 4 spatial dimensions\cite{burkard1999coupled}. This is in close agreement with a purely numerically evaluated 3D integral of the same elements but is significantly faster to evaluate when varying device parameters. 

The magnetic field Hamiltonian $H_B$ is given by a uniform magnetic field along the $z$-direction to lift the degeneracy between the $T_+$ and $T_-$ states by a Zeeman shift of $E_z=g \mu_B B_z$ where $B_z$ is the magnetic field strength along the $z$-direction. Additionally, as in experimental devices, it can be assumed that there is some magnetic field gradient term along $z$ generated by a micromagnet\cite{mcneil2010localized} or g-factor engineering\cite{ruskov2018electron,harvey2019spin}, that couples the $S$ and $T_0$ states for controlled rotations about the $x-$axis of the Bloch sphere in the encoded qubit space. This term was included for completeness, and was not found to substantially impact the results discussed. A magnetic gradient can also be applied along $x$, i.e. in the plane of the Si-SiGe quantum well, which couples the $S$ and $T_\pm$ states.

\subsection{Phonon Bath Hamiltonian} \label{sec:Hbath}

The phonon bath is assumed to be equivalent to that of bulk phonons in silicon, i.e. confinement effects due to the heterostructure on the phonons are not considered,
\begin{equation}
	H_{\text{bath}}=\sum_{\mathbf{q},s}\hbar \omega_{\mathbf{q},s}\left(a^\dagger_{\mathbf{q},s} a_{\mathbf{q},s} +\frac{1}{2}\right). 
\end{equation}

\noindent Here $\mathbf{q}$ are the phonon wave vectors across the first Brillouin zone, $s=l,t_1,t_2$ denotes the acoustic phonon mode, one longitudinal and two transverse, $\omega_{\mathbf{q},s}=q v_s$ is phonon angular frequency, $q=|\mathbf{q}|$, $v_s$ are the averaged speeds of sound for the different phonon modes in silicon; $v_{l}=\unit[9\times 10^3]{m/s}$, $v_{t_1}=v_{t_2}=\unit[5.4\times 10^3]{m/s}$\cite{kornich2018phonon}, and $a^\dagger_{\mathbf{q},s}(a_{\mathbf{q},s})$ are the bosonic creation (annihilation) operators for a phonon of wave vector $\mathbf{q}$ and mode $s$.

\subsection{Electron-Phonon Interaction Hamiltonian} \label{sec:HI}

Lastly, we use the deformation potential approach to model the electron-phonon interaction \cite{mahan2013many}. As such, the electron-phonon interaction Hamiltonian, $H_{\text{int}}$, for silicon is given from the displacement operator $\mathbf{u}$ as\cite{kornich2018phonon}
\begin{equation}
	H_{\text{int}}=\Xi_d \text{Tr }\mathbf{\epsilon}+\Xi_u\epsilon_{zz}, 
\end{equation}

\noindent where the strain operator $\mathbf{\epsilon}$ is defined as
\begin{equation}
	\epsilon_{i,j}=\frac{1}{2}\left(\frac{\partial u_i}{\partial r_j}+\frac{\partial u_j}{\partial r_i}\right),
\end{equation}
and the deformation potential constants are $\Xi_d=5$ eV and $\Xi_u=8.77$ eV \cite{yu2010vibrational}.
\noindent The displacement operator is given by
\begin{equation}
	\mathbf{u}=\sum_{\mathbf{q},s}\sqrt{\frac{\hbar}{2 \rho V \omega_{\mathbf{q},s}}}\mathbf{e}_{\mathbf{q},s}\left(a_{\mathbf{q},s}\mp_s a^\dagger_{-\mathbf{q},s}\right)e^{i\mathbf{q}\cdot\mathbf{r} }, 
\end{equation}

\noindent where $\rho=\unit[2.33]{g/cm^3}$ is the density of silicon, $V$ is the volume of the crystal, and $\mathbf{e}_{\mathbf{q},s}$ are the normalized wave vectors for each phonon mode $s=\{l,t_1,t_2\}$ for the longitudinal and two transverse acoustic modes respectively. These vectors are chosen such that $\mathbf{e}_{\mathbf{q},l}=\mathbf{q}/q$, $\mathbf{e}_{-\mathbf{q},t_1}=-\mathbf{e}_{\mathbf{q},t_1}$ and $\mathbf{e}_{-\mathbf{q},t_2}=\mathbf{e}_{\mathbf{q},t_1}$. Here also $\mp_s$ is a variable sign operator dependent on the phonon mode $s$ defined such that $\mp_l=\mp_{t_1}=-$ and $\mp_{t_1}=+$\cite{kornich2018phonon}.

From Eq.~(\ref{eq:Htotal}) and the basis states Eq.~(\ref{eq:basisStates}), the finite temperature operations of a given DQD system can be calculated. Hamiltonian spanning a Hilbert space up to 14 states may be considered. These state include the three triplet states with both electrons in the groundstate, the groundstate $(1,1)$ and $(2,0)/(0,2)$ charge configuration singlet states, along with all triplet states and the $(1,1)$ charge configuration singlet state with an electron in the first excited orbital state for both the left and right QD (excited along the $x$-axis). This allows us to study a DQD system at temperatures up to $T \sim \Delta E_{\text{orb}}/k_B \sim \unit[1.0]{K}$, for an experimentally reasonable range of confinement lengths $\unit[50]{nm} \geq l_c\geq \unit[20]{nm}$\cite{zajac2016scalable}.

\begin{figure*}
	[t] 
	\includegraphics[width=0.4\linewidth]{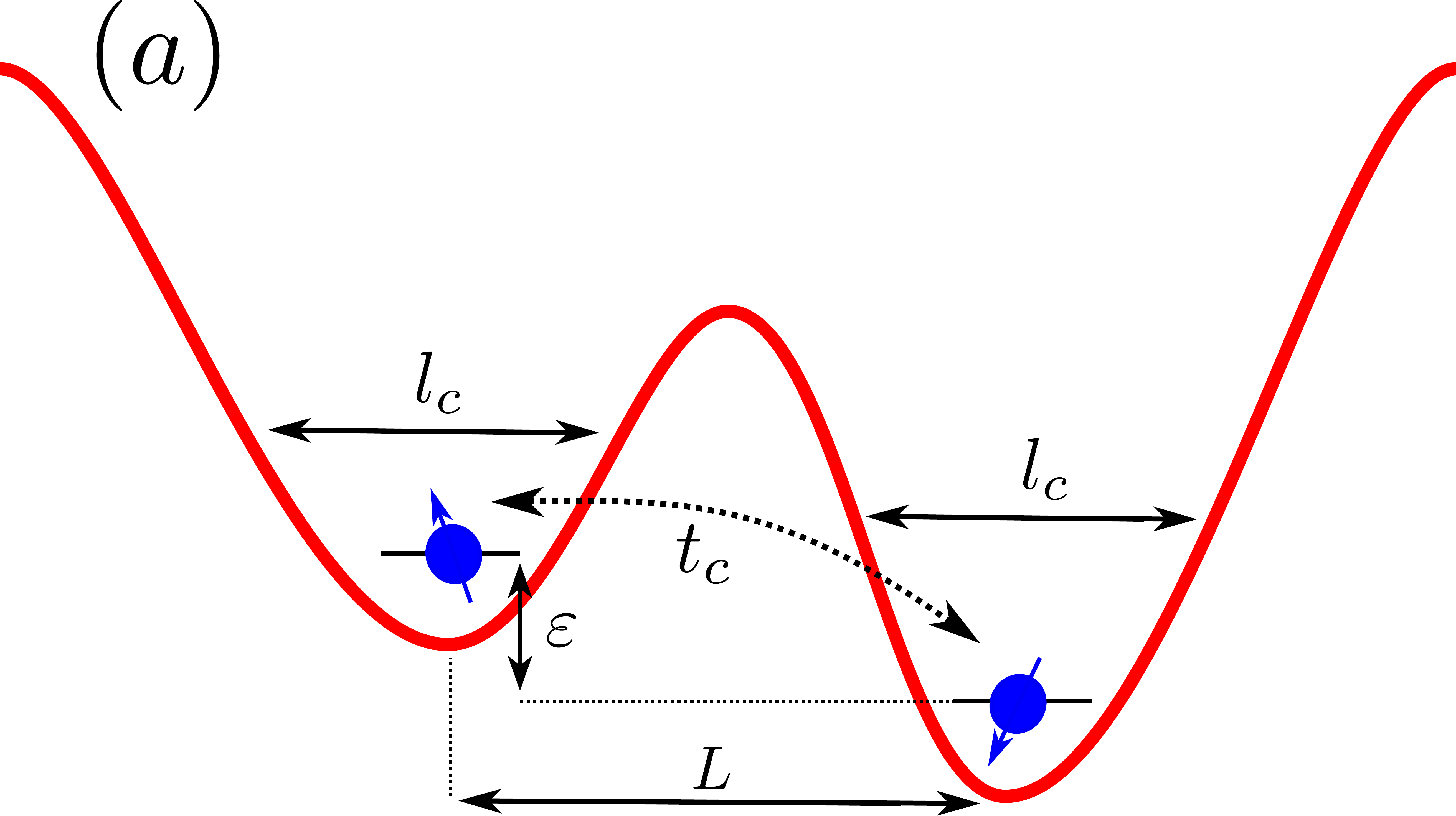} \hspace{0.05\linewidth}
	\includegraphics[width=0.4\linewidth]{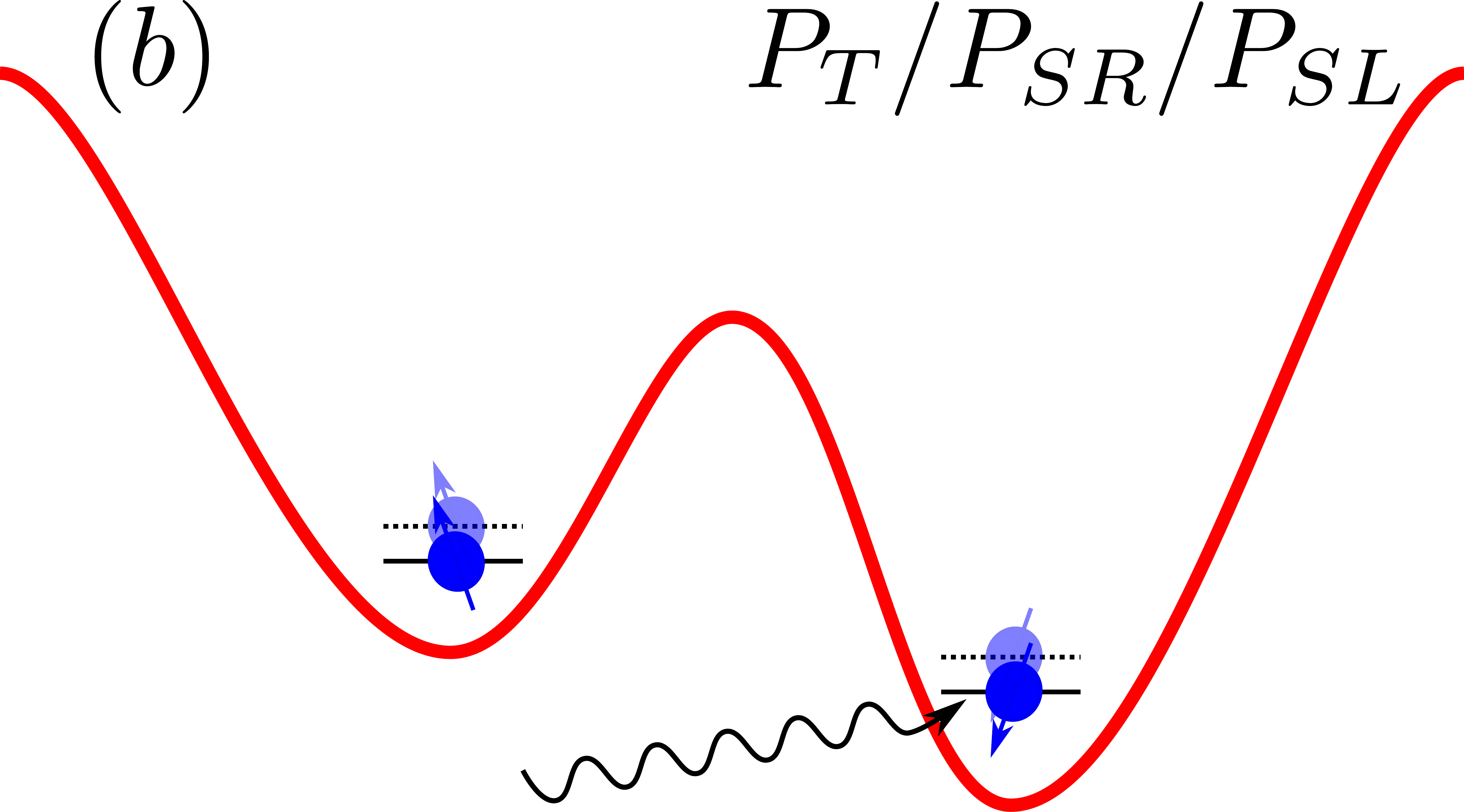} 
	\includegraphics[width=0.4\linewidth]{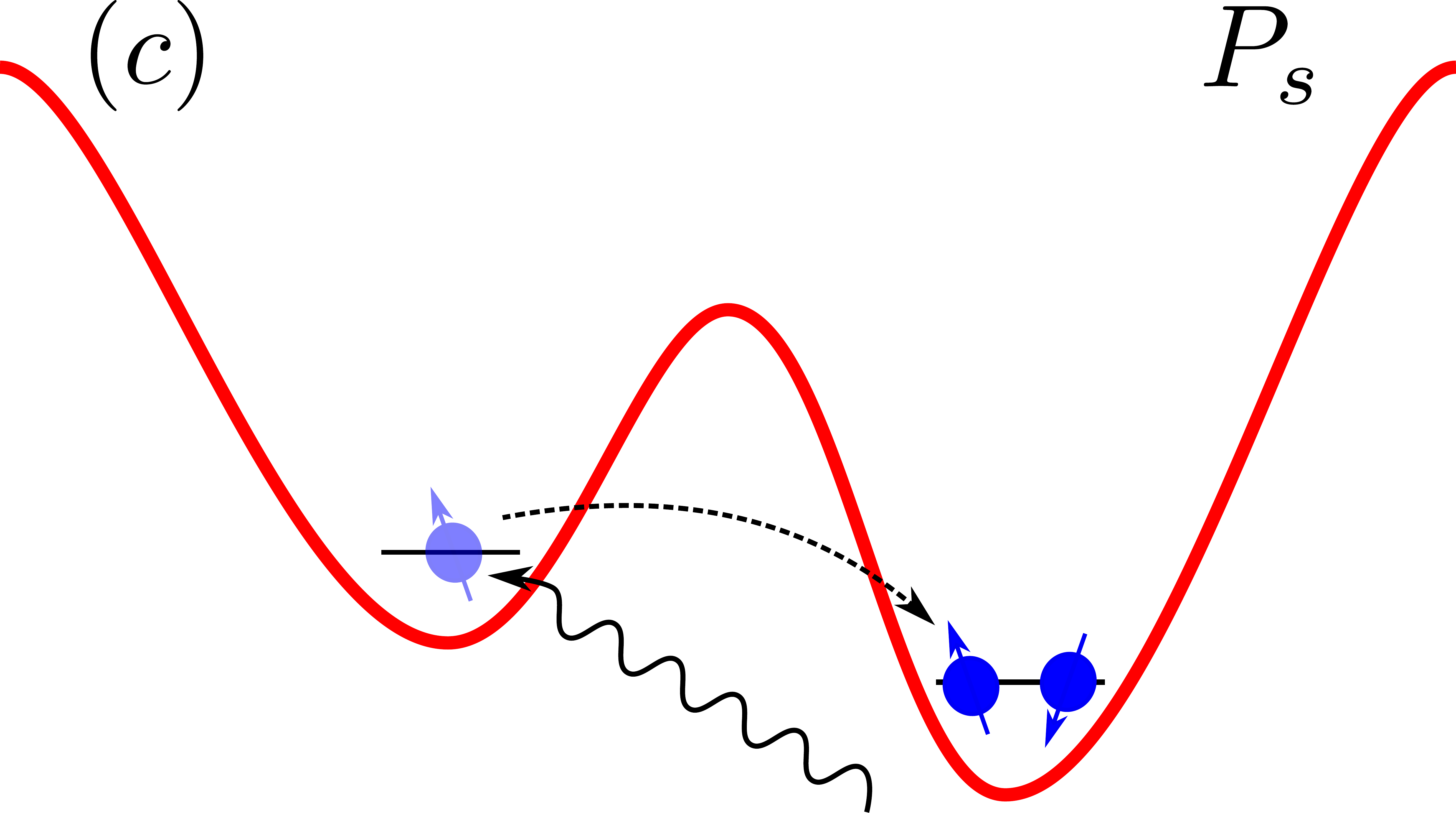} \hspace{0.05\linewidth}
	\includegraphics[width=0.4\linewidth]{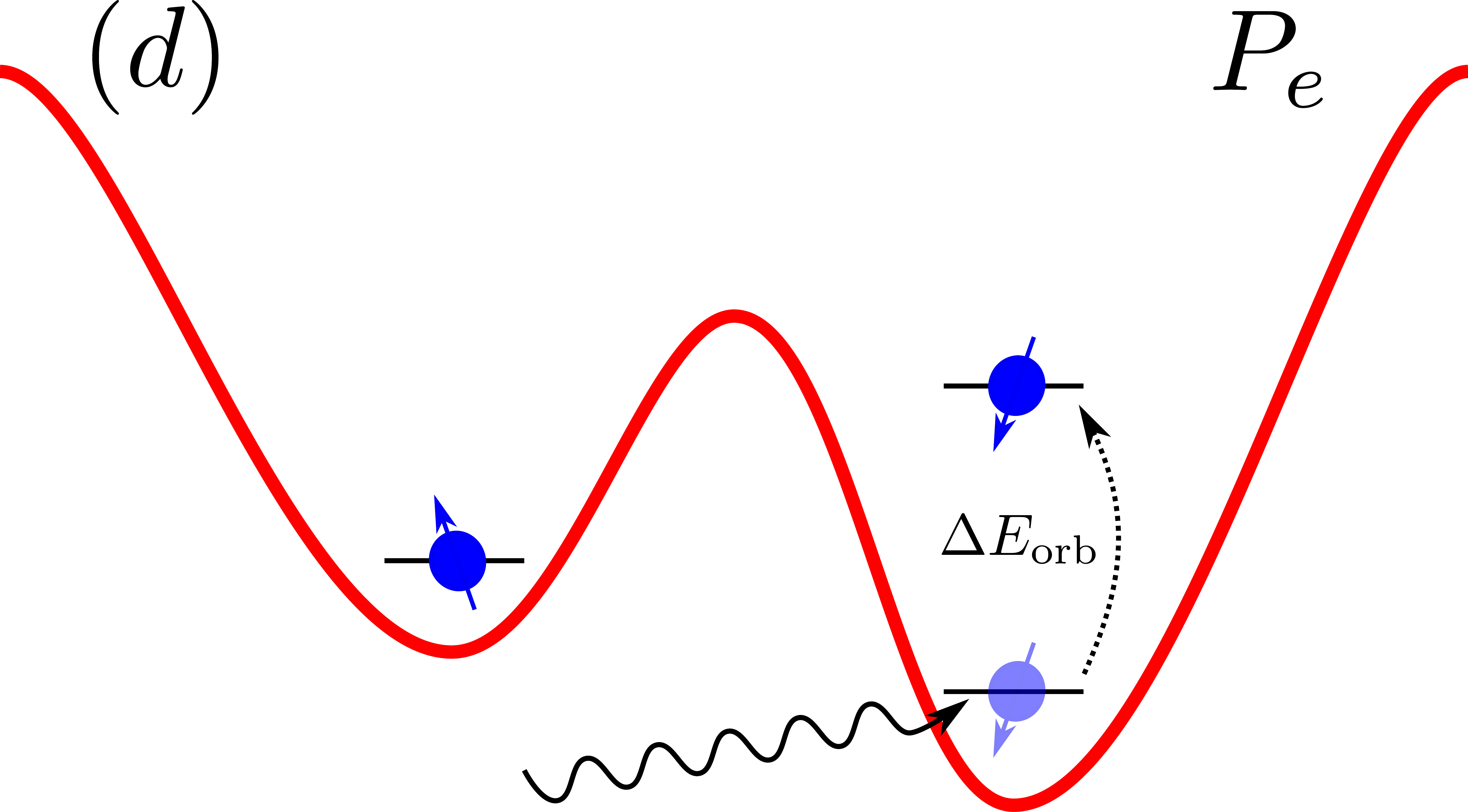} 
	\caption{(a) Two electron (blue) DQD potential (red) diagram depicting the confinement length $l_c$, inter-dot distance $L$, tunnel coupling $t_c$ and detuning $\varepsilon$. (b) Effect of the on-diagonal spin-phonon coupling terms $P_T,P_{SR}$ and $P_{SL}$. The phonon (black wave) interacts with an electron in the DQD system, shifting the groundstate energy of the spin state. (c) Effect of the off-diagonal spin-phonon coupling terms $P_s$. The phonon interacts with an electron in the DQD system such that the electron tunnels from the left to the right dot, shifting the charge configuration of the DQD system from $(1,1)$ to $(0,2)$. (d) Effect of the off-diagonal spin-phonon coupling terms $P_e$. The phonon interacts with an electron in the DQD system such that the electron is excited by the orbital energy $\Delta E_{\text{orb}}$ to the first excited orbital state.} 
\label{fig:DQD_Cartons} \end{figure*}

\subsection{Effective Hamiltonian} \label{sec:HFull}

For most this work, the following Hamiltonian of a DQD spin qubit system coupled to a phonon bath in the Hilbert space spanned by $\{\ket{T_0},\ket{S},\ket{S_{R}},\ket{T_0^*},\ket{S_0^*}\}$ is used,
\begin{widetext}
	\begin{equation}
		H_{\text{model}}=\left( 
		\begin{matrix}
			P_t & \frac{\delta b_z}{2} & 0 & P_e & 0 \\
			\frac{\delta b_z}{2} & V_+ - V_- + P_t & - \sqrt{2} t_c + P_s & 0 & P_e \\
			0 & - \sqrt{2} t_c + P_s^\dagger & U_c-\varepsilon - V_- + P_{SR} & 0 & 0 \\
			P_e^\dagger & 0 & 0 & \Delta E_{\text{orb}} + V_-^*-V_- +P_t^* & \frac{\delta b_z}{2} \\
			0 & P_e^\dagger & 0 & \frac{\delta b_z}{2} & \Delta E_{\text{orb}} +V_+^*-V_- + P_t^* 
		\end{matrix}
		\right), 
	\label{eq:H5x5} \end{equation}
\end{widetext}

\noindent where $V_\pm$ are the Coulomb energy splittings between the bonding and anti-bonding orbital states, $U$ is the electron charging energy of the QDs, $\delta b_z$ is the magnetic field gradient along $z$-axis used to mix the qubit states, $t_c$ is the tunnel coupling between the two dots, $\varepsilon$ is the detuning energy and all the $P_i$ terms are spin-phonon coupling terms. This Hilbert space is energetically well separated from other orbital excitations and $\delta b_x$, a magnetic field gradient along $x$-axis that mixes the $S$ and $T_\pm$ states, is initially omitted. The spin-phonon coupling terms are given as
\begin{subequations}
	\begin{equation}
		P_t^{(*)}=\bra{\Psi_{T_0^{(*)}}}H_{int}\ket{\Psi_{T_0^{(*)}}}=\bra{\Psi_{S^{(*)}}}H_{int}\ket{\Psi_{S^{(*)}}}, 
	\end{equation}
	\begin{equation}
		P_s=\bra{\Psi_{S}}H_{int}\ket{\Psi_{S_R}}, 
	\end{equation}
	\begin{equation}
		P_{SR}=\bra{\Psi_{S_R}}H_{int}\ket{\Psi_{S_R}}, 
	\end{equation}
	\begin{equation}
		P_{e}=\bra{\Psi_{T_0}}H_{int}\ket{\Psi_{T_0^*}}=\bra{\Psi_{S}}H_{int}\ket{\Psi_{S^*}}, 
	\end{equation}
	\label{eq:SpinPhonoTerms} \end{subequations}

\noindent where $\ket{\Psi_{i^*}}$ is the wavefunction of the spin state $i$ with the electron in the right hand QD in the first excited orbital state. The qubit Hamiltonian Eq.~(\ref{eq:HQubitRaw}) and the included spin-phonon coupling terms of Eq.~(\ref{eq:H5x5}) are depicted in Fig.~\ref{fig:DQD_Cartons}. For simplicity, when discussing the impact of the on-diagonal spin-phonon coupling elements, $P_t$ and $P_{SR}$, instead of considering their impact independently, the impact of $\tilde{P}_{SR}=P_{SR}-P_t$ will be given. This is equivalent to renormalizing Eq.~(\ref{eq:H5x5}) to $H_{\text{model}}-P_t \mathbb{I}$, where $\mathbb{I}$ is the identity matrix.

\section{Master Equation Analysis} \label{sec:MasterEq}

In order to study gate fidelities of exchange pulse operations at finite temperatures, a master equation method is employed\cite{kornich2014phonon,kornich2018phonon,landi2021non}. To do so, the rotation to diagonalize the system Hamiltonian without the phononic interaction terms is found and applied to the total system Hamiltonian such that Eq.~(\ref{eq:Htotal}) can be written as
\begin{equation}
	\tilde{\mathcal{H}}=\sum_{i=1}^N E_i \ket{i}\bra{i}+\tilde{H}_{\text{int}}+H_{\text{bath}}, 
\label{eq:RotH} \end{equation}

\noindent where $E_i$ are the eigenenergies of Eq.~(\ref{eq:H5x5}) without phononic interaction, $N$ is the size of the Hilbert space of the DQD system considered and $\tilde{H}_{\text{int}}$ is the rotated DQD-phonon interaction Hamiltonian. The interaction term of the rotated system can be separated as follows
\begin{subequations}
	\begin{equation}
		\tilde{H}_{\text{int}}=\sum_{\mathbf{q},s}A_{\mathbf{q},s}\otimes B_{\mathbf{q},s}, 
	\end{equation}
	\begin{equation}
		A_{\mathbf{q},s}=\sum_{i,j=1}^N R_{\gamma\chi}(\mathbf{q},s) \ket{i}\bra{j}, 
	\end{equation}
	\begin{equation}
		B_{\mathbf{q},s}=\left(a_{\mathbf{q},s}\mp_s a^\dagger_{-\mathbf{q},s}\right), 
	\end{equation}
\end{subequations}

\noindent where $R_{\gamma\chi}(\mathbf{q,s})$ are the rotated DQD-phonon matrix elements. To derive the master equation, Eq.~(\ref{eq:RotH}) is further transformed to an interacting frame such that the components of the interaction term of Eq.~(\ref{eq:RotH}) are given by
\begin{subequations}
	\begin{equation}
		\begin{split}
			A_{\mathbf{q},s}(t)=&e^{i H_q t}A_{\mathbf{q},s}e^{-i H_q t}\\=&\sum_{i,j=1}^N e^{i (\Delta_{i,j})t} R_{ij}(\mathbf{q},s) \ket{i}\bra{j}, 
		\end{split}
	\end{equation}
	\begin{equation}
		\begin{split}
			B_{\mathbf{q},s}(t)=&e^{i H_{\text{bath}} t}B_{\mathbf{q},s}e^{-i H_{\text{bath}} t}\\=&\left(a_{\mathbf{q},s} e^{-i \omega_{\mathbf{q},s} t} \mp_s a^\dagger_{-\mathbf{q},s} e^{i \omega_{\mathbf{q},s} t}\right), 
		\end{split}
	\end{equation}
\end{subequations}

\noindent where $\Delta_{i,j}=E_i-E_j$. In the interacting picture, the master equation is given by
\begin{equation}
	\begin{split}
		\partial_t \rho_q = -\sum_{\mathbf{q},\mathbf{q}',s}\int_0^\infty dt' C_{\mathbf{q},\mathbf{q}'}(t)\left[A_{\mathbf{q},s}(t),A_{\mathbf{q}',s}(t-t')\rho_q(t)\right]&\\-C_{\mathbf{q}',\mathbf{q}}(t)\left[A_{\mathbf{q}',s}(t-t')\rho_q(t),A_{\mathbf{q},s}(t)\right],& 
	\end{split}
\end{equation}

\noindent where $\rho_q$ is the density matrix of the DQD qubit system and
\begin{equation}
	\begin{split}
		C_{\mathbf{q},\mathbf{q}'}(t)=&\text{Tr }(B_{\mathbf{q},s}(t)B_{\mathbf{q}',s}(0)\rho_{\text{bath}})\\
		=&\delta_{\mathbf{q},\mathbf{q}'}\left((1+n_B (\omega_{\mathbf{q}}))e^{-i \omega_{\mathbf{q}} t}+n_B (\omega_{\mathbf{q}})e^{i \omega_{\mathbf{q}} t})\right), 
	\end{split}
\end{equation}

\noindent where $\rho_{\text{bath}}$ is the density matrix of the phonon bath, and by employing the temperature $T$ dependent bosonic distribution function 
\begin{equation}
	n_B (\omega_{\mathbf{q}})=\frac{1}{e^{\frac{\hbar \omega_{\mathbf{q}}}{k_B T}}-1}. 
\end{equation}

\noindent After applying a rotating wave approximation, the standard form of a master equation is recovered
\begin{equation}
	\begin{split}
		&\partial_t \rho_q=-\sum_{a,b} i \sigma_{a,b}\left[L_{a,b},\rho_q (t)\right]\\
		&+\sum_{a,b,c,d} \gamma_{a,b,c,d}\left(L_{a,b} \rho_q (t) L_{c,d}^\dagger -\frac{1}{2}\{L_{c,d}^\dagger L_{a,b},\rho_q (t) \}\right), 
	\end{split}
\end{equation}

\noindent with jump operators $L_{a,b}=\ket{a}\bra{b}$, Lamb shift coefficients

\begin{equation}
	\begin{split}
		\sigma_{a,b}=\frac{-i}{2}\sum_{\mathbf{q},\mathbf{q}'}\sum_c \delta_{E_a,E_b} \bra{c}A_{\mathbf{q}'}\ket{b}(\bra{c}A_{\mathbf{q}}^\dagger\ket{a})^*&\\
		\int_{-\infty}^{\infty} dt C_{\mathbf{q},\mathbf{q}'}(t) \text{sgn}(t)e^{i \Delta_{a,c} t},& 
	\end{split}
\end{equation}

\noindent and decay coefficients
\begin{equation}
	\begin{split}
		\gamma_{a,b,c,d}=\delta_{\Delta_{d,c},\Delta_{b,c}}\sum_{\mathbf{q},\mathbf{q}'}\bra{a}A_{\mathbf{q}'}\ket{b}(\bra{c}A_{\mathbf{q}}^\dagger\ket{d})^*&\\
		\int_{-\infty}^{\infty} dt C_{\mathbf{q},\mathbf{q}'}(t) e^{i \Delta_{b,a} t}.& 
	\end{split}
\end{equation}

\noindent The decay and Lamb shift coefficients are calculated numerically.

\section{Exchange Gate Operation} \label{sec:Gate}

To determine the viability of high temperature operations on encoded spin qubits in Si-SiGe, the robustness of exchange operations on a single DQD singlet-triplet qubit coupled to a phonon bath at finite temperature has been investigated. All calculations are of a $\pi$-phase gate given by pulsing the detuning of the two dots such that the exchange interaction takes the initial state, assumed throughout at $\ket{\uparrow\downarrow}\rightarrow\ket{\downarrow\uparrow}$. Focus is made on the $\pi$-phase gate, as it sweeps out the longest arc around the Bloch sphere whilst remaining a necessary and useful gate for quantum information purposes. Phonon induced errors for different phase gates, such as $\pi/2$ and $\pi/4$, are found to scale linearly with respect to phase gate angle. Initially, a perfect square wave pulse is considered, however other pulse shapes are investigated in Sec.~\ref{sec:RampedGate} and Sec.~\ref{sec:GaussGate}. The quality of these high temperature gate operations is quantified by the fidelity of the calculated state relative to the state of the same DQD system, experiencing the same exchange pulse without any spin-phonon interaction. While the chosen fidelity metric is generally expected to be smaller compared to when fidelity is measured against the perfect desired output state, it allows for the isolation of the spin-phonon induced errors during operations.


\subsection{Temperature Dependence} \label{sec:T}
\begin{figure}
	[t] 
	\includegraphics[width=0.49\linewidth]{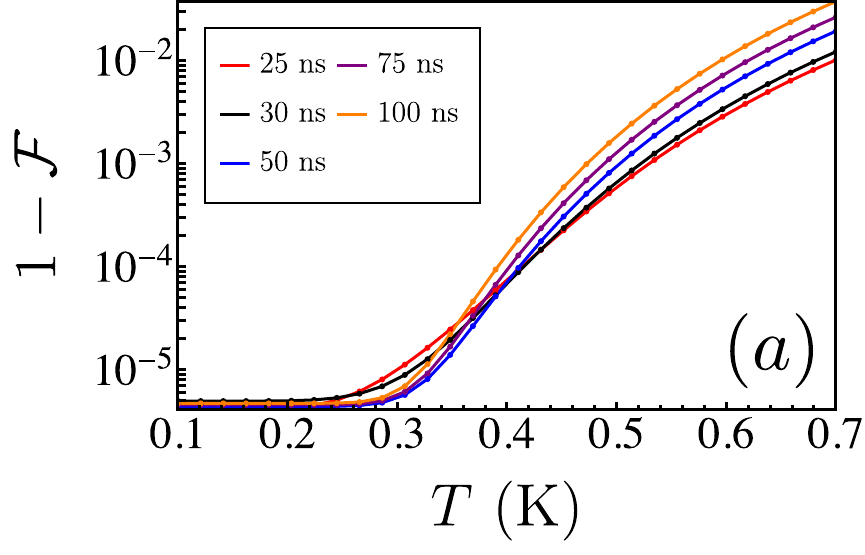} 
        \includegraphics[width=0.49\linewidth]{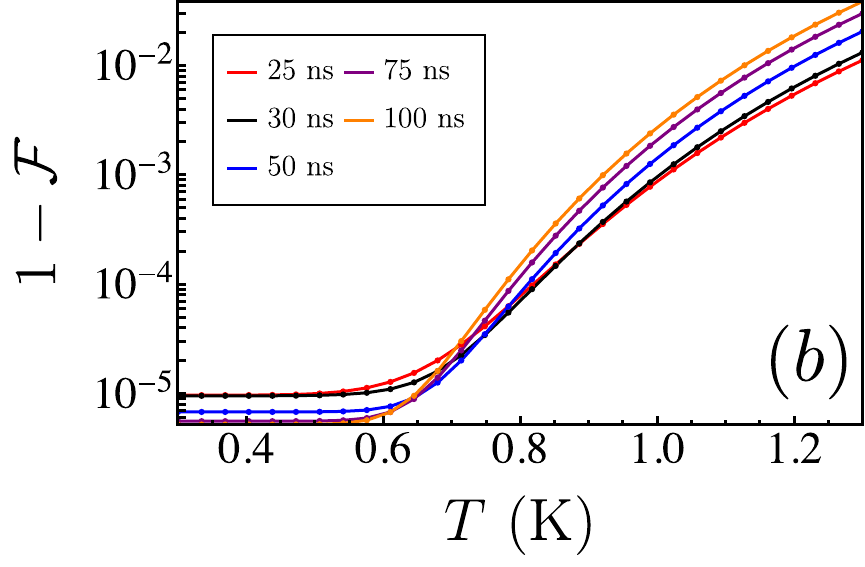} 
	\caption{Phonon-induced infidelity of a $\pi$-phase gate as a function of temperature for a range of square exchange pulse lengths for a DQD device with (a) $\Delta E_{\text{orb}}=\unit[460.2]{\mu eV}$ (b) $\Delta E_{\text{orb}}=\unit[1007]{\mu eV}$.} 
\label{fig:SquTDep} \end{figure}

Initially a DQD device in a Si-SiGe quantum well with the following parameters is considered: $l_z=\unit[6]{nm}$, inter-dot distance of $L=\unit[150]{nm}$, confinement length of $l_c=\unit[30]{nm}$, external global magnetic field of $B=\unit[20]{mT}$ and magnetic field gradient $\delta b_z=\unit[5]{mT}$ along the direction of the applied external field\cite{petta2005coherent,maune2012coherent,mills2022two,philips2022universal,kornich2018phonon}, i.e the $z$-direction. These parameters correspond to a orbital splitting of $\Delta E_{\text{orb}}=\unit[460.2]{\mu eV}$, as given by Eq.~(\ref{eq:E_orb}). With these device parameters, the fidelity of gate operations as a function of temperature is investigated for $\pi$-phase exchange pulses of varying lengths of time from $\unit[25-100]{ns}$, each at a different gate detuning $\varepsilon$, as shown in Fig.~\ref{fig:SquTDep}(a). Here a counter-intuitive behavior in the total phonon induced gate infidelity with temperature is observed. Naturally, it is expected that the slower gates will perform worse than the faster ones, as they are exposed to the phonon bath for longer. While this behavior is observed at higher temperatures of $\gtrsim\unit[500]{mK}$, below $\lesssim\unit[350]{mK}$ the slower gates tend to perform better than the faster ones. 

For comparison, in Fig.~\ref{fig:SquTDep}(b), the same calculations are performed for a device with inter-dot distance of $L=\unit[100]{nm}$ and confinement length of $l_c=\unit[20]{nm}$, corresponding to an orbital splitting of $\Delta E_{\text{orb}}=\unit[1007]{\mu eV}$. This is closer to recent state-of-the-art Si-SiGe spin qubit experiments\cite{mills2022two,philips2022universal}. Here, the same behavior as in Fig.~(\ref{fig:SquTDep})(a), is exhibited, with the temperature dependent degradation of fidelity starting at $\gtrsim\unit[700]{mK}$. Therefore the phenomenology of the orbital contribution to high temperature gate infidelities discussed for smaller orbital splittings ($\unit[300-500]{\mu eV}$) translates well for systems with larger orbital splittings ($\gtrsim\unit[1000]{\mu eV}$). Thus, for consistency, the focus for the rest of this work will be only on orbital splittings $\sim\unit[300-500]{\mu eV}$, well under our presumed valley splitting.

\begin{figure}
	[t] 
	\raggedright
	\includegraphics[width=0.49\linewidth]{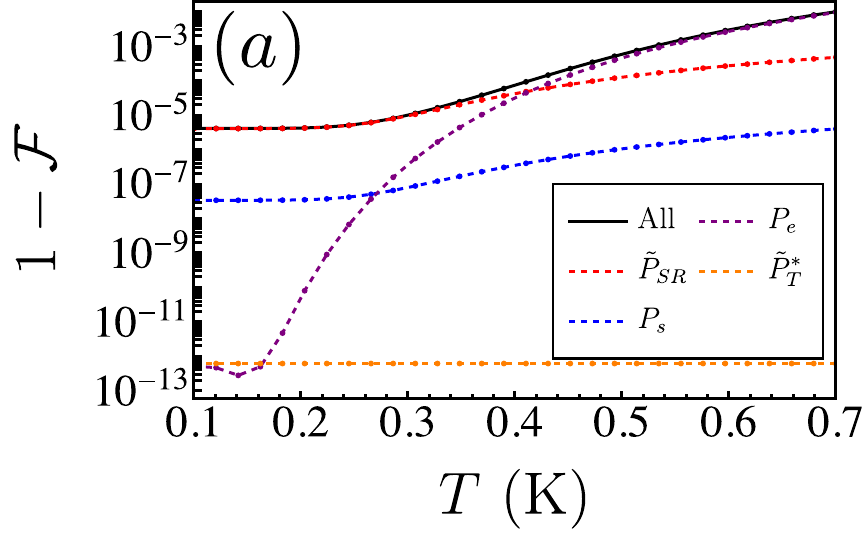} 
	\includegraphics[width=0.49\linewidth]{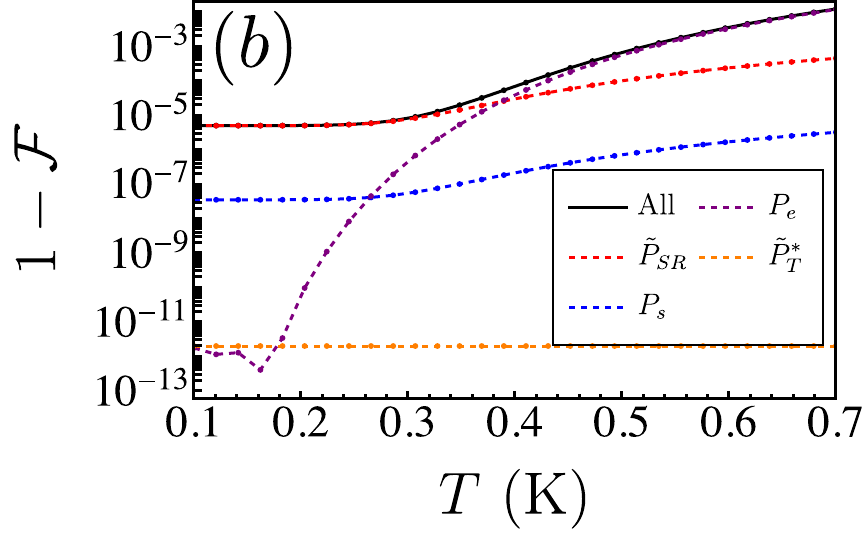} 
	\includegraphics[width=0.49\linewidth]{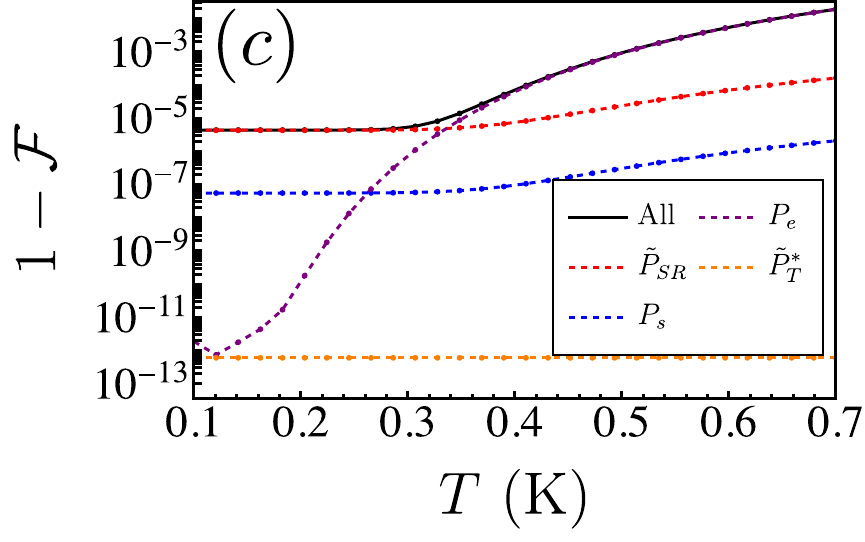} 
	\includegraphics[width=0.49\linewidth]{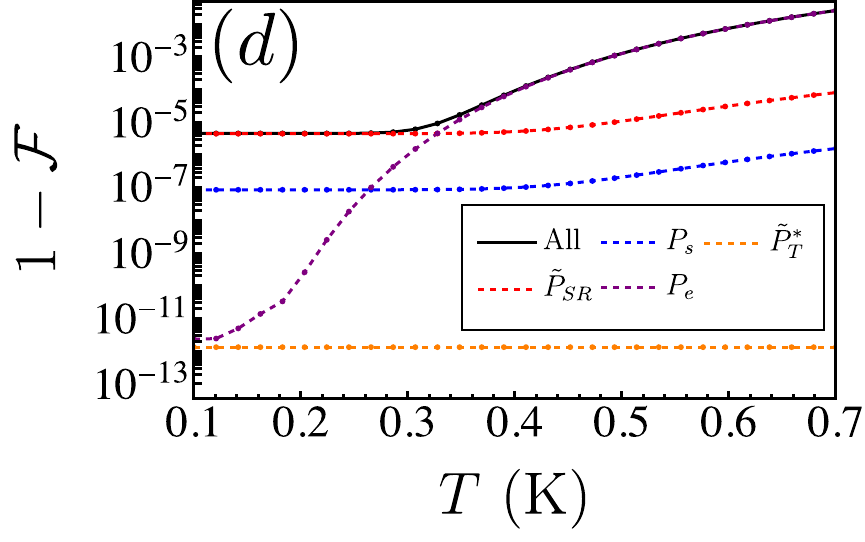} 
	\caption{Phonon-induced infidelity of a $\pi$-phase gate as a function of temperature for a (a) $\unit[25]{ns}$ pulse (b) $\unit[30]{ns}$ pulse (c) $\unit[50]{ns}$ pulse and (d) $\unit[75]{ns}$ pulse. Each plot shows total gate infidelity (solid black), and the contributing infidelities of each spin-phonon coupling term (colored dashed lines) for a DQD system with $\Delta E_{\text{orb}}=\unit[460.2]{\mu eV}$.} 
\label{fig:SquTDep_Breakdown} \end{figure}

The temperature dependence show in Fig.~\ref{fig:SquTDep} can be understood when the contribution of each spin-phonon coupling term is considered in isolation as shown in Fig.~\ref{fig:SquTDep_Breakdown}. In the case of all the gate times considered at low temperature, the dominant spin-phonon term is $\tilde{P}_{SR}$. At low temperatures, $\tilde{P}_{SR}$ induces a temperature independent energy shift. This shift becomes temperature dependent above $\sim\unit[300]{mK}$. At these higher temperatures, the dominant spin-phonon term is $P_{e}$, where orbital excitations play a larger role in the total gate error. The crossover point in temperature from which $\tilde{P}_{SR}$ dominates to where $P_{e}$ dominates is inversely proportional to gate time. However, for shorter pulses of $\unit[25-30]{ns}$, this crossover happens after the point where $\tilde{P}_{SR}$ is temperature dependent. This results in the observed crossover in behavior in Fig~\ref{fig:SquTDep}. All other spin-phonon coupling terms considered do not contribute significantly due to the assumed device parameters and temperature, particularly $P_T^*$ which will be omitted from analysis hereafter.

With this initial understanding of the spin-phonon induced errors in exchange gates, the qubit device parameters are investigated for potential optimal operating regimes. These include the inter-dot distance $L$ and confinement length $l_c$.

\subsection{Inter-dot Distance} \label{sec:L}

\begin{figure}
	[t]
	\raggedright
	\includegraphics[width=0.49\linewidth]{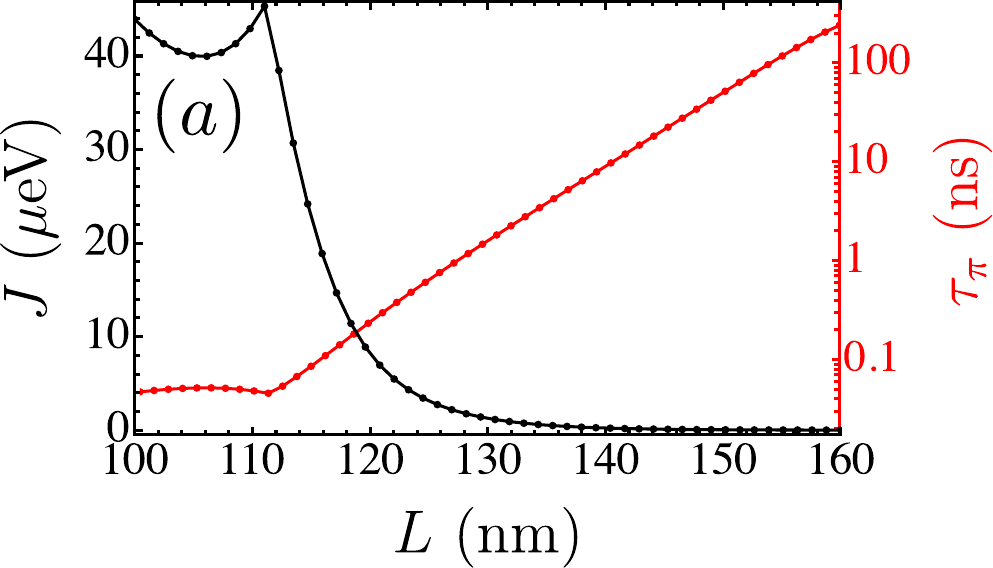}
	\includegraphics[width=0.45\linewidth]{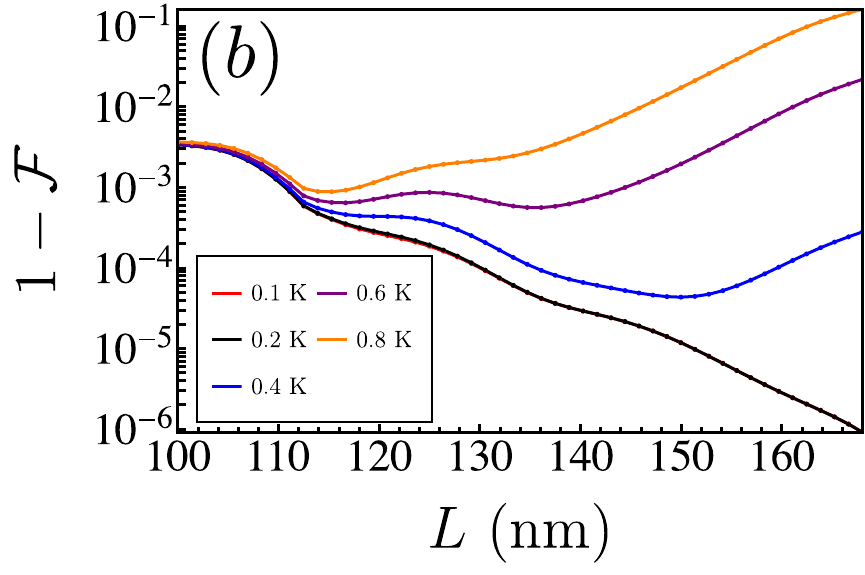}
	\includegraphics[width=0.45\linewidth]{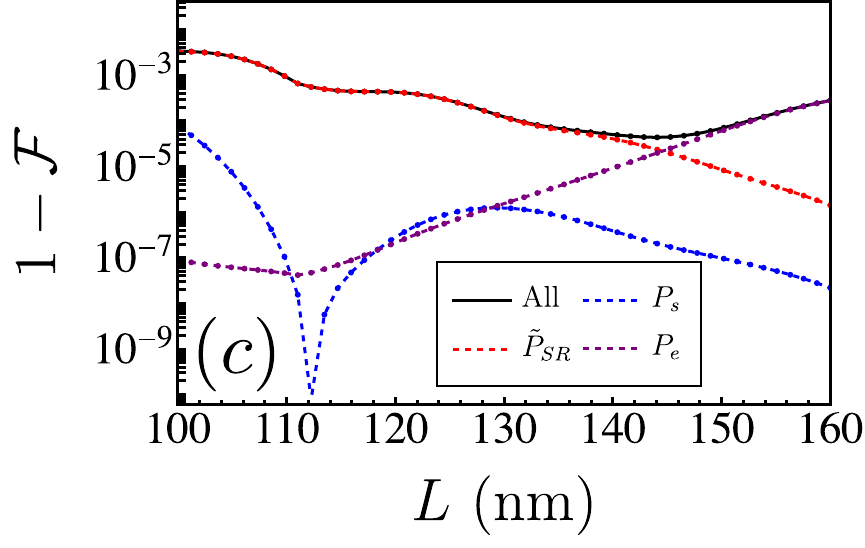}
	\hspace{1.2em}
	\includegraphics[width=0.45\linewidth]{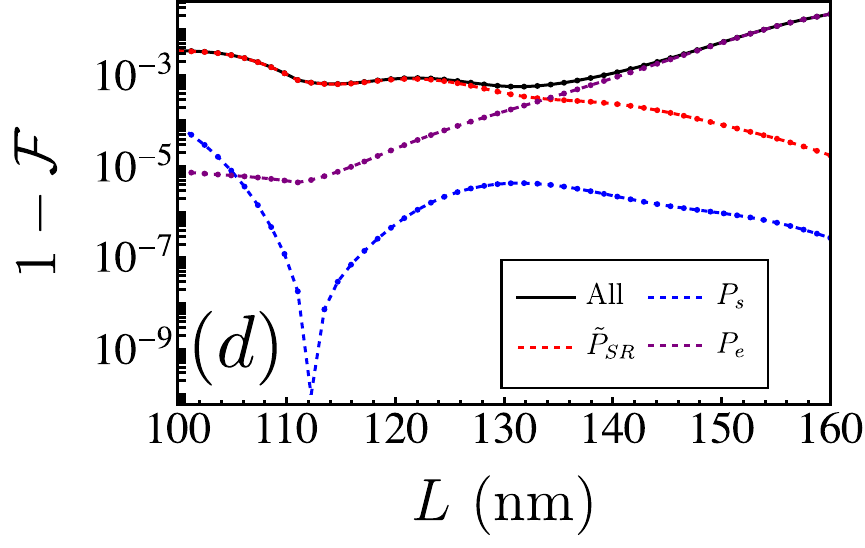}
	\caption{Phonon-induced infidelity of a $\pi$-phase gate as a function of $L$ at constant $l_c=\unit[30]{nm}$ and $\varepsilon=\unit[4]{meV}$. (a) The exchange interaction $J$ in black and corresponding time $\tau_\pi$ to perform a $\pi$-phase gate exchange gate in red at $\varepsilon=\unit[4]{meV}$ with $L$. (b) Phonon-induced gate infidelity with $L$ at constant $\varepsilon$ for temperatures $\unit[100-800]{mK}$. (c-d) Overall accumulated gate infidelity in solid black, and the contributing infidelities of each spin-phonon coupling term in the colored dashed lines at (c) $\unit[400]{mK}$} and (d) $\unit[600]{mK}$. 
\label{fig:DotDistanceTime} \end{figure}

At constant $l_c=\unit[30]{nm}$ and $\varepsilon=\unit[4]{meV}$, the relationship between the exchange interaction strength, $J$, and therefore the gate time for a $\pi$-pulse, $\tau_\pi$, as $L$ is increased is straightforward. The further apart the dots are the weaker the interaction between the electrons becomes, due to a vanishing wavefunction overlap,
\begin{equation}
	s=e^{-\frac{L^2}{4 l_c^2}}, 
\label{eq:overlapLlc} \end{equation}

\noindent and therefore the time to perform the gate at the fixed $\varepsilon$ increases exponentially, see Fig.~\ref{fig:DotDistanceTime}(a). This is reflected in the spin-phonon induced infidelity at various temperatures as is shown in Fig.~\ref{fig:DotDistanceTime}(b). At lower temperatures of $\unit[100-300]{mK}$, overall as $L$ increases, so too does the exchange gate quality. At higher temperatures of $\unit[600-800]{mK}$, at large $L$ the gate infidelity exponentially grows. Note also the kink in the applied exchange gate at constant $\varepsilon$ at low $L$. This is due to a changing of the sign of the $U_c-\varepsilon-V_-$ term on the diagonal of Eq.~\ref{eq:H5x5} given by the varying of the Coulomb terms $U_c-V_-$ as $L$ is varied, relative to the fixed $\varepsilon$.

This can be explained in the same manner as the crossovers in gate infidelity as a function of temperature. When comparing the contributing spin-phonon terms as a function of $L$ at various temperatures, as in Fig.~\ref{fig:DotDistanceTime}(c) and (d), it is clear that the temperature independent shift of the $\tilde{P}_{SR}$ term that dominates at low temperatures, is suppressed at larger $L$. Then at higher temperatures the $P_e$ term dominates. Furthermore, at low $L$, $\tau_\pi$ is so short that the effect of temperature is negligible, and a constant gate infidelity is observed. Finally, it is seen that the contribution of $P_s$ goes to 0 at the point $U_c-\varepsilon-V_-=0$, as previously discussed.

\begin{figure}
	[t]
	\raggedright
	\includegraphics[width=0.475\linewidth]{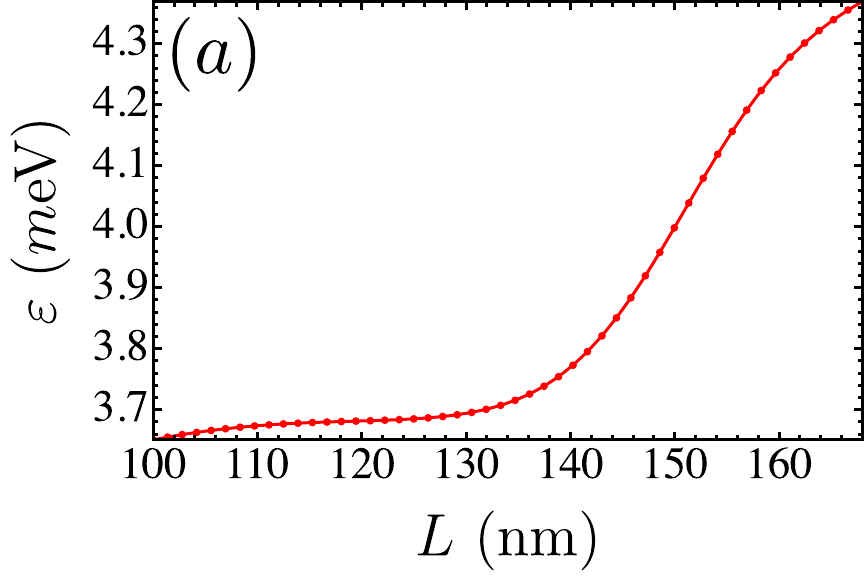}
	\hspace{0.1em}
	\includegraphics[width=0.49\linewidth]{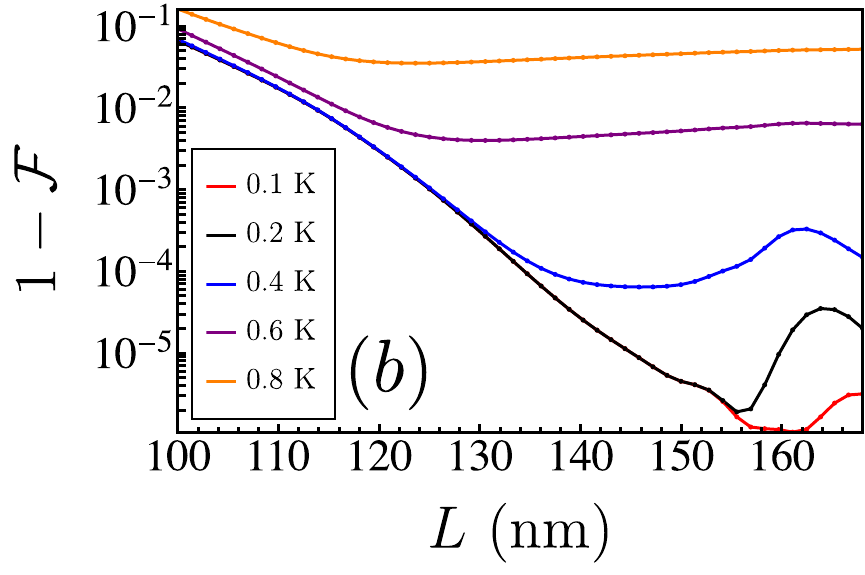} 
	\includegraphics[width=0.49\linewidth]{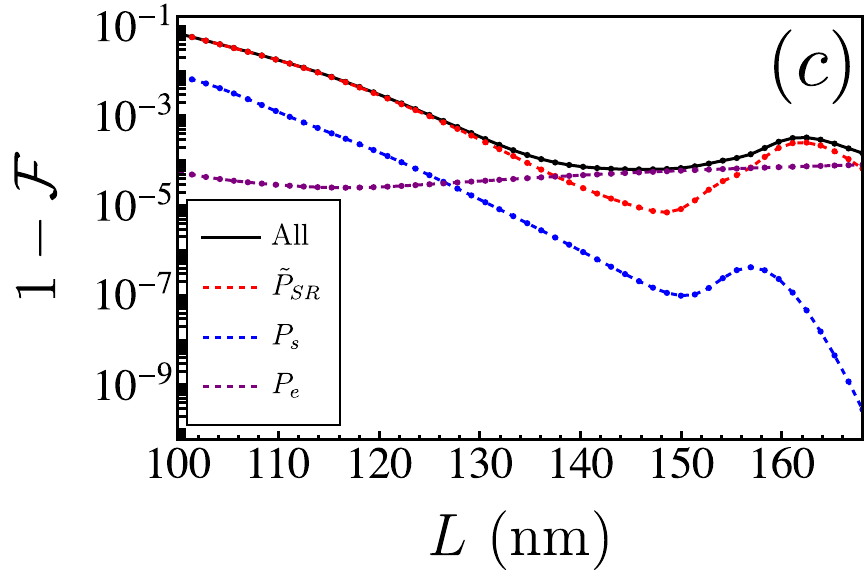} 
	\includegraphics[width=0.49\linewidth]{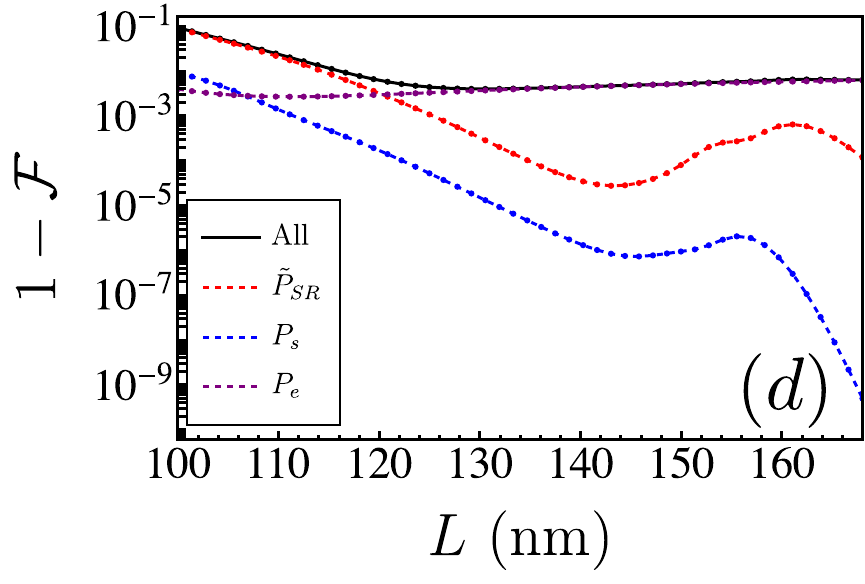} 
	\caption{Phonon-induced infidelity of a $\pi$-phase gate with $L$ at constant $l_c=\unit[30]{nm}$ and $\tau_\pi=\unit[50]{ns}$. (a) The detuning $\varepsilon$ to perform a $\pi$-phase gate exchange gate in red at $\tau_\pi=\unit[50]{ns}$ with $L$. (b) Phonon-induced gate infidelity with $L$ at constant $\tau_\pi$ for temperatures $\unit[100-800]{mK}$. (c-d) Overall accumulated gate infidelity in solid black, and the contributing infidelities of each spin-phonon coupling term in the colored dashed lines at (c) $\unit[400]{mK}$ and (d) $\unit[600]{mK}$.} 
\label{fig:DotDistanceE} \end{figure}

Equally, the effect of phonon-induced errors during gate operations as functions of device parameters can be studied at constant pulse time, with the pulse detuning adjusted accordingly to achieve a constant gate time. This is shown in Fig.~\ref{fig:DotDistanceE}. Fixing $\tau_\pi=\unit[50]{ns}$, the required $\varepsilon$ are shown in Fig.~\ref{fig:DotDistanceE}(a) to achieve a $\pi$-phase gate. The observed effect of temperature and device parameter on the calculated gate infidelities at fixed $\tau_\pi$ is similar to that at fixed $\varepsilon$, i.e. as temperature is increased the effect of the $P_e$ term dominates and gate infidelity is no longer the dominant term as the wavefunction overlap Eq.~(\ref{eq:overlapLlc}) is reduced.

However, as can be seen in Fig.~\ref{fig:DotDistanceE}(c), at large $L>\unit[150]{nm}$ there is a point where the on-diagonal $\tilde{P}_{SR}$ dominates again. This is due to the difference between the charging energy $U_c$ and how $\varepsilon$ is varied with $L$. As the wavefunction overlap of the two electrons decreases, the energy cost of adding a second electron into a QD increases due to stronger localization of the existing electron in the dot. Simultaneously, the stronger localization of the electrons in their respective dots requires significant increases in $\varepsilon$ to keep $J$ and therefore $\tau_\pi$ constant. At small $L$, the difference $U_c-\varepsilon$ increases as $L$ is increased, however at large $L$, $U_c-\varepsilon$ tends to 0 as $L$ is increased. This is illustrated by the sharp change in $\varepsilon$ for constant $J$ in Fig.~\ref{fig:DotDistanceE}(a). The observed re-emergence of $\tilde{P}_{SR}$ at large $L$, and low temperatures corresponds to the suppression of $U_c-\varepsilon$, as they share the same matrix element in the total Hamiltonian.

\subsection{Confinement Length} \label{sec:lc}

\begin{figure}
	[t]
	\raggedright
	\includegraphics[width=0.49\linewidth]{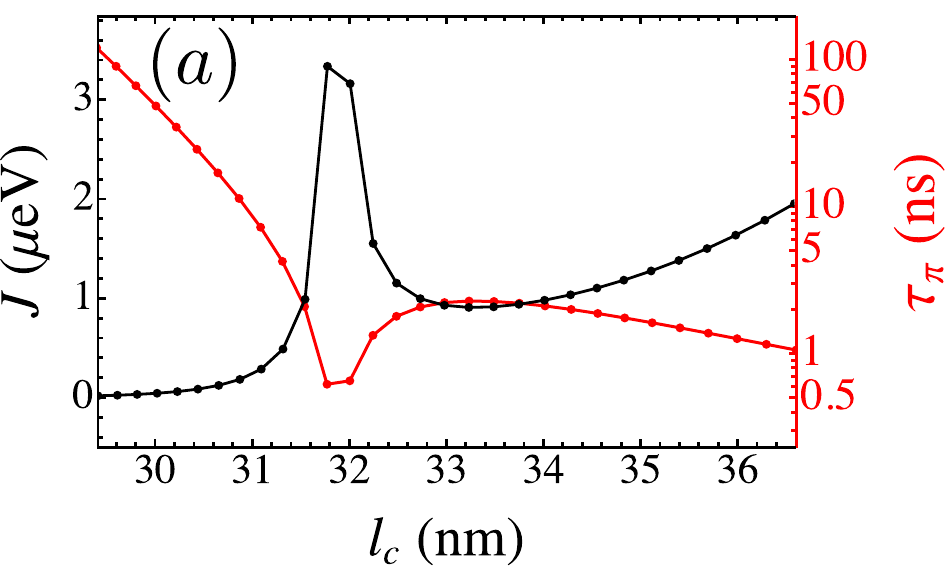} 
	\includegraphics[width=0.45\linewidth]{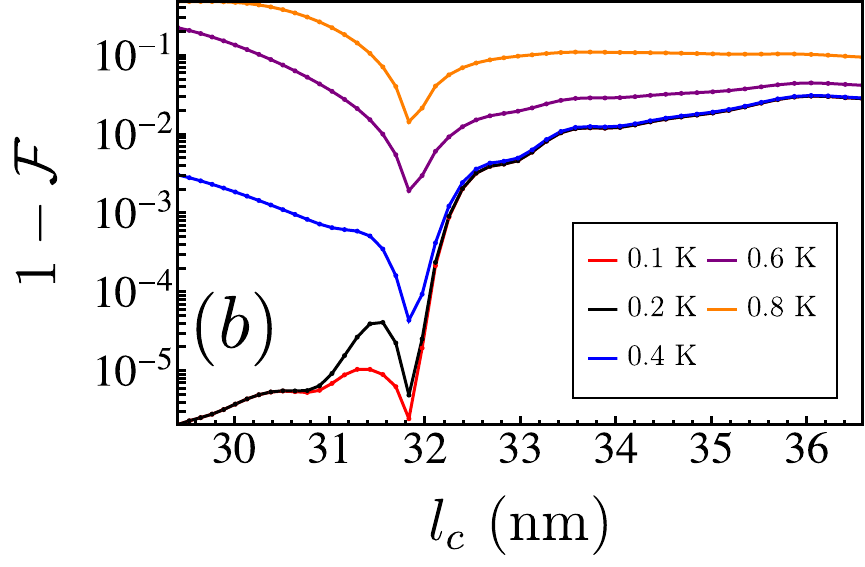} 
	\includegraphics[width=0.45\linewidth]{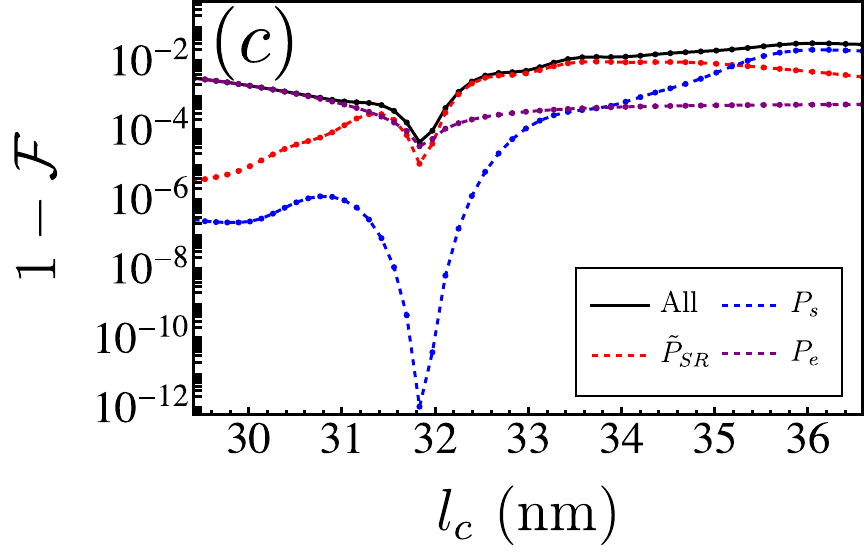}
	\hspace{0.8em}
	\includegraphics[width=0.45\linewidth]{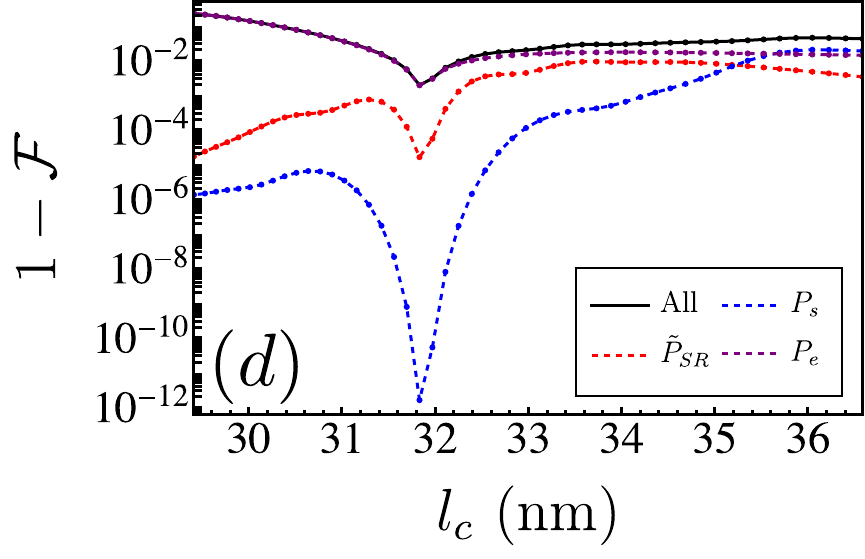} 
	\caption{Phonon-induced infidelity of a $\pi$-phase gate as a function of $l_c$ at constant $L=\unit[150]{nm}$ and $\varepsilon=\unit[4]{meV}$. (a) The exchange interaction $J$ in black and corresponding time $\tau_\pi$ to perform a $\pi$-phase gate exchange gate in red at $\varepsilon=\unit[4]{meV}$ with $l_c$. (b) Phonon-induced gate infidelity with $l_c$ at constant $\varepsilon$ for temperatures $\unit[100-800]{mK}$. (c-d) Overall accumulated gate infidelity in solid black, and the contributing infidelities of each spin-phonon coupling term in the colored dashed lines at (c) $\unit[400]{mK}$ and (d) $\unit[600]{mK}$.} 
\label{fig:ConfLenTime} \end{figure}

\begin{figure}
	[b]
	\raggedright
	\includegraphics[width=0.475\linewidth]{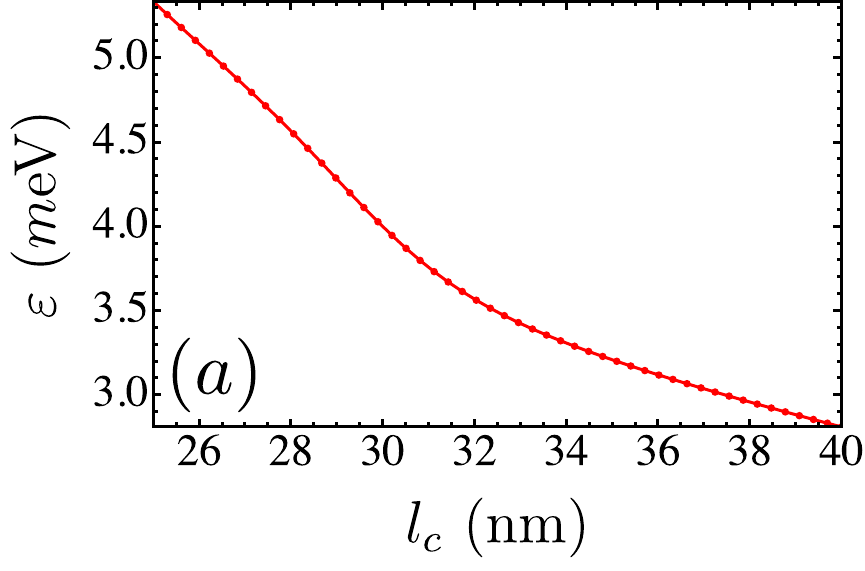}
	\hspace{0.1em}
	\includegraphics[width=0.49\linewidth]{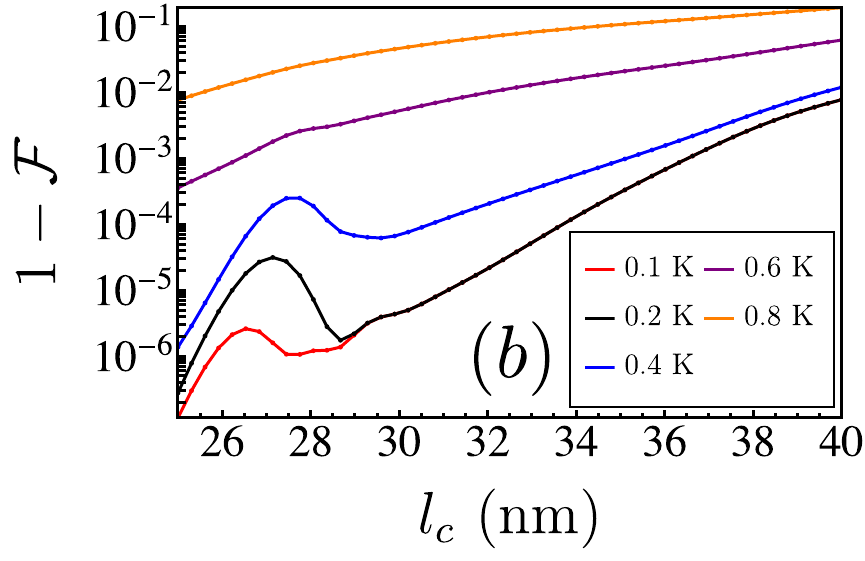} 
	\includegraphics[width=0.49\linewidth]{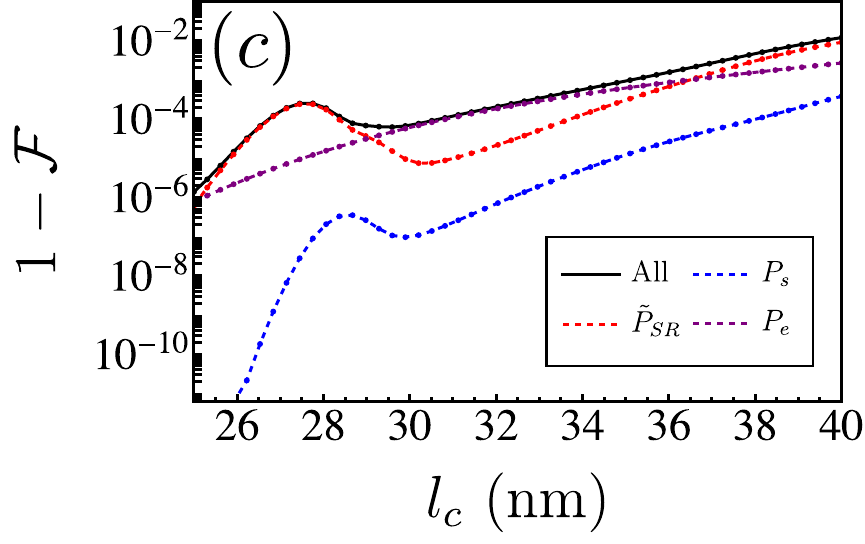} 
	\includegraphics[width=0.49\linewidth]{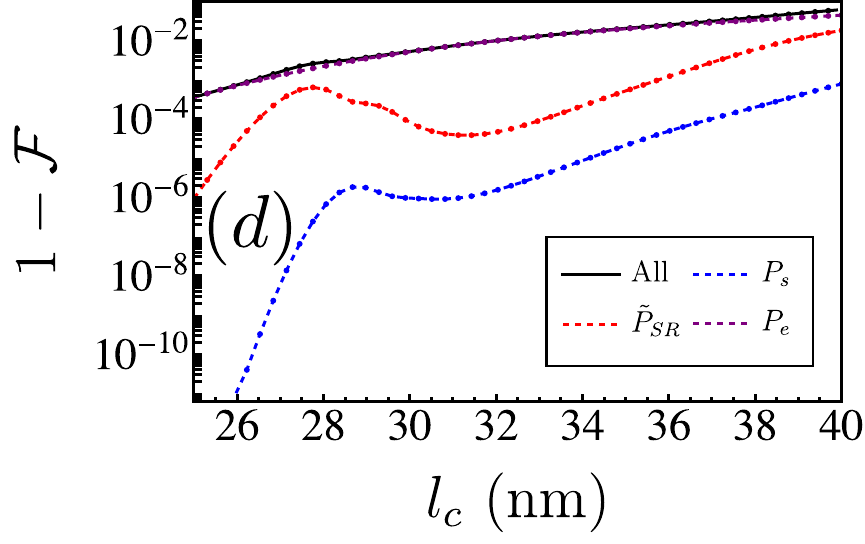} 
	\caption{Phonon-induced infidelity of a $\pi$-phase gate with $l_c$ at constant $L=\unit[150]{nm}$ and $\tau_\pi=\unit[50]{ns}$. (a) The detuning $\varepsilon$ to perform a $\pi$-phase gate exchange gate in red at $\tau_\pi=\unit[50]{ns}$ with $l_c$. (b) Phonon-induced gate infidelity with $l_c$ at constant $\tau_\pi$ for temperatures $\unit[100-800]{mK}$. (c-d) Overall accumulated gate infidelity in solid black, and the contributing infidelities of each spin-phonon coupling term in the colored dashed lines at (c) $\unit[400]{mK}$ and (d) $\unit[600]{mK}$.} 
\label{fig:ConfLengthE} \end{figure}

At constant $L=\unit[150]{nm}$ and $\varepsilon=\unit[4]{meV}$, the corresponding behavior of $J$ as $l_c$ is varied is the inverse of that of $L$. Here, not only does decreasing $l_c$ increase the overlap between the two electron wavefunctions, similarly to increasing $L$, but the shape of the constituent wavefunctions are affected as well. As such, spin-phonon induced gate infidelities as a function of $l_c$, as shown in Fig.~\ref{fig:ConfLenTime}, paints a similar picture as spin-phonon induced gate infidelities as a function of $L$, as shown in Fig.~\ref{fig:DotDistanceTime}. At high $l_c$, the effect of temperature is somewhat weak. Increased wavefunction overlap and flattened electron wavefunctions allows for the charge scattering spin-phonon term $P_s$ to eventually dominate. At smaller $J$, it is again clear that at temperatures below $\sim\unit[400]{mK}$ the on-diagonal $\tilde{P}_{SR}$ phonon emission term dominates, but at higher temperatures the $P_e$ phonon absorption term dominates. A similar peak in the applied $J$ and therefore a minimum in $\tau_\pi$ as $U_c-\varepsilon-V_-\rightarrow0$, is also seen as a function of $L$, with a similar effect on the contributions of all spin-phonon coupling elements at this point.

Fig.~\ref{fig:ConfLengthE} shows the phonon induced errors of gates performed with variable $l_c$ at constant gate time ($\tau_\pi=\unit[50]{ns}$), and thus variable $\varepsilon$. Here, the same behavior is observed at constant gate time as in Sec.~\ref{sec:L}. As the confinement length is increased the two electron wavefunction overlap also increases and the charging energy of each dot decreases significantly. The result is an overall increase in exchange gate infidelity, primarily due to the $P_e$ phonon interactions, with a peak at low $l_c$ due to the vanishing difference between $U_c$ and $\varepsilon$. Additionally, as in Fig.~\ref{fig:DotDistanceE}, we see a low temperature peak in the infidelity due to a peak in the $\tilde{P}_{SR}$ interaction given by the interplay of the Coulomb terms and applied $\varepsilon$.

\begin{figure}
	[b]
	\raggedright
	\includegraphics[width=0.49\linewidth]{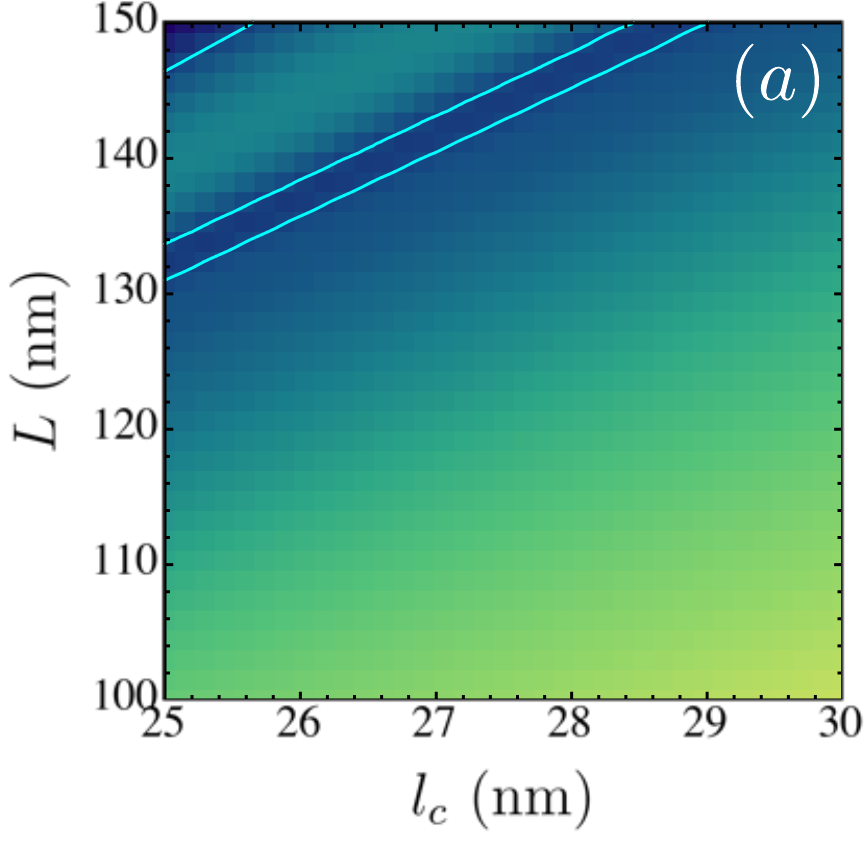} 
	\includegraphics[width=0.49\linewidth]{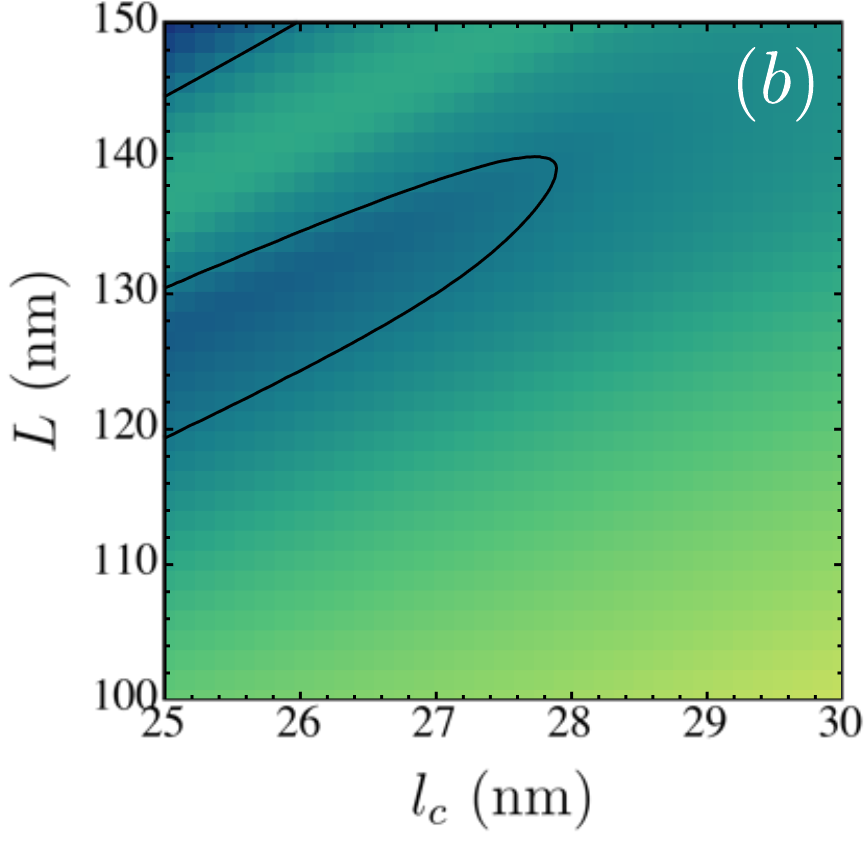} 
	\includegraphics[width=0.49\linewidth]{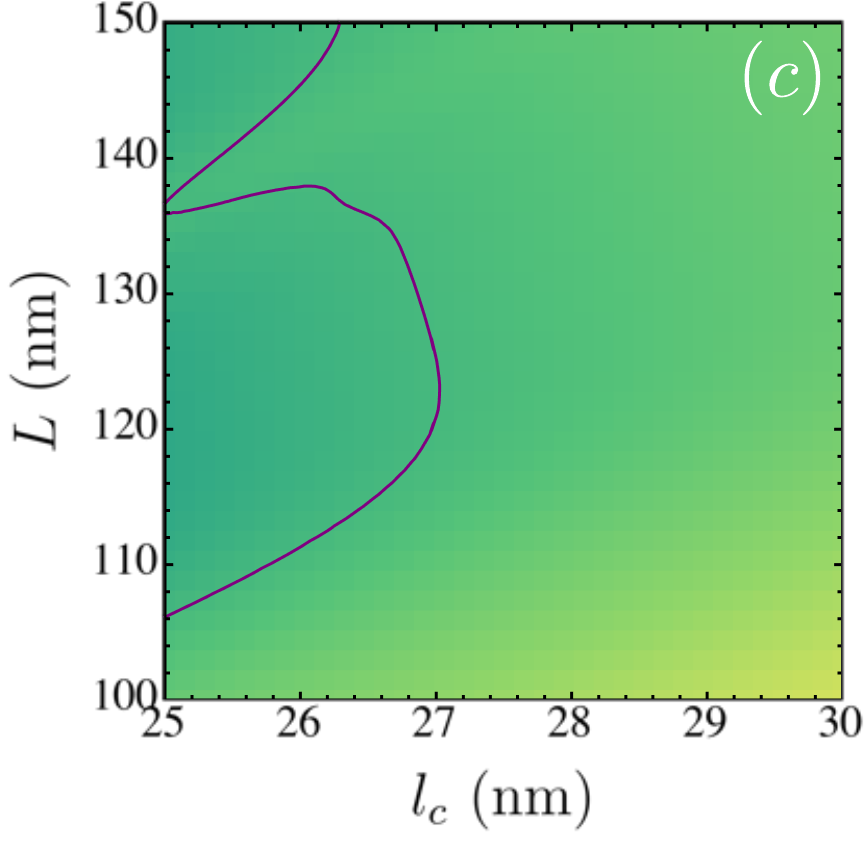}
	\includegraphics[width=0.49\linewidth]{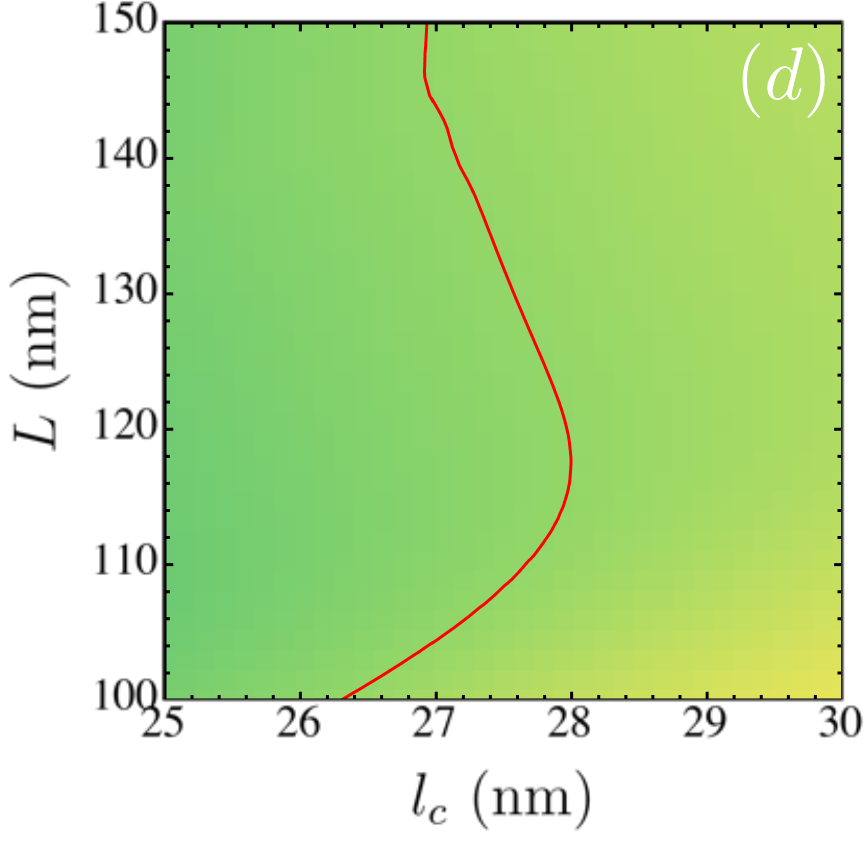}
        \centering
        \includegraphics[width=0.8\linewidth]{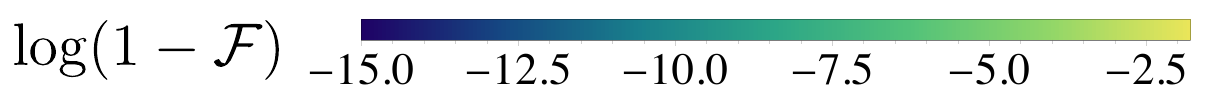} 
	\caption{Phonon-induced infidelity of a $\pi$-phase gate with variable $l_c$ and $L$ at (a) $\unit[200]{mK}$ (b) $\unit[400]{mK}$ (c) $\unit[600]{mK}$ (d) $\unit[800]{mK}$ and constant $\tau_\pi=\unit[50]{ns}$. The contours show the boundaries at which characteristic infidelity minima may be achieved at each respective temperature. The boundaries are given at $10^{-13}$ (cyan), $10^{-11}$ (black), $10^{-7}$ (purple) and $10^{-4}$ (red).}
\label{fig:lc_vs_L} \end{figure}

Lastly, for completeness, the phonon-induced infidelities of exchange $\pi$-pulses of time $\tau_\pi=\unit[50]{ns}$ are shown at various temperatures as a function of both $L$ and $l_c$ in Fig.~\ref{fig:lc_vs_L}. Here a combination of trends for varying $L$ and $l_c$ individually can be seen, however at low temperatures ($\unit[200]{mK}$) a clear minimum in gate infidelity is observed at a ratio $L/l_c\sim 5.3$. This minimum in infidelity is preceded by a peak in infidelity at $L/l_c\sim 5.6$ which corresponds to the low temperature $\tilde{P}_{SR}$ resurgence discussed for Fig.~\ref{fig:DotDistanceE} and Fig.~\ref{fig:ConfLengthE}. However, as temperature is increased and the $P_e$ phonon interaction terms become dominant, the ratio of $L/l_c$ at which an infidelity minimum is observed decreases and broadens, generally favoring a DQD system with a smaller $l_c$ and therefore larger $\Delta E$.

\subsection{Ramped Exchange Pulse} \label{sec:RampedGate}

\begin{figure}
	[t] 
	\raggedright
	\includegraphics[width=0.49\linewidth]{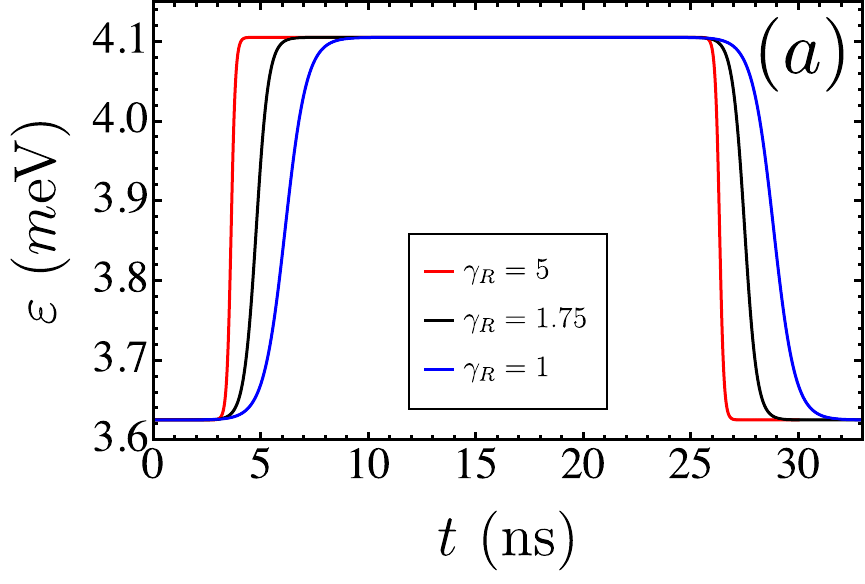} 
	\includegraphics[width=0.49\linewidth]{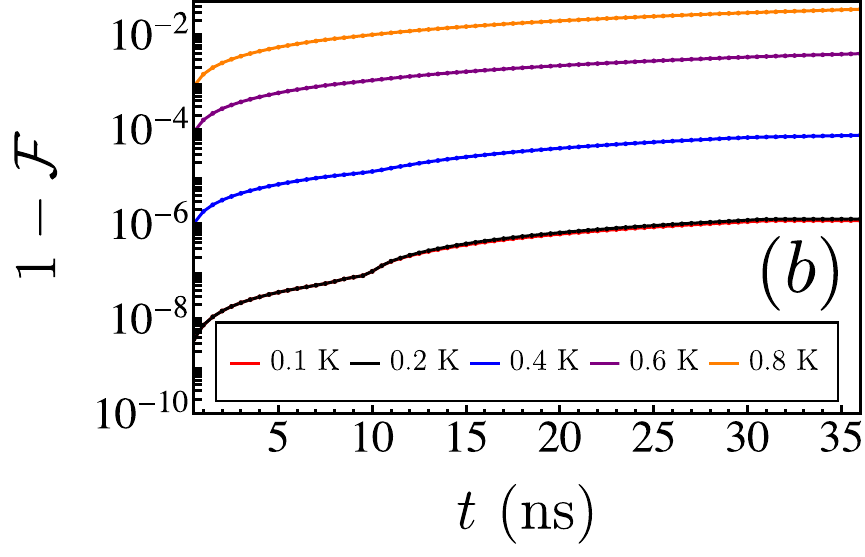} 
	\includegraphics[width=0.49\linewidth]{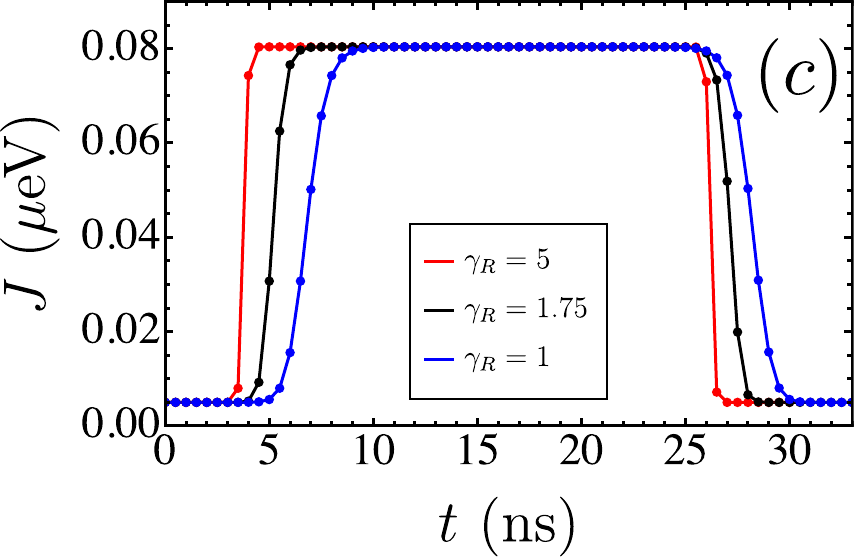} 
	\includegraphics[width=0.49\linewidth]{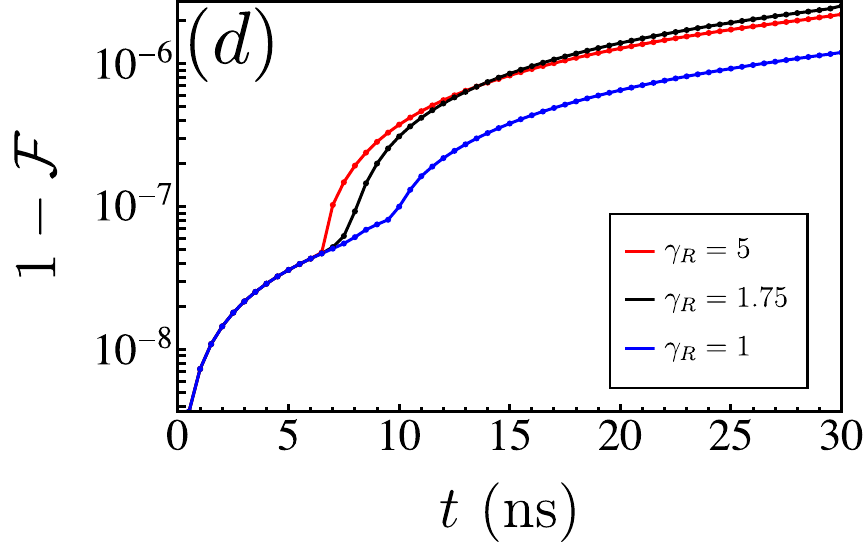} 
	\caption{Spin-phonon induced errors of ramped square wave detuning pulses. (a) The smooth ramping of square wave detuning $\pi$-pulses of varying sharpness given by the constant $\gamma_R$. (b) Time dependent spin-phonon induced infidelity of the pulse $\gamma_R=1$ at temperatures from $\unit[100]{mK}$ to $\unit[800]{mK}$. (c) The exchange interaction with time of each of the tested ramped square wave pulses. (d) A comparison of the variably smooth pulse shapes at $\unit[200]{mK}$ demonstrating the resistance of the smoother pulse shapes to spin-phonon induced errors.} 
\label{fig:RampedPulses} \end{figure}

Thus far, all exchange pulses considered have been perfect square wave detuning pulses. A next natural step in the characterization of phonon induced errors during exchange gates on DQD spin qubits is to simulate the effect of ramping the QD detuning from some idle detuning $\varepsilon_0$ where the exchange interaction is small relative to some active gate detuning $\varepsilon_g$ where the desired exchange interaction is achieved, similar to what is experimentally done. For a time dependent ramping model, the following piece-wise pulse is used
\begin{equation}
	\varepsilon (t) = 
	\begin{cases}
		\varepsilon_0 & t\leq 0, \\
		\frac{\varepsilon_g+\varepsilon_0+(\varepsilon_g-\varepsilon_0)\tanh \left[\gamma_R t -\pi \right]}{2} & 0\leq t\leq \frac{\tau_g}{2}, \\
		\frac{\varepsilon_g+\varepsilon_0+(\varepsilon_g-\varepsilon_0)\tanh \left[-\gamma_R (t-\tau_g) +\pi \right]}{2} & \frac{\tau_g}{2}\leq t 
	\end{cases}
	\label{eq:RmpdSQWV} \end{equation}

\noindent where $\gamma_R$ is a constant that determines the steepness of the ramped tails of the pulse and $\tau_g$ is the total gate time, in the case of a $\pi$-exchange pulse. This form of time dependent detuning is equivalent to a smooth ramped square wave with a single parameter $\gamma_R$ to tune its sharpness. As can be seen in Fig.~\ref{fig:RampedPulses}(a) as $\gamma_R$ is decreased, the smoother the ramping of $\varepsilon(\tau)$ to $\varepsilon_g$.

The evolution of the Lindbladian of a gate with a time dependent pulse, and therefore time dependent Hamiltonian, was treated in a step-wise manner, with a new set decay rates calculated at each step. In Fig.~\ref{fig:RampedPulses}, this is done with a time step of $\unit[0.5]{ns}$, which was found to be sufficient for a smooth evolution of all the relevant processes during the pulse. From the time dependent simulations, two interesting features emerge. Firstly, at temperatures around $\unit[100-200]{mK}$ there is a clear and smooth increase in the spin-phonon coupling at the exchange maximum of the pulse. This, can be seen Fig.~\ref{fig:RampedPulses}(b), demonstrating a vulnerability of DQD qubits to spin-phonon coupling during exchange gates. Secondly, as the pulse becomes smoother the effect of the spin-phonon induced errors during the exchange gate is suppressed, demonstrating that simple pulse shaping is a viable tool for spin-phonon error mitigation. This is demonstrated in Fig.~\ref{fig:RampedPulses}(d).

The reason for the improved gate fidelity with the smoother pulse ramping is given by the dominance of the on-diagonal $\tilde{P}_{SR}$ term. In Eq.~(\ref{eq:H5x5}) this term populates the same matrix element of the effective Hamiltonian as the pulsed detuning $\varepsilon (t)$. This can be shown analytically, if a Schrieffer-Wolff transformation of Eq.~(\ref{eq:H5x5}) is done to isolate the qubit subspace as in Ref.~\cite{kornich2018phonon}, assuming the $\delta b_z$, $t_c$ and all spin-phonon coupling elements to be small. To third order, the transformed Hamiltonian $\tilde{H}_{SW}$ may be written as follows
\begin{equation}
	\tilde{H}_{SW} = \tilde{H}_{0}+\tilde{H}_{1P}+\tilde{H}_{2P}, 
\end{equation}

\noindent where $\tilde{H}_{0}$ is the phonon independent Hamiltonian, $\tilde{H}_{1P}$ is the single-phonon process Hamiltonian and $\tilde{H}_{2P}$ is the two-phonon process the Hamiltonian. These transformed spin-phonon coupling Hamiltonians are as follows
\begin{subequations}
	\begin{equation}
		\tilde{H}_{1P}=\frac{\tilde{P}_{SR} t_c^2}{(\varepsilon(t)-U+\Delta V)^2}\sigma_z=\Gamma_{1P}(t)\tilde{P}_{SR}\sigma_z, 
	\label{eq:SW1P} \end{equation}
	\begin{equation}
		\begin{split}
			\tilde{H}_{2P}=&\frac{\delta b_z \Delta V^2 P_e P_e^\dagger}{4 \Delta E_{\text{orb}}^2(\Delta E_{\text{orb}}-\Delta V)^2}\sigma_x+\\&\frac{ \Delta V P_e P_e^\dagger}{\Delta E_{\text{orb}}(\Delta E_{\text{orb}}-\Delta V)}\sigma_z, 
		\end{split}
		\label{eq:SW2P} \end{equation}
\end{subequations}

\noindent where $\sigma_i$ is the $i^{\text{th}}$ Pauli matrix acting in the qubit subspace. Here the contribution of the $P_s^{(\dagger)}$ terms has been omitted due to their relatively small contribution. Whats notable about these two spin-phonon coupling Hamiltonians is that in the single-phonon process Hamiltonian, the $\tilde{P}_{SR}$ term of Eq.~(\ref{eq:SW1P}) is dependent on $\varepsilon (t)$ while the two-phonon process Hamiltonian, the $P_{e}^{(\dagger)}$ terms of Eq.~(\ref{eq:SW2P}) are not. The prior term accounts for the effect of the pulse shape on the spin-phonon induced infidelity at lower temperatures, while the latter accounts for the pulse shape independent behavior seen at higher temperature, as can be seen in Fig.~\ref{fig:RampedPulses}(b) at $T>\unit[500]{mK}$. This time dependent coupling term is labeled $\Gamma_{1P}(t)$=

The time dependent scaling term of Eq.~(\ref{eq:SW1P}) can directly be optimized to ensure protection against spin-phonon coupling at $T<\unit[500]{mK}$. During an exchange gate evolution defined as the unitary
\begin{equation}
	\mathcal{U}=e^{-\frac{i\hbar}{2}\sigma_z\int_0^{\tau_g} dt J(t)}, 
\end{equation}

\noindent acting in the qubit space, where $J(t)$ is the time dependent exchange pulse, the single spin-phonon errors given in Eq.~(\ref{eq:SW1P}) may be minimized by selecting a pulse such that the integral
\begin{equation}
	\int_0^{\tau_g} dt \tilde{H}_{1P}(t)\propto\int_0^{\tau_g} dt \frac{ t_c^2}{(\varepsilon(t)-U+\Delta V)^2}, 
\label{eq:minint} \end{equation}

\noindent is minimized. This gives a natural advantage to smoother pulses.

\begin{figure}
	[b] 
	\raggedright
	\includegraphics[width=0.49\linewidth]{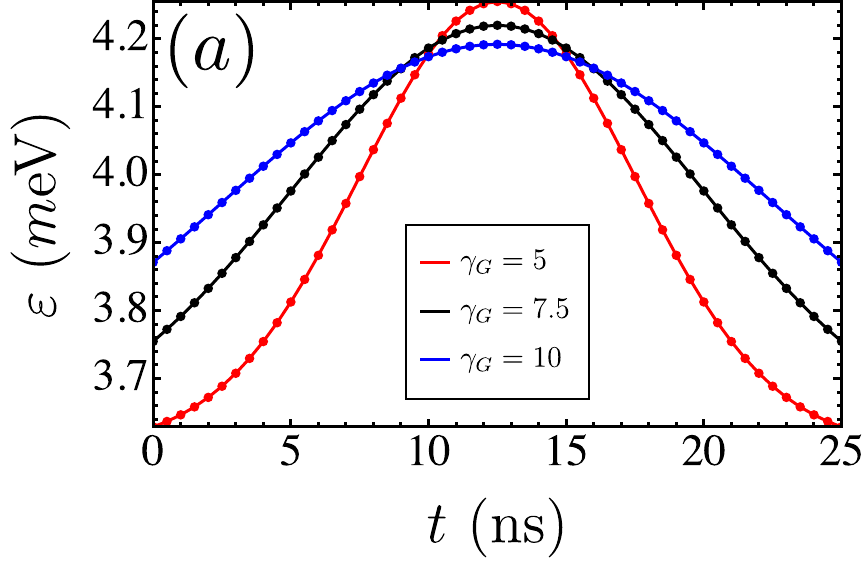} 
	\includegraphics[width=0.49\linewidth]{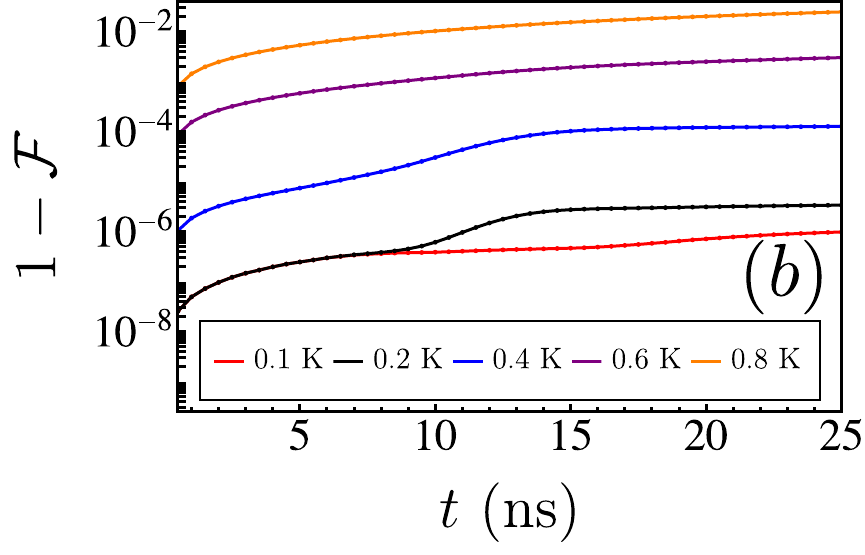} 
	\includegraphics[width=0.49\linewidth]{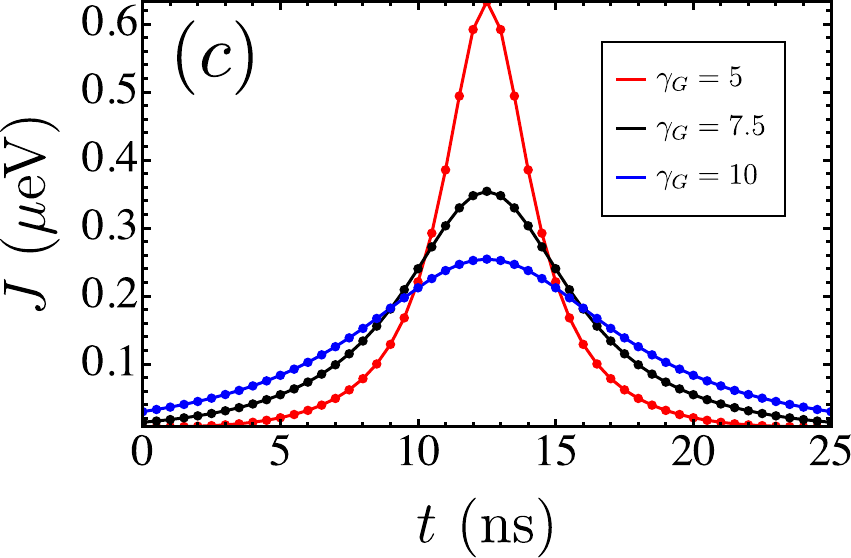} 
	\includegraphics[width=0.49\linewidth]{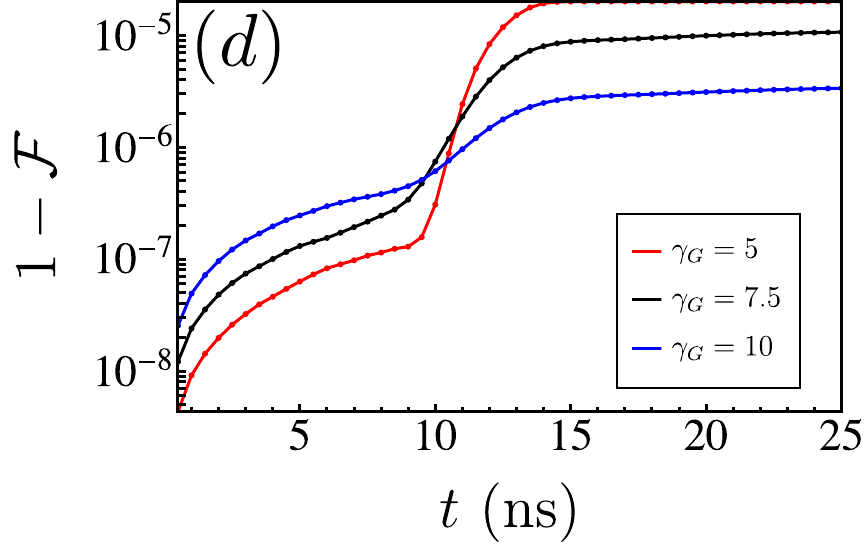} 
	\caption{Spin-phonon induced errors of Gaussian detuning pulses. (a) The Gaussian $\pi$-pulses of varying sharpness given by the constant $\gamma_G$. (b) Time dependent spin-phonon induced infidelity of the pulse $\gamma_G=10$ at temperatures from $\unit[100]{mK}$ to $\unit[800]{mK}$. (c) The exchange interaction with time of each of the tested Gaussian pulses. (d) A comparison of the variably smooth pulse shapes at $\unit[200]{mK}$ demonstrating the resistance of the smoother pulse shape $\gamma_G=10$ to spin-phonon induced errors.} 
\label{fig:GausPulses} \end{figure}

\subsection{Gaussian Exchange Pulse} \label{sec:GaussGate}

To further probe the result in Fig.~\ref{fig:RampedPulses}(d) demonstrating an advantage of smoother pulse shapes in robustness from spin-phonon induced errors, Gaussian pulse shapes are also tested. Here the detuning pulse shape is given as
\begin{equation}
	\varepsilon (t) = (\varepsilon_g-\varepsilon_0) e^\frac{-(t-\tau_g/2)^2}{4 \gamma_G^2}, 
\end{equation}

\noindent where $\gamma_G$ is some constant of the sharpness of the Gaussian pulse used, inversely to the ramped square wave of Eq.~(\ref{eq:RmpdSQWV}), as $\gamma_G$ increases, so too does the smoothness of the pulse. Here, unlike in the ramped pulses, to insure a constant gate comparison, $\varepsilon_0$ and $\varepsilon_g$ are varied slightly with $\gamma_G$, as is seen in Fig.~\ref{fig:GausPulses}(a). This is such that the integral $\int_0^{\tau_g}dt J(\varepsilon,t)=\pi$ is constant over the bounded time for each pulse tested, resulting in the corresponding exchange pulses given in Fig.~\ref{fig:GausPulses}(c). Fig.~\ref{fig:GausPulses}(b) shows shows the time dependent infidelities of the $\gamma_G=10$ exchange pulse at varying temperatures. A similar but smoother behavior in time and temperature is observed for these Gaussian exchange pulses compared to the ramped square wave pulses in Fig.~\ref{fig:RampedPulses}(b). Fig.~\ref{fig:GausPulses}(d) shows the effect of $\gamma_G$ on the overall infidelity at a fixed temperature $\unit[200]{mK}$. Here is it clear that the smoother pulses in Fig.~\ref{fig:GausPulses}(c) corresponding to larger values of $\gamma_G$ result in more robust gates.

\begin{figure}
	[t] 
	\raggedright
	\includegraphics[width=0.49\linewidth]{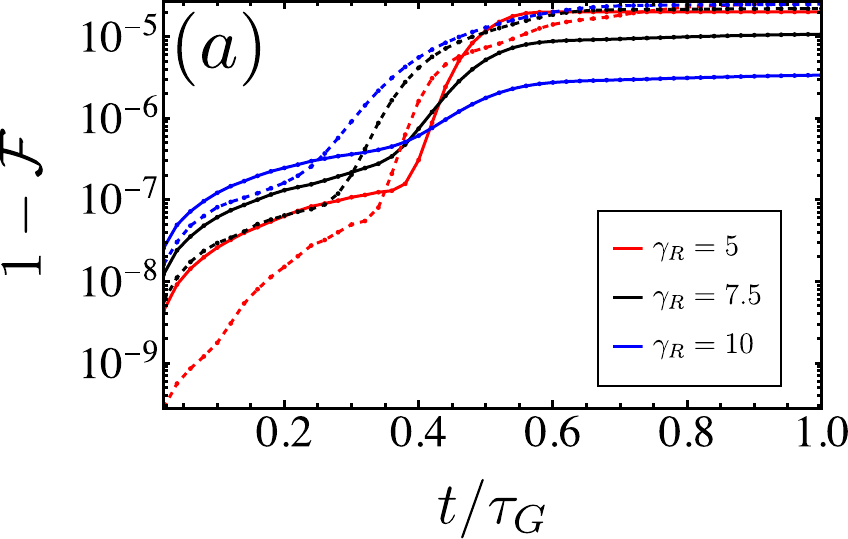} 
	\includegraphics[width=0.49\linewidth]{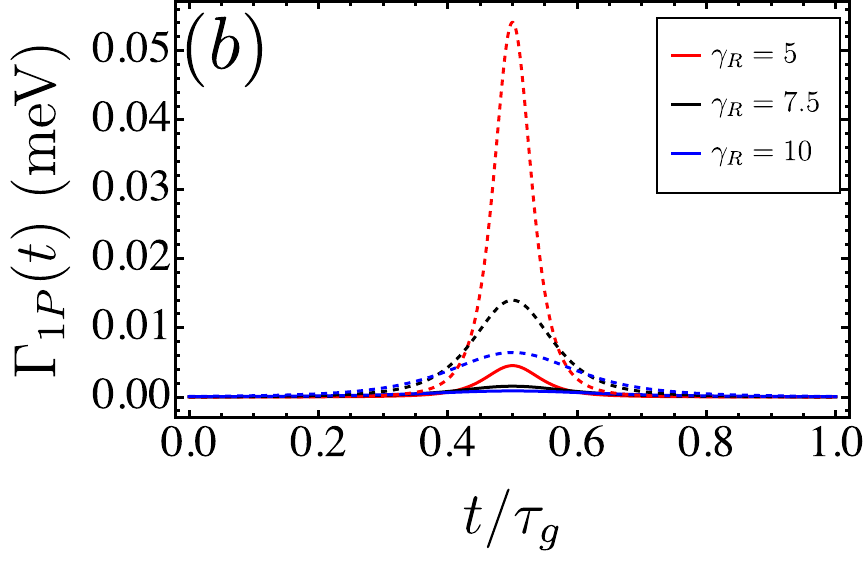} 
	\caption{(a)Comparison of the spin-phonon infidelity benefit of smooth Gaussian pulses for $\tau_g=\unit[25]{ns}$ (solid) and $\tau_g=\unit[10]{ns}$ (dashed). (b) The associated time dependent coupling term $\Gamma_{1P}(t)$ to the dominant spin-phonon process, demonstrating significantly higher coupling for the shorter pulse despite its smoothness, which results in larger values for the integral given in Eq.~(\ref{eq:minint}).} 
\label{fig:Gaus_Compare} \end{figure}

There is, of course, a trade-off in the calculated gate fidelity gains by smoothing the pulses, depending on the overall gate time. This is shown in Fig.~\ref{fig:Gaus_Compare}(a), demonstrating that the overall gate fidelity gains of smooth Gaussian pulses is hampered at short ($\tau_\pi=\unit[10]{ns}$) pulses. This is due to the larger spin-phonon coupling peak, and so an overall larger integral over the gate time as in Eq.~(\ref{eq:minint}). This is shown in Fig.~\ref{fig:Gaus_Compare}(b).

\section{Exchange Gates at the Symmetric Operating Point} \label{sec:SOP}

It is common to control the exchange interaction in a DQD system by varying the detuning of the two dots. However, it is equally valid to vary the exchange instead by directly varying the overlap of the electron wavefunctions as to keep the detuning of the two dots $\sim \unit[0]{meV}$. This is know as the symmetric operating point (SOP), where the qubit is protected against first order charge noise and so is a desirable method of operation\cite{reed2016reduced}. At the SOP the exchange is controlled by varying the barrier plunger between the two dots. In our model, the tunnel barrier cannot be directly addressed. Instead, exchange pulses at the SOP will be simulated by modulating the confinement length $l_c$ of both dots simultaneously at $\varepsilon=\unit[0]{meV}$ to implement gates instead of modulating $\varepsilon$.  This is done at fixed $L=\unit[150]{nm}$ such that the resulting calculated infidelities may be compared to the equivalent system where exchange is pulsed by varying $\varepsilon$ as shown in Fig.~\ref{fig:SquTDep}. This approximation gives a good idea of the effect of spin-phonon coupling as the tunnel barrier of a DQD system is varied.

\begin{figure}
	[t] 
	\raggedright
	\includegraphics[width=0.49\linewidth]{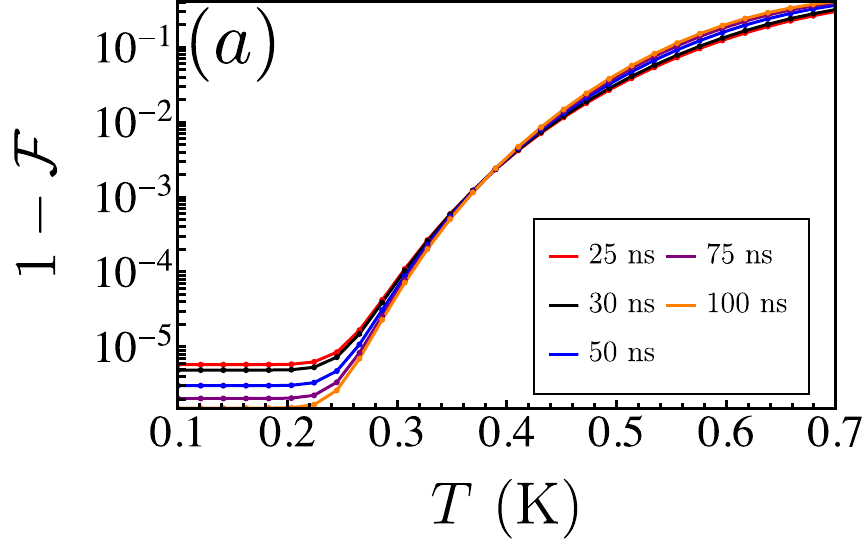} 
	\includegraphics[width=0.49\linewidth]{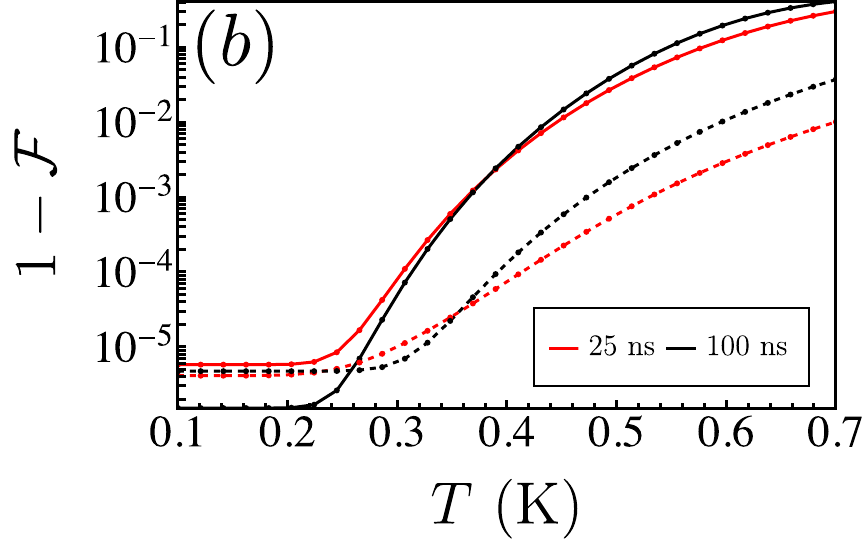} 
	\caption{Spin-phonon induced errors of exchange pulses at the SOP. (a) Phonon-induced infidelity of a $\pi$-phase gate with temperature for a range of square exchange pulse lengths around the SOP by varying $l_c$ at $L=\unit[150]{nm}$} demonstrating a narrower spread and greater susceptibility to phonon-induced errors at higher ($T>\unit[300]{mK}$) compared to similar exchange pulses by detuning. (b) Comparison of the susceptibility of exchange pulses to phonon induced errors around the SOP (solid) vs detuning pulses (dashed), demonstrating the sharp drop off in the benefit of pulsing around the SOP at $T>\unit[300]{mK}$. 
\label{fig:SOP_Compare} \end{figure}

The effect of spin-phonon interactions on SOP exchange gate fidelities are show in Fig.~\ref{fig:SOP_Compare}(a). Here, a similar decay in $\pi$-pulse gate fidelities as a function of gate times is observed as in the detuning pulse case, with two notable differences. At low temperatures $<\unit[200]{mK}$ the longer SOP pulses are more up to an order of magnitude resilient to spin-phonon interactions compared to equivalent to detuning pulses. This is due to the charging energies of the dots dampening the effect of $U_c+\tilde{P}_{SR}$ when $\varepsilon=0$ in Eq.~\ref{eq:H5x5}. However, at temperatures $>\unit[200]{mK}$ there is a sharp increase in the calculated infidelity for all gate times compared to detuning pulses, as is shown in Fig.~\ref{fig:SOP_Compare}(b). This is due to a much more pronounced influence of the $P_e$ spin-phonon coupling term with temperature, as given by the varying of orbital energy difference $\Delta E_{\text{orb}}$ as $l_c$ is pulsed. While this effect may be somewhat exaggerated by the simulated method of SOP exchange operations, i.e. modulating the $l_c$ of the two dots as opposed to a tunnel barrier gate, any method of controlling exchange that also compromises $\Delta E_{\text{orb}}$ is expected to be limited at high temperature due to spin-phonon coupling.
 
\section{Gate Leakage} \label{sec:Leakage}

An added benefit of the master equation approach employed here is the inclusion of states outside of the computation space, from which spin-phonon induced leakage can be calculated. The calculated results have been presented as the evolution of a single encoded singlet-triplet qubit, a qubit type susceptible to leakage to spin states outside of this subspace, such as the $T_\pm$ states. Here the Hamiltonian is extended from Eq.~(\ref{eq:H5x5}) to include the $T_\pm$ states and their orbital excited states, however the method of calculating the evolved gate from which spin-phonon induced leakage can be deduced is identical except for a larger Hilbert space to that of the gate infidelity calculations. The full Hilbert space considered is $\{\ket{T_0},\ket{S},\ket{T_\pm},\ket{S_{R/L}},\ket{T_0^*},\ket{S^*},\ket{T_\pm^*}\}$. Additionally, a magnetic field gradient term $\delta b_x$ coupling the computational $S/T_0$ spin states to the $T_\pm$ states is also included. Tuning this term will amplify the calculated spin-phonon interaction induced leakage. Leakage is quantified as
\begin{figure}
	[t] 
	\raggedright
	\includegraphics[width=0.49\linewidth]{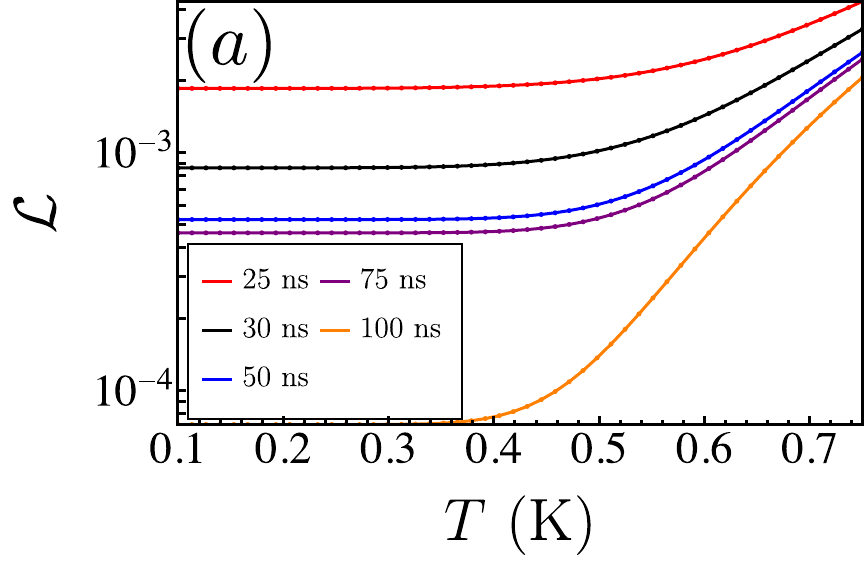} 
	\includegraphics[width=0.49\linewidth]{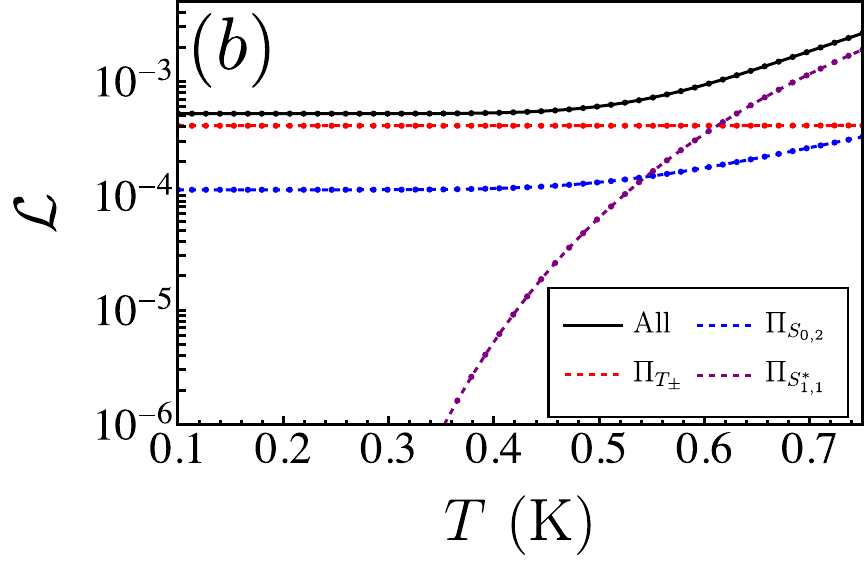} 
	\caption{(a) Total Spin-phonon induced spin-state leakage as a function of temperature during of detuning controlled exchange $\pi$-pulses of lengths varying from $\unit[25-100]{ns}$. Here conversely to the overall gate fidelity, the longer pulses are more resilient to spin-state leakage as a function of temperature. (b) The Spin-phonon induced spin-state leakage of the $\unit[50]{ns}$, broken down into the contributions from all included spin states outside the logical subspace. As with the spin-phonon induced infidelity, at high temperatures excitations to higher orbital states dominates. Here $\delta b_x=\unit[100]{mT}$.} 
\label{fig:LeakPlot} \end{figure}

\begin{equation}
	\mathcal{L}=1-\Tr(\Pi_{Q}\rho), 
\end{equation}

\noindent where $\Pi_{Q}$ is the projector onto the qubit subspace. Excitations to higher orbital states are also treated as leakage here. Although this is not strictly spin-state leakage, as the spin component is unaffected, the device behavior will change uncontrollably with orbital excitations, and so is equally undesirable.

\begin{figure}
	[t]
	\raggedright
	\includegraphics[width=0.9\linewidth,left]{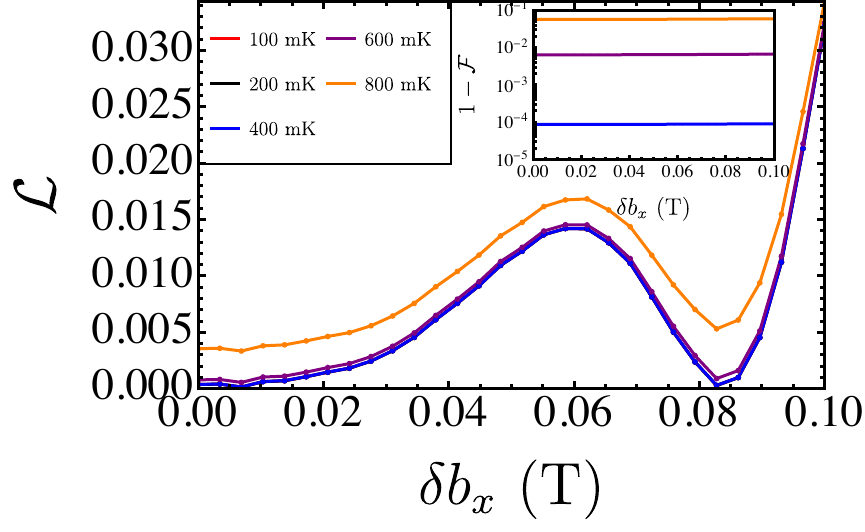} 
	\caption{Total spin state leakage as a function of spin mixing magnetic field gradient $\delta b_x$ of a $\unit[50]{ns}$ detuning controlled exchange $\pi$-pulse. Here the spin-phonon coupling offers only an offset in the leakage behavior as temperature is increased. This is mirrored by the inset, where the pulse infidelity remains static in $\delta b_x$ relative to an equivalent pulse without spin-phonon coupling. The oscillations in the leakage are due to $S-T_\pm$ Rabi oscillation resonances in $\delta b_x$ and the selected $\unit[50]{ns}$ pulse time.} 
\label{fig:LeakdB} \end{figure}

In Fig.~\ref{fig:LeakPlot}(a), the spin-phonon induced leakage as a function of temperature for a selection of exchange pulse lengths is given. Interestingly, unlike in similar calculation of the overall state fidelity of the exact same pulses as given in Fig.~\ref{fig:SquTDep}, at higher temperatures, the longer exchange pulses are more resilient to leakage, despite the overall gate fidelity being worse. This demonstrates that leakage is not the primary error caused by spin-phonon interaction, but due to the DQD Hamiltonian itself. Fig.~\ref{fig:LeakPlot}(b) shows that leakage broken down into its spin state components. Leakage to the $T_\pm$ states is mostly static in temperature, until $>\unit[500]{mK}$, where spin-phonon induced leakage to excited orbital states dominates, and spin-phonon charge scattering to the $S_{R/L}$ state starts to scale with temperature, inhibiting coupling to the $T_\pm$ states.

The lack of impact of spin-phonon interactions on the calculated exchange pulse leakage is also shown in Fig.~\ref{fig:LeakdB}. The added magnetic field gradient $\delta b_x$ can be seen as the dominant term on induced leakage, with spin-phonon interactions offering only an offset in the total behavior with temperature. This DQD device dependent leakage of exchange operations implies that the calculated exchange gate infidelities remain addressable with conventional quantum error correcting schemes, i.e. ones only concerned with the qubit subspace, as temperatures are increased, as induced errors remain within the spin-space albeit not always within the groundstate orbital.

\section{Experimental Signatures and Considerations} \label{sec:Experiment}
\begin{figure}
	[b] 
	\includegraphics[width=0.9\linewidth,left]{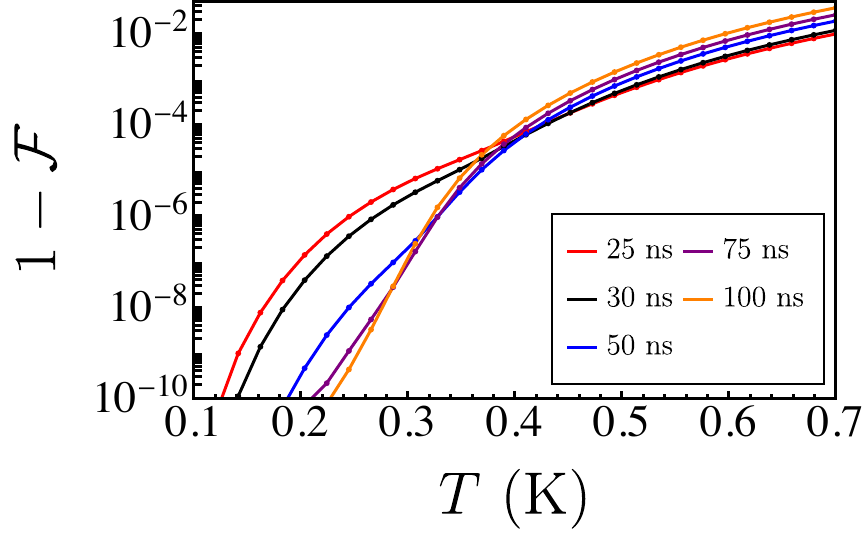} 
	\caption{Phonon-induced infidelity of a $\pi$-phase gate as a function of temperature for a range of square exchange pulse lengths relative to the same gate at $\unit[20]{mK}$.} 
\label{fig:TDep20mk} \end{figure}

Thus far all gate infidelities shown have been calculated demonstrating the effect of spin-phonon interactions compared to the the evolution of the same Hamiltonian without spin-phonon interactions. However, even as temperature tends to $\unit[0]{K}$ spin-phonon interaction, particularly phonon emission, cannot be switched off in experiment. Prior examples offer an insight into the exact effect of phonons on a DQD system, however these signatures may not be directly measured experimentally. For a better comparison with what would be observed experimentally, instead of calculating infidelity relative to no spin-phonon interaction, the infidelity relative to some achievable small temperature of $\unit[20]{mK}$ may be calculated. This is shown in Fig.~\ref{fig:TDep20mk}, demonstrating gate infidelity of perfect square wave pulses of varying $\tau_\pi$ as a function of temperature. The main difference between Fig.~\ref{fig:TDep20mk} and Fig.~\ref{fig:SquTDep}, which show the same relationship in infidelity relative to no spin-phonon interaction, is the normalization of the temperature independent shift due to the $\tilde{P}_{SR}$ term. However, the main signature of the behavior, the crossover in gate fidelity from long to short pulses at $\sim\unit[300]{mK}$ due to the dominance of the $P_e$ over $\tilde{P}_{SR}$ matrix elements remains. All such crossovers observed when calculating infidelity relative to no spin-phonon interactions are expected to remain a valuable experimental signature of high temperature exchange gates. Additionally, the temperature dependent behavior below the crossover showing negligibly small gate infidelities are likely to be washed out experimentally by other noise sources such as charge noise which can also be temperature dependent\cite{petit2018spin}.

\begin{figure}
	[t]
    \centering
	\includegraphics[width=0.9\linewidth,left]{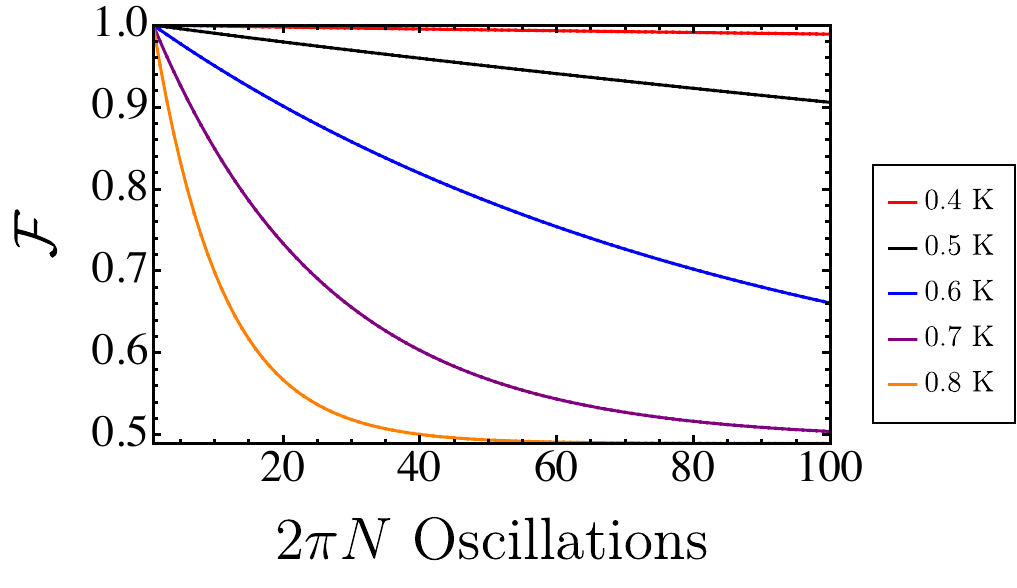} 
	\caption{Fidelity of a number $N$ of full $2\pi$ Rabi oscillations with spin-phonon coupling with temperature. Here a detuning pulse equivalent to a $\pi$-pulse of $\tau_\pi=\unit[25]{ns}$ is assumed.} 
\label{fig:Rabi} \end{figure}

Another useful experiment to determine the impact of finite temperature spin-phonon interactions on gate operations is to observe the state fidelity of DQD undergoing exchange operations with the number of full Rabi oscillations. A simulation to this effect is given in Fig.~\ref{fig:Rabi}. Here a DQD is evolved at a detuning equivalent to a $\pi$-pulse of $\tau_\pi=\unit[25]{ns}$ and the state fidelity as a function of the total number of full $2\pi$ Rabi oscillations is given. A clear exponential decay in fidelity is observed, with a $\sim T^4$ sensitivity to temperature as is expected from previous dephasing calculations\cite{kornich2018phonon}.

Although Fig.~\ref{fig:TDep20mk} and Fig.~\ref{fig:Rabi} demonstrate possible experimental signatures of the effects discussed throughout, there are other experimental obstacles to overcome when operating high temperature spin qubits. This includes the diminishing Pauli spin blockade (PSB) readout visibility, due to the broadening of the single electron transistor (SET) peaks. However, this can be somewhat circumvented by alternate post-processing of the SET signal\cite{petit2020universal}. Additionally, alternative SET architectures show in SiMOS devices for improved resolution at high temperatures may be employed on a SiGe chip\cite{huang2021high}.

\section{Conclusion} \label{sec:Conclusion}

Here an extensive study of the effects of spin-phonon interaction of exchange gate fidelities in a Si-SiGe heterostructure DQD system has been provided. Our results suggests that qubit operations around $\unit[200-300]{mK}$ are theoretically viable, impacting gate fidelities by at most $10^{-5}-10^{-4}$. This is consistent with recent experimental results\cite{undseth2023hotter}. Our results show a distinct crossover in the dominant infidelity inducing spin-phonon matrix elements as a function of temperature, the point of which is determined by length of the exchange pulse used. Pulse shapes were also discussed, showing that smoother pulses tend to provided some level of resilience against the spin-phonon coupling during a gate. However, the fidelity gains from smoother pulses were also dependent on the length and maximum exchange energy of the implemented gate, giving a lower limit in pulse time to which smoother pulses offer a greater fidelity. Expected leakage errors due to spin-phonon interactions are also probed with the master equation approach, demonstrating a resilience to relatively high temperatures. This implies and a compatibility with hot Si-SiGe qubits and conventional quantum error correcting codes. Some examples of possible experimental signatures of the discussed behavior are given, showing that the discussed temperature dependent crossover may be observed, however the calculated behavior at $<\unit[200]{mK}$ is small enough to not be resolvable. Finally, although these calculations were done for SiGe spin qubits, similar results are expected for SiMOS spin qubits given equivalent planar confinement potentials.

A natural next step in extending these calculations is the inclusion of valley states. The addition of a valley degree of freedom would greatly increase the Hilbert space,  but would offer further insight into the possible spin-phonon error channels. Ultimately, when focusing on the lowest energy excitations, a model including valleys would resemble that of Eq.~(\ref{eq:H5x5}), and so some similar behavior as that investigated here is expected. Additionally, a thermalized phonon bath has been assumed throughout. While this is a helpful assumption, there is scope for the investigation of other phonon phenomena that could limit spin qubit operations. These include driven phonon sources due to the back action of a nearby single electron transistor used for spin readout.

\section{Acknowledgments} \label{ref:Acknowledgements}

We acknowledge helpful discussions with U. G\"ung\"ord\"u, A. Little, A. Mills, R. Ruskov, and YP. Shim.

\bibliography{HotGate}

\begin{thebibliography}{43}%
\makeatletter
\providecommand \@ifxundefined [1]{%
 \@ifx{#1\undefined}
}%
\providecommand \@ifnum [1]{%
 \ifnum #1\expandafter \@firstoftwo
 \else \expandafter \@secondoftwo
 \fi
}%
\providecommand \@ifx [1]{%
 \ifx #1\expandafter \@firstoftwo
 \else \expandafter \@secondoftwo
 \fi
}%
\providecommand \natexlab [1]{#1}%
\providecommand \enquote  [1]{``#1''}%
\providecommand \bibnamefont  [1]{#1}%
\providecommand \bibfnamefont [1]{#1}%
\providecommand \citenamefont [1]{#1}%
\providecommand \href@noop [0]{\@secondoftwo}%
\providecommand \href [0]{\begingroup \@sanitize@url \@href}%
\providecommand \@href[1]{\@@startlink{#1}\@@href}%
\providecommand \@@href[1]{\endgroup#1\@@endlink}%
\providecommand \@sanitize@url [0]{\catcode `\\12\catcode `\$12\catcode `\&12\catcode `\#12\catcode `\^12\catcode `\_12\catcode `\%12\relax}%
\providecommand \@@startlink[1]{}%
\providecommand \@@endlink[0]{}%
\providecommand \url  [0]{\begingroup\@sanitize@url \@url }%
\providecommand \@url [1]{\endgroup\@href {#1}{\urlprefix }}%
\providecommand \urlprefix  [0]{URL }%
\providecommand \Eprint [0]{\href }%
\providecommand \doibase [0]{https://doi.org/}%
\providecommand \selectlanguage [0]{\@gobble}%
\providecommand \bibinfo  [0]{\@secondoftwo}%
\providecommand \bibfield  [0]{\@secondoftwo}%
\providecommand \translation [1]{[#1]}%
\providecommand \BibitemOpen [0]{}%
\providecommand \bibitemStop [0]{}%
\providecommand \bibitemNoStop [0]{.\EOS\space}%
\providecommand \EOS [0]{\spacefactor3000\relax}%
\providecommand \BibitemShut  [1]{\csname bibitem#1\endcsname}%
\let\auto@bib@innerbib\@empty
\bibitem [{\citenamefont {Burkard}\ \emph {et~al.}(2023)\citenamefont {Burkard}, \citenamefont {Ladd}, \citenamefont {Pan}, \citenamefont {Nichol},\ and\ \citenamefont {Petta}}]{burkard2021semiconductor}%
  \BibitemOpen
  \bibfield  {author} {\bibinfo {author} {\bibfnamefont {G.}~\bibnamefont {Burkard}}, \bibinfo {author} {\bibfnamefont {T.~D.}\ \bibnamefont {Ladd}}, \bibinfo {author} {\bibfnamefont {A.}~\bibnamefont {Pan}}, \bibinfo {author} {\bibfnamefont {J.~M.}\ \bibnamefont {Nichol}},\ and\ \bibinfo {author} {\bibfnamefont {J.~R.}\ \bibnamefont {Petta}},\ }\bibfield  {title} {\bibinfo {title} {Semiconductor spin qubits},\ }\href {https://journals.aps.org/rmp/abstract/10.1103/RevModPhys.95.025003} {\bibfield  {journal} {\bibinfo  {journal} {Reviews of Modern Physics}\ }\textbf {\bibinfo {volume} {95}},\ \bibinfo {pages} {025003} (\bibinfo {year} {2023})}\BibitemShut {NoStop}%
\bibitem [{\citenamefont {Petit}\ \emph {et~al.}(2022{\natexlab{a}})\citenamefont {Petit}, \citenamefont {Russ}, \citenamefont {Eenink}, \citenamefont {Lawrie}, \citenamefont {Clarke}, \citenamefont {Vandersypen},\ and\ \citenamefont {Veldhorst}}]{petit2020high}%
  \BibitemOpen
  \bibfield  {author} {\bibinfo {author} {\bibfnamefont {L.}~\bibnamefont {Petit}}, \bibinfo {author} {\bibfnamefont {M.}~\bibnamefont {Russ}}, \bibinfo {author} {\bibfnamefont {G.~H.}\ \bibnamefont {Eenink}}, \bibinfo {author} {\bibfnamefont {W.~I.}\ \bibnamefont {Lawrie}}, \bibinfo {author} {\bibfnamefont {J.~S.}\ \bibnamefont {Clarke}}, \bibinfo {author} {\bibfnamefont {L.~M.}\ \bibnamefont {Vandersypen}},\ and\ \bibinfo {author} {\bibfnamefont {M.}~\bibnamefont {Veldhorst}},\ }\bibfield  {title} {\bibinfo {title} {Design and integration of single-qubit rotations and two-qubit gates in silicon above one kelvin},\ }\href {https://www.nature.com/articles/s43246-022-00304-9} {\bibfield  {journal} {\bibinfo  {journal} {Communications Materials}\ }\textbf {\bibinfo {volume} {3}},\ \bibinfo {pages} {82} (\bibinfo {year} {2022}{\natexlab{a}})}\BibitemShut {NoStop}%
\bibitem [{\citenamefont {Vandersypen}\ \emph {et~al.}(2017)\citenamefont {Vandersypen}, \citenamefont {Bluhm}, \citenamefont {Clarke}, \citenamefont {Dzurak}, \citenamefont {Ishihara}, \citenamefont {Morello}, \citenamefont {Reilly}, \citenamefont {Schreiber},\ and\ \citenamefont {Veldhorst}}]{vandersypen2017interfacing}%
  \BibitemOpen
  \bibfield  {author} {\bibinfo {author} {\bibfnamefont {L.}~\bibnamefont {Vandersypen}}, \bibinfo {author} {\bibfnamefont {H.}~\bibnamefont {Bluhm}}, \bibinfo {author} {\bibfnamefont {J.}~\bibnamefont {Clarke}}, \bibinfo {author} {\bibfnamefont {A.}~\bibnamefont {Dzurak}}, \bibinfo {author} {\bibfnamefont {R.}~\bibnamefont {Ishihara}}, \bibinfo {author} {\bibfnamefont {A.}~\bibnamefont {Morello}}, \bibinfo {author} {\bibfnamefont {D.}~\bibnamefont {Reilly}}, \bibinfo {author} {\bibfnamefont {L.}~\bibnamefont {Schreiber}},\ and\ \bibinfo {author} {\bibfnamefont {M.}~\bibnamefont {Veldhorst}},\ }\bibfield  {title} {\bibinfo {title} {Interfacing spin qubits in quantum dots and donors—hot, dense, and coherent},\ }\href {https://www.nature.com/articles/s41534-017-0038-y} {\bibfield  {journal} {\bibinfo  {journal} {npj Quantum Information}\ }\textbf {\bibinfo {volume} {3}},\ \bibinfo {pages} {34} (\bibinfo {year} {2017})}\BibitemShut {NoStop}%
\bibitem [{\citenamefont {Yang}\ \emph {et~al.}(2020)\citenamefont {Yang}, \citenamefont {Leon}, \citenamefont {Hwang}, \citenamefont {Saraiva}, \citenamefont {Tanttu}, \citenamefont {Huang}, \citenamefont {Camirand~Lemyre}, \citenamefont {Chan}, \citenamefont {Tan}, \citenamefont {Hudson} \emph {et~al.}}]{yang2020operation}%
  \BibitemOpen
  \bibfield  {author} {\bibinfo {author} {\bibfnamefont {C.~H.}\ \bibnamefont {Yang}}, \bibinfo {author} {\bibfnamefont {R.}~\bibnamefont {Leon}}, \bibinfo {author} {\bibfnamefont {J.}~\bibnamefont {Hwang}}, \bibinfo {author} {\bibfnamefont {A.}~\bibnamefont {Saraiva}}, \bibinfo {author} {\bibfnamefont {T.}~\bibnamefont {Tanttu}}, \bibinfo {author} {\bibfnamefont {W.}~\bibnamefont {Huang}}, \bibinfo {author} {\bibfnamefont {J.}~\bibnamefont {Camirand~Lemyre}}, \bibinfo {author} {\bibfnamefont {K.~W.}\ \bibnamefont {Chan}}, \bibinfo {author} {\bibfnamefont {K.}~\bibnamefont {Tan}}, \bibinfo {author} {\bibfnamefont {F.~E.}\ \bibnamefont {Hudson}}, \emph {et~al.},\ }\bibfield  {title} {\bibinfo {title} {Operation of a silicon quantum processor unit cell above one kelvin},\ }\href {https://www.nature.com/articles/s41586-020-2171-6?www.nature.com} {\bibfield  {journal} {\bibinfo  {journal} {Nature}\ }\textbf {\bibinfo {volume} {580}},\ \bibinfo {pages} {350} (\bibinfo {year} {2020})}\BibitemShut {NoStop}%
\bibitem [{\citenamefont {Petit}\ \emph {et~al.}(2022{\natexlab{b}})\citenamefont {Petit}, \citenamefont {Russ}, \citenamefont {Eenink}, \citenamefont {Lawrie}, \citenamefont {Clarke}, \citenamefont {Vandersypen},\ and\ \citenamefont {Veldhorst}}]{petit2022design}%
  \BibitemOpen
  \bibfield  {author} {\bibinfo {author} {\bibfnamefont {L.}~\bibnamefont {Petit}}, \bibinfo {author} {\bibfnamefont {M.}~\bibnamefont {Russ}}, \bibinfo {author} {\bibfnamefont {G.~H.}\ \bibnamefont {Eenink}}, \bibinfo {author} {\bibfnamefont {W.~I.}\ \bibnamefont {Lawrie}}, \bibinfo {author} {\bibfnamefont {J.~S.}\ \bibnamefont {Clarke}}, \bibinfo {author} {\bibfnamefont {L.~M.}\ \bibnamefont {Vandersypen}},\ and\ \bibinfo {author} {\bibfnamefont {M.}~\bibnamefont {Veldhorst}},\ }\bibfield  {title} {\bibinfo {title} {Design and integration of single-qubit rotations and two-qubit gates in silicon above one kelvin},\ }\href {https://www.nature.com/articles/s43246-022-00304-9} {\bibfield  {journal} {\bibinfo  {journal} {Communications Materials}\ }\textbf {\bibinfo {volume} {3}},\ \bibinfo {pages} {82} (\bibinfo {year} {2022}{\natexlab{b}})}\BibitemShut {NoStop}%
\bibitem [{\citenamefont {Huang}\ \emph {et~al.}(2021)\citenamefont {Huang}, \citenamefont {Lim}, \citenamefont {Leon}, \citenamefont {Yang}, \citenamefont {Hudson}, \citenamefont {Escott}, \citenamefont {Saraiva}, \citenamefont {Dzurak},\ and\ \citenamefont {Laucht}}]{huang2021high}%
  \BibitemOpen
  \bibfield  {author} {\bibinfo {author} {\bibfnamefont {J.~Y.}\ \bibnamefont {Huang}}, \bibinfo {author} {\bibfnamefont {W.~H.}\ \bibnamefont {Lim}}, \bibinfo {author} {\bibfnamefont {R.~C.}\ \bibnamefont {Leon}}, \bibinfo {author} {\bibfnamefont {C.~H.}\ \bibnamefont {Yang}}, \bibinfo {author} {\bibfnamefont {F.~E.}\ \bibnamefont {Hudson}}, \bibinfo {author} {\bibfnamefont {C.~C.}\ \bibnamefont {Escott}}, \bibinfo {author} {\bibfnamefont {A.}~\bibnamefont {Saraiva}}, \bibinfo {author} {\bibfnamefont {A.~S.}\ \bibnamefont {Dzurak}},\ and\ \bibinfo {author} {\bibfnamefont {A.}~\bibnamefont {Laucht}},\ }\bibfield  {title} {\bibinfo {title} {A high-sensitivity charge sensor for silicon qubits above 1 k},\ }\href {https://pubs.acs.org/doi/full/10.1021/acs.nanolett.1c01003?casa_token=3k5HTNV_BAoAAAAA%3APa3Yrclhkd4gRkml6ps0afw1JeTqQKZvHJSYFw-bP3_aiuhgbjhKgDlX3R_0oTJxWaaW7rTKe0AsTUQ} {\bibfield  {journal} {\bibinfo  {journal} {Nano Letters}\ }\textbf {\bibinfo {volume} {21}},\ \bibinfo {pages} {6328} (\bibinfo
  {year} {2021})}\BibitemShut {NoStop}%
\bibitem [{\citenamefont {Undseth}\ \emph {et~al.}(2023)\citenamefont {Undseth}, \citenamefont {Pietx-Casas}, \citenamefont {Raymenants}, \citenamefont {Mehmandoost}, \citenamefont {M{\k{a}}dzik}, \citenamefont {Philips}, \citenamefont {De~Snoo}, \citenamefont {Michalak}, \citenamefont {Amitonov}, \citenamefont {Tryputen} \emph {et~al.}}]{undseth2023hotter}%
  \BibitemOpen
  \bibfield  {author} {\bibinfo {author} {\bibfnamefont {B.}~\bibnamefont {Undseth}}, \bibinfo {author} {\bibfnamefont {O.}~\bibnamefont {Pietx-Casas}}, \bibinfo {author} {\bibfnamefont {E.}~\bibnamefont {Raymenants}}, \bibinfo {author} {\bibfnamefont {M.}~\bibnamefont {Mehmandoost}}, \bibinfo {author} {\bibfnamefont {M.~T.}\ \bibnamefont {M{\k{a}}dzik}}, \bibinfo {author} {\bibfnamefont {S.~G.}\ \bibnamefont {Philips}}, \bibinfo {author} {\bibfnamefont {S.~L.}\ \bibnamefont {De~Snoo}}, \bibinfo {author} {\bibfnamefont {D.~J.}\ \bibnamefont {Michalak}}, \bibinfo {author} {\bibfnamefont {S.~V.}\ \bibnamefont {Amitonov}}, \bibinfo {author} {\bibfnamefont {L.}~\bibnamefont {Tryputen}}, \emph {et~al.},\ }\bibfield  {title} {\bibinfo {title} {Hotter is easier: unexpected temperature dependence of spin qubit frequencies},\ }\href {https://journals.aps.org/prx/abstract/10.1103/PhysRevX.13.041015} {\bibfield  {journal} {\bibinfo  {journal} {Physical Review X}\ }\textbf {\bibinfo {volume} {13}},\ \bibinfo {pages}
  {041015} (\bibinfo {year} {2023})}\BibitemShut {NoStop}%
\bibitem [{\citenamefont {Burkard}\ \emph {et~al.}(1999{\natexlab{a}})\citenamefont {Burkard}, \citenamefont {Loss}, \citenamefont {DiVincenzo},\ and\ \citenamefont {Smolin}}]{burkard1999physical}%
  \BibitemOpen
  \bibfield  {author} {\bibinfo {author} {\bibfnamefont {G.}~\bibnamefont {Burkard}}, \bibinfo {author} {\bibfnamefont {D.}~\bibnamefont {Loss}}, \bibinfo {author} {\bibfnamefont {D.~P.}\ \bibnamefont {DiVincenzo}},\ and\ \bibinfo {author} {\bibfnamefont {J.~A.}\ \bibnamefont {Smolin}},\ }\bibfield  {title} {\bibinfo {title} {Physical optimization of quantum error correction circuits},\ }\href {https://journals.aps.org/prb/abstract/10.1103/PhysRevB.60.11404} {\bibfield  {journal} {\bibinfo  {journal} {Physical Review B}\ }\textbf {\bibinfo {volume} {60}},\ \bibinfo {pages} {11404} (\bibinfo {year} {1999}{\natexlab{a}})}\BibitemShut {NoStop}%
\bibitem [{\citenamefont {Philips}\ \emph {et~al.}(2022)\citenamefont {Philips}, \citenamefont {M{\k{a}}dzik}, \citenamefont {Amitonov}, \citenamefont {de~Snoo}, \citenamefont {Russ}, \citenamefont {Kalhor}, \citenamefont {Volk}, \citenamefont {Lawrie}, \citenamefont {Brousse}, \citenamefont {Tryputen} \emph {et~al.}}]{philips2022universal}%
  \BibitemOpen
  \bibfield  {author} {\bibinfo {author} {\bibfnamefont {S.~G.}\ \bibnamefont {Philips}}, \bibinfo {author} {\bibfnamefont {M.~T.}\ \bibnamefont {M{\k{a}}dzik}}, \bibinfo {author} {\bibfnamefont {S.~V.}\ \bibnamefont {Amitonov}}, \bibinfo {author} {\bibfnamefont {S.~L.}\ \bibnamefont {de~Snoo}}, \bibinfo {author} {\bibfnamefont {M.}~\bibnamefont {Russ}}, \bibinfo {author} {\bibfnamefont {N.}~\bibnamefont {Kalhor}}, \bibinfo {author} {\bibfnamefont {C.}~\bibnamefont {Volk}}, \bibinfo {author} {\bibfnamefont {W.~I.}\ \bibnamefont {Lawrie}}, \bibinfo {author} {\bibfnamefont {D.}~\bibnamefont {Brousse}}, \bibinfo {author} {\bibfnamefont {L.}~\bibnamefont {Tryputen}}, \emph {et~al.},\ }\bibfield  {title} {\bibinfo {title} {Universal control of a six-qubit quantum processor in silicon},\ }\href {https://www.nature.com/articles/s41586-022-05117-x} {\bibfield  {journal} {\bibinfo  {journal} {Nature}\ }\textbf {\bibinfo {volume} {609}},\ \bibinfo {pages} {919} (\bibinfo {year} {2022})}\BibitemShut {NoStop}%
\bibitem [{\citenamefont {Kornich}\ \emph {et~al.}(2018)\citenamefont {Kornich}, \citenamefont {Kloeffel},\ and\ \citenamefont {Loss}}]{kornich2018phonon}%
  \BibitemOpen
  \bibfield  {author} {\bibinfo {author} {\bibfnamefont {V.}~\bibnamefont {Kornich}}, \bibinfo {author} {\bibfnamefont {C.}~\bibnamefont {Kloeffel}},\ and\ \bibinfo {author} {\bibfnamefont {D.}~\bibnamefont {Loss}},\ }\bibfield  {title} {\bibinfo {title} {Phonon-assisted relaxation and decoherence of singlet-triplet qubits in si/sige quantum dots},\ }\href {https://quantum-journal.org/papers/q-2018-05-28-70/} {\bibfield  {journal} {\bibinfo  {journal} {Quantum}\ }\textbf {\bibinfo {volume} {2}},\ \bibinfo {pages} {70} (\bibinfo {year} {2018})}\BibitemShut {NoStop}%
\bibitem [{\citenamefont {Kornich}\ \emph {et~al.}(2014)\citenamefont {Kornich}, \citenamefont {Kloeffel},\ and\ \citenamefont {Loss}}]{kornich2014phonon}%
  \BibitemOpen
  \bibfield  {author} {\bibinfo {author} {\bibfnamefont {V.}~\bibnamefont {Kornich}}, \bibinfo {author} {\bibfnamefont {C.}~\bibnamefont {Kloeffel}},\ and\ \bibinfo {author} {\bibfnamefont {D.}~\bibnamefont {Loss}},\ }\bibfield  {title} {\bibinfo {title} {Phonon-mediated decay of singlet-triplet qubits in double quantum dots},\ }\href {https://journals.aps.org/prb/abstract/10.1103/PhysRevB.89.085410} {\bibfield  {journal} {\bibinfo  {journal} {Physical Review B}\ }\textbf {\bibinfo {volume} {89}},\ \bibinfo {pages} {085410} (\bibinfo {year} {2014})}\BibitemShut {NoStop}%
\bibitem [{\citenamefont {Golovach}\ \emph {et~al.}(2004)\citenamefont {Golovach}, \citenamefont {Khaetskii},\ and\ \citenamefont {Loss}}]{golovach2004phonon}%
  \BibitemOpen
  \bibfield  {author} {\bibinfo {author} {\bibfnamefont {V.~N.}\ \bibnamefont {Golovach}}, \bibinfo {author} {\bibfnamefont {A.}~\bibnamefont {Khaetskii}},\ and\ \bibinfo {author} {\bibfnamefont {D.}~\bibnamefont {Loss}},\ }\bibfield  {title} {\bibinfo {title} {Phonon-induced decay of the electron spin in quantum dots},\ }\href {https://journals.aps.org/prl/abstract/10.1103/PhysRevLett.93.016601} {\bibfield  {journal} {\bibinfo  {journal} {Physical Review Letters}\ }\textbf {\bibinfo {volume} {93}},\ \bibinfo {pages} {016601} (\bibinfo {year} {2004})}\BibitemShut {NoStop}%
\bibitem [{\citenamefont {Borhani}\ \emph {et~al.}(2006)\citenamefont {Borhani}, \citenamefont {Golovach},\ and\ \citenamefont {Loss}}]{borhani2006spin}%
  \BibitemOpen
  \bibfield  {author} {\bibinfo {author} {\bibfnamefont {M.}~\bibnamefont {Borhani}}, \bibinfo {author} {\bibfnamefont {V.~N.}\ \bibnamefont {Golovach}},\ and\ \bibinfo {author} {\bibfnamefont {D.}~\bibnamefont {Loss}},\ }\bibfield  {title} {\bibinfo {title} {Spin decay in a quantum dot coupled to a quantum point contact},\ }\href {https://journals.aps.org/prb/abstract/10.1103/PhysRevB.73.155311} {\bibfield  {journal} {\bibinfo  {journal} {Physical Review B}\ }\textbf {\bibinfo {volume} {73}},\ \bibinfo {pages} {155311} (\bibinfo {year} {2006})}\BibitemShut {NoStop}%
\bibitem [{\citenamefont {He}\ \emph {et~al.}(2023)\citenamefont {He}, \citenamefont {Chan},\ and\ \citenamefont {Wang}}]{he2023theory}%
  \BibitemOpen
  \bibfield  {author} {\bibinfo {author} {\bibfnamefont {G.}~\bibnamefont {He}}, \bibinfo {author} {\bibfnamefont {G.~X.}\ \bibnamefont {Chan}},\ and\ \bibinfo {author} {\bibfnamefont {X.}~\bibnamefont {Wang}},\ }\bibfield  {title} {\bibinfo {title} {Theory on electron--phonon spin dephasing in gaas multi-electron double quantum dots},\ }\href {https://onlinelibrary.wiley.com/doi/full/10.1002/qute.202200074} {\bibfield  {journal} {\bibinfo  {journal} {Advanced Quantum Technologies}\ }\textbf {\bibinfo {volume} {6}},\ \bibinfo {pages} {2200074} (\bibinfo {year} {2023})}\BibitemShut {NoStop}%
\bibitem [{\citenamefont {Kuroyama}\ \emph {et~al.}(2023)\citenamefont {Kuroyama}, \citenamefont {Matsuo}, \citenamefont {Tarucha},\ and\ \citenamefont {Tokura}}]{kuroyama2023phonon}%
  \BibitemOpen
  \bibfield  {author} {\bibinfo {author} {\bibfnamefont {K.}~\bibnamefont {Kuroyama}}, \bibinfo {author} {\bibfnamefont {S.}~\bibnamefont {Matsuo}}, \bibinfo {author} {\bibfnamefont {S.}~\bibnamefont {Tarucha}},\ and\ \bibinfo {author} {\bibfnamefont {Y.}~\bibnamefont {Tokura}},\ }\bibfield  {title} {\bibinfo {title} {Phonon-mediated spin dynamics in a two-electron double quantum dot under a phonon temperature gradient},\ }\href {https://journals.aps.org/prb/abstract/10.1103/PhysRevB.108.115308} {\bibfield  {journal} {\bibinfo  {journal} {Physical Review B}\ }\textbf {\bibinfo {volume} {108}},\ \bibinfo {pages} {115308} (\bibinfo {year} {2023})}\BibitemShut {NoStop}%
\bibitem [{\citenamefont {Krychowski}\ and\ \citenamefont {Lipi{\'n}ski}(2023)}]{krychowski2023electron}%
  \BibitemOpen
  \bibfield  {author} {\bibinfo {author} {\bibfnamefont {D.}~\bibnamefont {Krychowski}}\ and\ \bibinfo {author} {\bibfnamefont {S.}~\bibnamefont {Lipi{\'n}ski}},\ }\bibfield  {title} {\bibinfo {title} {Electron--phonon interaction and electronic correlations in transport through electrostatically and tunnel coupled quantum dots},\ }\href {https://www.sciencedirect.com/science/article/pii/S0304885323010661} {\bibfield  {journal} {\bibinfo  {journal} {Journal of Magnetism and Magnetic Materials}\ }\textbf {\bibinfo {volume} {588}},\ \bibinfo {pages} {171416} (\bibinfo {year} {2023})}\BibitemShut {NoStop}%
\bibitem [{\citenamefont {Wardrop}\ and\ \citenamefont {Doherty}(2014)}]{wardrop2014exchange}%
  \BibitemOpen
  \bibfield  {author} {\bibinfo {author} {\bibfnamefont {M.~P.}\ \bibnamefont {Wardrop}}\ and\ \bibinfo {author} {\bibfnamefont {A.~C.}\ \bibnamefont {Doherty}},\ }\bibfield  {title} {\bibinfo {title} {Exchange-based two-qubit gate for singlet-triplet qubits},\ }\href {https://journals.aps.org/prb/abstract/10.1103/PhysRevB.90.045418} {\bibfield  {journal} {\bibinfo  {journal} {Physical Review B}\ }\textbf {\bibinfo {volume} {90}},\ \bibinfo {pages} {045418} (\bibinfo {year} {2014})}\BibitemShut {NoStop}%
\bibitem [{\citenamefont {Andrews}\ \emph {et~al.}(2019)\citenamefont {Andrews}, \citenamefont {Jones}, \citenamefont {Reed}, \citenamefont {Jones}, \citenamefont {Ha}, \citenamefont {Jura}, \citenamefont {Kerckhoff}, \citenamefont {Levendorf}, \citenamefont {Meenehan}, \citenamefont {Merkel} \emph {et~al.}}]{andrews2019quantifying}%
  \BibitemOpen
  \bibfield  {author} {\bibinfo {author} {\bibfnamefont {R.~W.}\ \bibnamefont {Andrews}}, \bibinfo {author} {\bibfnamefont {C.}~\bibnamefont {Jones}}, \bibinfo {author} {\bibfnamefont {M.~D.}\ \bibnamefont {Reed}}, \bibinfo {author} {\bibfnamefont {A.~M.}\ \bibnamefont {Jones}}, \bibinfo {author} {\bibfnamefont {S.~D.}\ \bibnamefont {Ha}}, \bibinfo {author} {\bibfnamefont {M.~P.}\ \bibnamefont {Jura}}, \bibinfo {author} {\bibfnamefont {J.}~\bibnamefont {Kerckhoff}}, \bibinfo {author} {\bibfnamefont {M.}~\bibnamefont {Levendorf}}, \bibinfo {author} {\bibfnamefont {S.}~\bibnamefont {Meenehan}}, \bibinfo {author} {\bibfnamefont {S.~T.}\ \bibnamefont {Merkel}}, \emph {et~al.},\ }\bibfield  {title} {\bibinfo {title} {Quantifying error and leakage in an encoded si/sige triple-dot qubit},\ }\href {https://idp.nature.com/authorize/casa?redirect_uri=https://www.nature.com/articles/s41565-019-0500-4&casa_token=mtai8W_-wQ0AAAAA:rU-S3b-nogvkwBK4elVM6X1CHJmb3TW2zy9_ASmYc-DAstT5kdHTlWI4a85hswxTmEZmcyKK66UCjARNHw}
  {\bibfield  {journal} {\bibinfo  {journal} {Nat. Nanotechnol.}\ }\textbf {\bibinfo {volume} {14}},\ \bibinfo {pages} {747} (\bibinfo {year} {2019})}\BibitemShut {NoStop}%
\bibitem [{\citenamefont {Mehl}\ \emph {et~al.}(2015)\citenamefont {Mehl}, \citenamefont {Bluhm},\ and\ \citenamefont {DiVincenzo}}]{mehl2015fault}%
  \BibitemOpen
  \bibfield  {author} {\bibinfo {author} {\bibfnamefont {S.}~\bibnamefont {Mehl}}, \bibinfo {author} {\bibfnamefont {H.}~\bibnamefont {Bluhm}},\ and\ \bibinfo {author} {\bibfnamefont {D.~P.}\ \bibnamefont {DiVincenzo}},\ }\bibfield  {title} {\bibinfo {title} {Fault-tolerant quantum computation for singlet-triplet qubits with leakage errors},\ }\href {https://journals.aps.org/prb/abstract/10.1103/PhysRevB.91.085419} {\bibfield  {journal} {\bibinfo  {journal} {Physical Review B}\ }\textbf {\bibinfo {volume} {91}},\ \bibinfo {pages} {085419} (\bibinfo {year} {2015})}\BibitemShut {NoStop}%
\bibitem [{\citenamefont {Fong}\ and\ \citenamefont {Wandzura}(2011)}]{fong2011universal}%
  \BibitemOpen
  \bibfield  {author} {\bibinfo {author} {\bibfnamefont {B.~H.}\ \bibnamefont {Fong}}\ and\ \bibinfo {author} {\bibfnamefont {S.~M.}\ \bibnamefont {Wandzura}},\ }\bibfield  {title} {\bibinfo {title} {Universal quantum computation and leakage reduction in the 3-qubit decoherence free subsystem},\ }\href {https://arxiv.org/abs/1102.2909} {\bibfield  {journal} {\bibinfo  {journal} {arXiv preprint arXiv:1102.2909}\ } (\bibinfo {year} {2011})}\BibitemShut {NoStop}%
\bibitem [{\citenamefont {Huang}\ \emph {et~al.}(2023)\citenamefont {Huang}, \citenamefont {Su}, \citenamefont {Lim}, \citenamefont {Feng}, \citenamefont {van Straaten}, \citenamefont {Severin}, \citenamefont {Gilbert}, \citenamefont {Stuyck}, \citenamefont {Tanttu}, \citenamefont {Serrano} \emph {et~al.}}]{huang2023high}%
  \BibitemOpen
  \bibfield  {author} {\bibinfo {author} {\bibfnamefont {J.~Y.}\ \bibnamefont {Huang}}, \bibinfo {author} {\bibfnamefont {R.~Y.}\ \bibnamefont {Su}}, \bibinfo {author} {\bibfnamefont {W.~H.}\ \bibnamefont {Lim}}, \bibinfo {author} {\bibfnamefont {M.}~\bibnamefont {Feng}}, \bibinfo {author} {\bibfnamefont {B.}~\bibnamefont {van Straaten}}, \bibinfo {author} {\bibfnamefont {B.}~\bibnamefont {Severin}}, \bibinfo {author} {\bibfnamefont {W.}~\bibnamefont {Gilbert}}, \bibinfo {author} {\bibfnamefont {N.~D.}\ \bibnamefont {Stuyck}}, \bibinfo {author} {\bibfnamefont {T.}~\bibnamefont {Tanttu}}, \bibinfo {author} {\bibfnamefont {S.}~\bibnamefont {Serrano}}, \emph {et~al.},\ }\bibfield  {title} {\bibinfo {title} {High-fidelity operation and algorithmic initialisation of spin qubits above one kelvin},\ }\href {https://arxiv.org/abs/2308.02111} {\bibfield  {journal} {\bibinfo  {journal} {arXiv preprint arXiv:2308.02111}\ } (\bibinfo {year} {2023})}\BibitemShut {NoStop}%
\bibitem [{\citenamefont {Brooks}\ and\ \citenamefont {Tahan}(2021)}]{brooks2021hybrid}%
  \BibitemOpen
  \bibfield  {author} {\bibinfo {author} {\bibfnamefont {M.}~\bibnamefont {Brooks}}\ and\ \bibinfo {author} {\bibfnamefont {C.}~\bibnamefont {Tahan}},\ }\bibfield  {title} {\bibinfo {title} {Hybrid exchange--measurement-based qubit operations in semiconductor double-quantum-dot qubits},\ }\href {https://journals.aps.org/prapplied/abstract/10.1103/PhysRevApplied.16.064019} {\bibfield  {journal} {\bibinfo  {journal} {Physical Review Applied}\ }\textbf {\bibinfo {volume} {16}},\ \bibinfo {pages} {064019} (\bibinfo {year} {2021})}\BibitemShut {NoStop}%
\bibitem [{\citenamefont {Brooks}\ and\ \citenamefont {Tahan}(2023)}]{brooks2023Meas}%
  \BibitemOpen
  \bibfield  {author} {\bibinfo {author} {\bibfnamefont {M.}~\bibnamefont {Brooks}}\ and\ \bibinfo {author} {\bibfnamefont {C.}~\bibnamefont {Tahan}},\ }\bibfield  {title} {\bibinfo {title} {Quantum computation by spin-parity measurements with encoded spin qubits},\ }\href {https://doi.org/10.1103/PhysRevB.108.035206} {\bibfield  {journal} {\bibinfo  {journal} {Physical Review B}\ }\textbf {\bibinfo {volume} {108}},\ \bibinfo {pages} {035206} (\bibinfo {year} {2023})}\BibitemShut {NoStop}%
\bibitem [{\citenamefont {Friesen}\ \emph {et~al.}(2007)\citenamefont {Friesen}, \citenamefont {Chutia}, \citenamefont {Tahan},\ and\ \citenamefont {Coppersmith}}]{friesen2007valley}%
  \BibitemOpen
  \bibfield  {author} {\bibinfo {author} {\bibfnamefont {M.}~\bibnamefont {Friesen}}, \bibinfo {author} {\bibfnamefont {S.}~\bibnamefont {Chutia}}, \bibinfo {author} {\bibfnamefont {C.}~\bibnamefont {Tahan}},\ and\ \bibinfo {author} {\bibfnamefont {S.}~\bibnamefont {Coppersmith}},\ }\bibfield  {title} {\bibinfo {title} {Valley splitting theory of si ge/ si/ si ge quantum wells},\ }\href {https://journals.aps.org/prb/abstract/10.1103/PhysRevB.75.115318} {\bibfield  {journal} {\bibinfo  {journal} {Physical Review B}\ }\textbf {\bibinfo {volume} {75}},\ \bibinfo {pages} {115318} (\bibinfo {year} {2007})}\BibitemShut {NoStop}%
\bibitem [{\citenamefont {Shi}\ \emph {et~al.}(2011)\citenamefont {Shi}, \citenamefont {Simmons}, \citenamefont {Prance}, \citenamefont {King~Gamble}, \citenamefont {Friesen}, \citenamefont {Savage}, \citenamefont {Lagally}, \citenamefont {Coppersmith},\ and\ \citenamefont {Eriksson}}]{shi2011tunable}%
  \BibitemOpen
  \bibfield  {author} {\bibinfo {author} {\bibfnamefont {Z.}~\bibnamefont {Shi}}, \bibinfo {author} {\bibfnamefont {C.}~\bibnamefont {Simmons}}, \bibinfo {author} {\bibfnamefont {J.}~\bibnamefont {Prance}}, \bibinfo {author} {\bibfnamefont {J.}~\bibnamefont {King~Gamble}}, \bibinfo {author} {\bibfnamefont {M.}~\bibnamefont {Friesen}}, \bibinfo {author} {\bibfnamefont {D.}~\bibnamefont {Savage}}, \bibinfo {author} {\bibfnamefont {M.}~\bibnamefont {Lagally}}, \bibinfo {author} {\bibfnamefont {S.}~\bibnamefont {Coppersmith}},\ and\ \bibinfo {author} {\bibfnamefont {M.}~\bibnamefont {Eriksson}},\ }\bibfield  {title} {\bibinfo {title} {Tunable singlet-triplet splitting in a few-electron si/sige quantum dot},\ }\href {https://pubs.aip.org/aip/apl/article-abstract/99/23/233108/282245/Tunable-singlet-triplet-splitting-in-a-few?redirectedFrom=fulltext} {\bibfield  {journal} {\bibinfo  {journal} {Applied Physics Letters}\ }\textbf {\bibinfo {volume} {99}} (\bibinfo {year} {2011})}\BibitemShut {NoStop}%
\bibitem [{\citenamefont {Zajac}\ \emph {et~al.}(2015)\citenamefont {Zajac}, \citenamefont {Hazard}, \citenamefont {Mi}, \citenamefont {Wang},\ and\ \citenamefont {Petta}}]{zajac2015reconfigurable}%
  \BibitemOpen
  \bibfield  {author} {\bibinfo {author} {\bibfnamefont {D.}~\bibnamefont {Zajac}}, \bibinfo {author} {\bibfnamefont {T.}~\bibnamefont {Hazard}}, \bibinfo {author} {\bibfnamefont {X.}~\bibnamefont {Mi}}, \bibinfo {author} {\bibfnamefont {K.}~\bibnamefont {Wang}},\ and\ \bibinfo {author} {\bibfnamefont {J.~R.}\ \bibnamefont {Petta}},\ }\bibfield  {title} {\bibinfo {title} {A reconfigurable gate architecture for si/sige quantum dots},\ }\href {https://pubs.aip.org/aip/apl/article/106/22/223507/27629} {\bibfield  {journal} {\bibinfo  {journal} {Applied Physics Letters}\ }\textbf {\bibinfo {volume} {106}} (\bibinfo {year} {2015})}\BibitemShut {NoStop}%
\bibitem [{\citenamefont {Losert}\ \emph {et~al.}(2023)\citenamefont {Losert}, \citenamefont {Eriksson}, \citenamefont {Joynt}, \citenamefont {Rahman}, \citenamefont {Scappucci}, \citenamefont {Coppersmith},\ and\ \citenamefont {Friesen}}]{losert2023practical}%
  \BibitemOpen
  \bibfield  {author} {\bibinfo {author} {\bibfnamefont {M.~P.}\ \bibnamefont {Losert}}, \bibinfo {author} {\bibfnamefont {M.}~\bibnamefont {Eriksson}}, \bibinfo {author} {\bibfnamefont {R.}~\bibnamefont {Joynt}}, \bibinfo {author} {\bibfnamefont {R.}~\bibnamefont {Rahman}}, \bibinfo {author} {\bibfnamefont {G.}~\bibnamefont {Scappucci}}, \bibinfo {author} {\bibfnamefont {S.~N.}\ \bibnamefont {Coppersmith}},\ and\ \bibinfo {author} {\bibfnamefont {M.}~\bibnamefont {Friesen}},\ }\bibfield  {title} {\bibinfo {title} {Practical strategies for enhancing the valley splitting in si/sige quantum wells},\ }\href {https://journals.aps.org/prb/abstract/10.1103/PhysRevB.108.125405} {\bibfield  {journal} {\bibinfo  {journal} {Physical Review B}\ }\textbf {\bibinfo {volume} {108}},\ \bibinfo {pages} {125405} (\bibinfo {year} {2023})}\BibitemShut {NoStop}%
\bibitem [{\citenamefont {Boykin}\ \emph {et~al.}(2004)\citenamefont {Boykin}, \citenamefont {Klimeck}, \citenamefont {Eriksson}, \citenamefont {Friesen}, \citenamefont {Coppersmith}, \citenamefont {Von~Allmen}, \citenamefont {Oyafuso},\ and\ \citenamefont {Lee}}]{boykin2004valley}%
  \BibitemOpen
  \bibfield  {author} {\bibinfo {author} {\bibfnamefont {T.~B.}\ \bibnamefont {Boykin}}, \bibinfo {author} {\bibfnamefont {G.}~\bibnamefont {Klimeck}}, \bibinfo {author} {\bibfnamefont {M.}~\bibnamefont {Eriksson}}, \bibinfo {author} {\bibfnamefont {M.}~\bibnamefont {Friesen}}, \bibinfo {author} {\bibfnamefont {S.}~\bibnamefont {Coppersmith}}, \bibinfo {author} {\bibfnamefont {P.}~\bibnamefont {Von~Allmen}}, \bibinfo {author} {\bibfnamefont {F.}~\bibnamefont {Oyafuso}},\ and\ \bibinfo {author} {\bibfnamefont {S.}~\bibnamefont {Lee}},\ }\bibfield  {title} {\bibinfo {title} {Valley splitting in strained silicon quantum wells},\ }\href {https://pubs.aip.org/aip/apl/article-abstract/84/1/115/115791/Valley-splitting-in-strained-silicon-quantum-wells} {\bibfield  {journal} {\bibinfo  {journal} {Applied Physics Letters}\ }\textbf {\bibinfo {volume} {84}},\ \bibinfo {pages} {115} (\bibinfo {year} {2004})}\BibitemShut {NoStop}%
\bibitem [{\citenamefont {Goswami}\ \emph {et~al.}(2007)\citenamefont {Goswami}, \citenamefont {Slinker}, \citenamefont {Friesen}, \citenamefont {McGuire}, \citenamefont {Truitt}, \citenamefont {Tahan}, \citenamefont {Klein}, \citenamefont {Chu}, \citenamefont {Mooney}, \citenamefont {Van Der~Weide} \emph {et~al.}}]{goswami2007controllable}%
  \BibitemOpen
  \bibfield  {author} {\bibinfo {author} {\bibfnamefont {S.}~\bibnamefont {Goswami}}, \bibinfo {author} {\bibfnamefont {K.}~\bibnamefont {Slinker}}, \bibinfo {author} {\bibfnamefont {M.}~\bibnamefont {Friesen}}, \bibinfo {author} {\bibfnamefont {L.}~\bibnamefont {McGuire}}, \bibinfo {author} {\bibfnamefont {J.}~\bibnamefont {Truitt}}, \bibinfo {author} {\bibfnamefont {C.}~\bibnamefont {Tahan}}, \bibinfo {author} {\bibfnamefont {L.}~\bibnamefont {Klein}}, \bibinfo {author} {\bibfnamefont {J.}~\bibnamefont {Chu}}, \bibinfo {author} {\bibfnamefont {P.}~\bibnamefont {Mooney}}, \bibinfo {author} {\bibfnamefont {D.~W.}\ \bibnamefont {Van Der~Weide}}, \emph {et~al.},\ }\bibfield  {title} {\bibinfo {title} {Controllable valley splitting in silicon quantum devices},\ }\href {https://www.nature.com/articles/nphys475} {\bibfield  {journal} {\bibinfo  {journal} {Nature Physics}\ }\textbf {\bibinfo {volume} {3}},\ \bibinfo {pages} {41} (\bibinfo {year} {2007})}\BibitemShut {NoStop}%
\bibitem [{\citenamefont {Burkard}\ \emph {et~al.}(1999{\natexlab{b}})\citenamefont {Burkard}, \citenamefont {Loss},\ and\ \citenamefont {DiVincenzo}}]{burkard1999coupled}%
  \BibitemOpen
  \bibfield  {author} {\bibinfo {author} {\bibfnamefont {G.}~\bibnamefont {Burkard}}, \bibinfo {author} {\bibfnamefont {D.}~\bibnamefont {Loss}},\ and\ \bibinfo {author} {\bibfnamefont {D.~P.}\ \bibnamefont {DiVincenzo}},\ }\bibfield  {title} {\bibinfo {title} {Coupled quantum dots as quantum gates},\ }\href {https://journals.aps.org/prb/abstract/10.1103/PhysRevB.59.2070} {\bibfield  {journal} {\bibinfo  {journal} {Physical Review B}\ }\textbf {\bibinfo {volume} {59}},\ \bibinfo {pages} {2070} (\bibinfo {year} {1999}{\natexlab{b}})}\BibitemShut {NoStop}%
\bibitem [{\citenamefont {McNeil}\ \emph {et~al.}(2010)\citenamefont {McNeil}, \citenamefont {Schneble}, \citenamefont {Kataoka}, \citenamefont {Ford}, \citenamefont {Kasama}, \citenamefont {Dunin-Borkowski}, \citenamefont {Feinberg}, \citenamefont {Harrison}, \citenamefont {Barnes}, \citenamefont {Tse} \emph {et~al.}}]{mcneil2010localized}%
  \BibitemOpen
  \bibfield  {author} {\bibinfo {author} {\bibfnamefont {R.~P.}\ \bibnamefont {McNeil}}, \bibinfo {author} {\bibfnamefont {R.~J.}\ \bibnamefont {Schneble}}, \bibinfo {author} {\bibfnamefont {M.}~\bibnamefont {Kataoka}}, \bibinfo {author} {\bibfnamefont {C.~J.}\ \bibnamefont {Ford}}, \bibinfo {author} {\bibfnamefont {T.}~\bibnamefont {Kasama}}, \bibinfo {author} {\bibfnamefont {R.~E.}\ \bibnamefont {Dunin-Borkowski}}, \bibinfo {author} {\bibfnamefont {J.~M.}\ \bibnamefont {Feinberg}}, \bibinfo {author} {\bibfnamefont {R.~J.}\ \bibnamefont {Harrison}}, \bibinfo {author} {\bibfnamefont {C.~H.}\ \bibnamefont {Barnes}}, \bibinfo {author} {\bibfnamefont {D.~H.}\ \bibnamefont {Tse}}, \emph {et~al.},\ }\bibfield  {title} {\bibinfo {title} {Localized magnetic fields in arbitrary directions using patterned nanomagnets},\ }\href {https://pubs.acs.org/doi/abs/10.1021/nl902949v?casa_token=CIPMpyqseZEAAAAA:od55cBDo7Po0yvRtqlONa7KyPmtgCv5z0UQ7Jg-c2zpje89j0lX0fiVIzebRCa_UdeLcEXMbluppjAI} {\bibfield  {journal} {\bibinfo
  {journal} {Nano Letts.}\ }\textbf {\bibinfo {volume} {10}},\ \bibinfo {pages} {1549} (\bibinfo {year} {2010})}\BibitemShut {NoStop}%
\bibitem [{\citenamefont {Ruskov}\ \emph {et~al.}(2018)\citenamefont {Ruskov}, \citenamefont {Veldhorst}, \citenamefont {Dzurak},\ and\ \citenamefont {Tahan}}]{ruskov2018electron}%
  \BibitemOpen
  \bibfield  {author} {\bibinfo {author} {\bibfnamefont {R.}~\bibnamefont {Ruskov}}, \bibinfo {author} {\bibfnamefont {M.}~\bibnamefont {Veldhorst}}, \bibinfo {author} {\bibfnamefont {A.~S.}\ \bibnamefont {Dzurak}},\ and\ \bibinfo {author} {\bibfnamefont {C.}~\bibnamefont {Tahan}},\ }\bibfield  {title} {\bibinfo {title} {Electron g-factor of valley states in realistic silicon quantum dots},\ }\href {https://journals.aps.org/prb/abstract/10.1103/PhysRevB.98.245424} {\bibfield  {journal} {\bibinfo  {journal} {Physical Review B}\ }\textbf {\bibinfo {volume} {98}},\ \bibinfo {pages} {245424} (\bibinfo {year} {2018})}\BibitemShut {NoStop}%
\bibitem [{\citenamefont {Harvey-Collard}\ \emph {et~al.}(2019)\citenamefont {Harvey-Collard}, \citenamefont {Jacobson}, \citenamefont {Bureau-Oxton}, \citenamefont {Jock}, \citenamefont {Srinivasa}, \citenamefont {Mounce}, \citenamefont {Ward}, \citenamefont {Anderson}, \citenamefont {Manginell}, \citenamefont {Wendt} \emph {et~al.}}]{harvey2019spin}%
  \BibitemOpen
  \bibfield  {author} {\bibinfo {author} {\bibfnamefont {P.}~\bibnamefont {Harvey-Collard}}, \bibinfo {author} {\bibfnamefont {N.~T.}\ \bibnamefont {Jacobson}}, \bibinfo {author} {\bibfnamefont {C.}~\bibnamefont {Bureau-Oxton}}, \bibinfo {author} {\bibfnamefont {R.~M.}\ \bibnamefont {Jock}}, \bibinfo {author} {\bibfnamefont {V.}~\bibnamefont {Srinivasa}}, \bibinfo {author} {\bibfnamefont {A.~M.}\ \bibnamefont {Mounce}}, \bibinfo {author} {\bibfnamefont {D.~R.}\ \bibnamefont {Ward}}, \bibinfo {author} {\bibfnamefont {J.~M.}\ \bibnamefont {Anderson}}, \bibinfo {author} {\bibfnamefont {R.~P.}\ \bibnamefont {Manginell}}, \bibinfo {author} {\bibfnamefont {J.~R.}\ \bibnamefont {Wendt}}, \emph {et~al.},\ }\bibfield  {title} {\bibinfo {title} {Spin-orbit interactions for singlet-triplet qubits in silicon},\ }\href {https://journals.aps.org/prl/abstract/10.1103/PhysRevLett.122.217702} {\bibfield  {journal} {\bibinfo  {journal} {Physical Review Letters}\ }\textbf {\bibinfo {volume} {122}},\ \bibinfo {pages} {217702}
  (\bibinfo {year} {2019})}\BibitemShut {NoStop}%
\bibitem [{\citenamefont {Mahan}(2013)}]{mahan2013many}%
  \BibitemOpen
  \bibfield  {author} {\bibinfo {author} {\bibfnamefont {G.~D.}\ \bibnamefont {Mahan}},\ }\href@noop {} {\emph {\bibinfo {title} {Many-particle physics}}}\ (\bibinfo  {publisher} {Springer Science \& Business Media},\ \bibinfo {year} {2013})\BibitemShut {NoStop}%
\bibitem [{\citenamefont {Yu}\ \emph {et~al.}(2010)\citenamefont {Yu}, \citenamefont {Cardona}, \citenamefont {Yu},\ and\ \citenamefont {Cardona}}]{yu2010vibrational}%
  \BibitemOpen
  \bibfield  {author} {\bibinfo {author} {\bibfnamefont {P.~Y.}\ \bibnamefont {Yu}}, \bibinfo {author} {\bibfnamefont {M.}~\bibnamefont {Cardona}}, \bibinfo {author} {\bibfnamefont {P.~Y.}\ \bibnamefont {Yu}},\ and\ \bibinfo {author} {\bibfnamefont {M.}~\bibnamefont {Cardona}},\ }\bibfield  {title} {\bibinfo {title} {Vibrational properties of semiconductors, and electron-phonon interactions},\ }\href {https://link.springer.com/chapter/10.1007/3-540-26475-2_3} {\bibfield  {journal} {\bibinfo  {journal} {Fundamentals of Semiconductors: Physics and Materials Properties}\ ,\ \bibinfo {pages} {107}} (\bibinfo {year} {2010})}\BibitemShut {NoStop}%
\bibitem [{\citenamefont {Zajac}\ \emph {et~al.}(2016)\citenamefont {Zajac}, \citenamefont {Hazard}, \citenamefont {Mi}, \citenamefont {Nielsen},\ and\ \citenamefont {Petta}}]{zajac2016scalable}%
  \BibitemOpen
  \bibfield  {author} {\bibinfo {author} {\bibfnamefont {D.}~\bibnamefont {Zajac}}, \bibinfo {author} {\bibfnamefont {T.}~\bibnamefont {Hazard}}, \bibinfo {author} {\bibfnamefont {X.}~\bibnamefont {Mi}}, \bibinfo {author} {\bibfnamefont {E.}~\bibnamefont {Nielsen}},\ and\ \bibinfo {author} {\bibfnamefont {J.~R.}\ \bibnamefont {Petta}},\ }\bibfield  {title} {\bibinfo {title} {Scalable gate architecture for a one-dimensional array of semiconductor spin qubits},\ }\href {https://journals.aps.org/prapplied/abstract/10.1103/PhysRevApplied.6.054013} {\bibfield  {journal} {\bibinfo  {journal} {Physical Review Applied}\ }\textbf {\bibinfo {volume} {6}},\ \bibinfo {pages} {054013} (\bibinfo {year} {2016})}\BibitemShut {NoStop}%
\bibitem [{\citenamefont {Landi}\ \emph {et~al.}(2022)\citenamefont {Landi}, \citenamefont {Poletti},\ and\ \citenamefont {Schaller}}]{landi2021non}%
  \BibitemOpen
  \bibfield  {author} {\bibinfo {author} {\bibfnamefont {G.~T.}\ \bibnamefont {Landi}}, \bibinfo {author} {\bibfnamefont {D.}~\bibnamefont {Poletti}},\ and\ \bibinfo {author} {\bibfnamefont {G.}~\bibnamefont {Schaller}},\ }\bibfield  {title} {\bibinfo {title} {Nonequilibrium boundary-driven quantum systems: Models, methods, and properties},\ }\href {https://journals.aps.org/rmp/abstract/10.1103/RevModPhys.94.045006} {\bibfield  {journal} {\bibinfo  {journal} {Reviews of Modern Physics}\ }\textbf {\bibinfo {volume} {94}},\ \bibinfo {pages} {045006} (\bibinfo {year} {2022})}\BibitemShut {NoStop}%
\bibitem [{\citenamefont {Petta}\ \emph {et~al.}(2005)\citenamefont {Petta}, \citenamefont {Johnson}, \citenamefont {Taylor}, \citenamefont {Laird}, \citenamefont {Yacoby}, \citenamefont {Lukin}, \citenamefont {Marcus}, \citenamefont {Hanson},\ and\ \citenamefont {Gossard}}]{petta2005coherent}%
  \BibitemOpen
  \bibfield  {author} {\bibinfo {author} {\bibfnamefont {J.~R.}\ \bibnamefont {Petta}}, \bibinfo {author} {\bibfnamefont {A.~C.}\ \bibnamefont {Johnson}}, \bibinfo {author} {\bibfnamefont {J.~M.}\ \bibnamefont {Taylor}}, \bibinfo {author} {\bibfnamefont {E.~A.}\ \bibnamefont {Laird}}, \bibinfo {author} {\bibfnamefont {A.}~\bibnamefont {Yacoby}}, \bibinfo {author} {\bibfnamefont {M.~D.}\ \bibnamefont {Lukin}}, \bibinfo {author} {\bibfnamefont {C.~M.}\ \bibnamefont {Marcus}}, \bibinfo {author} {\bibfnamefont {M.~P.}\ \bibnamefont {Hanson}},\ and\ \bibinfo {author} {\bibfnamefont {A.~C.}\ \bibnamefont {Gossard}},\ }\bibfield  {title} {\bibinfo {title} {Coherent manipulation of coupled electron spins in semiconductor quantum dots},\ }\href {https://www.science.org/doi/abs/10.1126/science.1116955?casa_token=XjeoiUuo6GIAAAAA:hddsQGEJXBsFsOLUncbz9G4CqB6edRAdk3lPbxeIw6MSxl8HEkfLyjER80FMQmzVArob60Yai-QoqAQ} {\bibfield  {journal} {\bibinfo  {journal} {Science}\ }\textbf {\bibinfo {volume} {309}},\ \bibinfo {pages}
  {2180} (\bibinfo {year} {2005})}\BibitemShut {NoStop}%
\bibitem [{\citenamefont {Maune}\ \emph {et~al.}(2012)\citenamefont {Maune}, \citenamefont {Borselli}, \citenamefont {Huang}, \citenamefont {Ladd}, \citenamefont {Deelman}, \citenamefont {Holabird}, \citenamefont {Kiselev}, \citenamefont {Alvarado-Rodriguez}, \citenamefont {Ross}, \citenamefont {Schmitz} \emph {et~al.}}]{maune2012coherent}%
  \BibitemOpen
  \bibfield  {author} {\bibinfo {author} {\bibfnamefont {B.~M.}\ \bibnamefont {Maune}}, \bibinfo {author} {\bibfnamefont {M.~G.}\ \bibnamefont {Borselli}}, \bibinfo {author} {\bibfnamefont {B.}~\bibnamefont {Huang}}, \bibinfo {author} {\bibfnamefont {T.~D.}\ \bibnamefont {Ladd}}, \bibinfo {author} {\bibfnamefont {P.~W.}\ \bibnamefont {Deelman}}, \bibinfo {author} {\bibfnamefont {K.~S.}\ \bibnamefont {Holabird}}, \bibinfo {author} {\bibfnamefont {A.~A.}\ \bibnamefont {Kiselev}}, \bibinfo {author} {\bibfnamefont {I.}~\bibnamefont {Alvarado-Rodriguez}}, \bibinfo {author} {\bibfnamefont {R.~S.}\ \bibnamefont {Ross}}, \bibinfo {author} {\bibfnamefont {A.~E.}\ \bibnamefont {Schmitz}}, \emph {et~al.},\ }\bibfield  {title} {\bibinfo {title} {Coherent singlet-triplet oscillations in a silicon-based double quantum dot},\ }\href
  {https://idp.nature.com/authorize/casa?redirect_uri=https://www.nature.com/articles/nature10707&casa_token=MaHl4TNjatIAAAAA:V9gKEAdX355GX0qzgVMFS3Mn8zmQsD3b6xQ4J8EnSfL2wdcUpYsiemM5vj3BfCchyQTaz4Q0y1bHADIt} {\bibfield  {journal} {\bibinfo  {journal} {Nature}\ }\textbf {\bibinfo {volume} {481}},\ \bibinfo {pages} {344} (\bibinfo {year} {2012})}\BibitemShut {NoStop}%
\bibitem [{\citenamefont {Mills}\ \emph {et~al.}(2022)\citenamefont {Mills}, \citenamefont {Guinn}, \citenamefont {Gullans}, \citenamefont {Sigillito}, \citenamefont {Feldman}, \citenamefont {Nielsen},\ and\ \citenamefont {Petta}}]{mills2022two}%
  \BibitemOpen
  \bibfield  {author} {\bibinfo {author} {\bibfnamefont {A.~R.}\ \bibnamefont {Mills}}, \bibinfo {author} {\bibfnamefont {C.~R.}\ \bibnamefont {Guinn}}, \bibinfo {author} {\bibfnamefont {M.~J.}\ \bibnamefont {Gullans}}, \bibinfo {author} {\bibfnamefont {A.~J.}\ \bibnamefont {Sigillito}}, \bibinfo {author} {\bibfnamefont {M.~M.}\ \bibnamefont {Feldman}}, \bibinfo {author} {\bibfnamefont {E.}~\bibnamefont {Nielsen}},\ and\ \bibinfo {author} {\bibfnamefont {J.~R.}\ \bibnamefont {Petta}},\ }\bibfield  {title} {\bibinfo {title} {Two-qubit silicon quantum processor with operation fidelity exceeding 99\%},\ }\href {https://www.science.org/doi/abs/10.1126/sciadv.abn5130} {\bibfield  {journal} {\bibinfo  {journal} {Science Advances}\ }\textbf {\bibinfo {volume} {8}},\ \bibinfo {pages} {eabn5130} (\bibinfo {year} {2022})}\BibitemShut {NoStop}%
\bibitem [{\citenamefont {Reed}\ \emph {et~al.}(2016)\citenamefont {Reed}, \citenamefont {Maune}, \citenamefont {Andrews}, \citenamefont {Borselli}, \citenamefont {Eng}, \citenamefont {Jura}, \citenamefont {Kiselev}, \citenamefont {Ladd}, \citenamefont {Merkel}, \citenamefont {Milosavljevic} \emph {et~al.}}]{reed2016reduced}%
  \BibitemOpen
  \bibfield  {author} {\bibinfo {author} {\bibfnamefont {M.}~\bibnamefont {Reed}}, \bibinfo {author} {\bibfnamefont {B.}~\bibnamefont {Maune}}, \bibinfo {author} {\bibfnamefont {R.}~\bibnamefont {Andrews}}, \bibinfo {author} {\bibfnamefont {M.}~\bibnamefont {Borselli}}, \bibinfo {author} {\bibfnamefont {K.}~\bibnamefont {Eng}}, \bibinfo {author} {\bibfnamefont {M.}~\bibnamefont {Jura}}, \bibinfo {author} {\bibfnamefont {A.}~\bibnamefont {Kiselev}}, \bibinfo {author} {\bibfnamefont {T.}~\bibnamefont {Ladd}}, \bibinfo {author} {\bibfnamefont {S.}~\bibnamefont {Merkel}}, \bibinfo {author} {\bibfnamefont {I.}~\bibnamefont {Milosavljevic}}, \emph {et~al.},\ }\bibfield  {title} {\bibinfo {title} {Reduced sensitivity to charge noise in semiconductor spin qubits via symmetric operation},\ }\href {https://journals.aps.org/prl/abstract/10.1103/PhysRevLett.116.110402} {\bibfield  {journal} {\bibinfo  {journal} {Physical Review Letters}\ }\textbf {\bibinfo {volume} {116}},\ \bibinfo {pages} {110402} (\bibinfo {year}
  {2016})}\BibitemShut {NoStop}%
\bibitem [{\citenamefont {Petit}\ \emph {et~al.}(2018)\citenamefont {Petit}, \citenamefont {Boter}, \citenamefont {Eenink}, \citenamefont {Droulers}, \citenamefont {Tagliaferri}, \citenamefont {Li}, \citenamefont {Franke}, \citenamefont {Singh}, \citenamefont {Clarke}, \citenamefont {Schouten} \emph {et~al.}}]{petit2018spin}%
  \BibitemOpen
  \bibfield  {author} {\bibinfo {author} {\bibfnamefont {L.}~\bibnamefont {Petit}}, \bibinfo {author} {\bibfnamefont {J.}~\bibnamefont {Boter}}, \bibinfo {author} {\bibfnamefont {H.}~\bibnamefont {Eenink}}, \bibinfo {author} {\bibfnamefont {G.}~\bibnamefont {Droulers}}, \bibinfo {author} {\bibfnamefont {M.}~\bibnamefont {Tagliaferri}}, \bibinfo {author} {\bibfnamefont {R.}~\bibnamefont {Li}}, \bibinfo {author} {\bibfnamefont {D.}~\bibnamefont {Franke}}, \bibinfo {author} {\bibfnamefont {K.}~\bibnamefont {Singh}}, \bibinfo {author} {\bibfnamefont {J.}~\bibnamefont {Clarke}}, \bibinfo {author} {\bibfnamefont {R.}~\bibnamefont {Schouten}}, \emph {et~al.},\ }\bibfield  {title} {\bibinfo {title} {Spin lifetime and charge noise in hot silicon quantum dot qubits},\ }\href {https://journals.aps.org/prl/abstract/10.1103/PhysRevLett.121.076801} {\bibfield  {journal} {\bibinfo  {journal} {Physical Review Letters}\ }\textbf {\bibinfo {volume} {121}},\ \bibinfo {pages} {076801} (\bibinfo {year} {2018})}\BibitemShut
  {NoStop}%
\bibitem [{\citenamefont {Petit}\ \emph {et~al.}(2020)\citenamefont {Petit}, \citenamefont {Eenink}, \citenamefont {Russ}, \citenamefont {Lawrie}, \citenamefont {Hendrickx}, \citenamefont {Philips}, \citenamefont {Clarke}, \citenamefont {Vandersypen},\ and\ \citenamefont {Veldhorst}}]{petit2020universal}%
  \BibitemOpen
  \bibfield  {author} {\bibinfo {author} {\bibfnamefont {L.}~\bibnamefont {Petit}}, \bibinfo {author} {\bibfnamefont {H.}~\bibnamefont {Eenink}}, \bibinfo {author} {\bibfnamefont {M.}~\bibnamefont {Russ}}, \bibinfo {author} {\bibfnamefont {W.}~\bibnamefont {Lawrie}}, \bibinfo {author} {\bibfnamefont {N.}~\bibnamefont {Hendrickx}}, \bibinfo {author} {\bibfnamefont {S.}~\bibnamefont {Philips}}, \bibinfo {author} {\bibfnamefont {J.}~\bibnamefont {Clarke}}, \bibinfo {author} {\bibfnamefont {L.}~\bibnamefont {Vandersypen}},\ and\ \bibinfo {author} {\bibfnamefont {M.}~\bibnamefont {Veldhorst}},\ }\bibfield  {title} {\bibinfo {title} {Universal quantum logic in hot silicon qubits},\ }\href {https://www.nature.com/articles/s41586-020-2170-7.pdf} {\bibfield  {journal} {\bibinfo  {journal} {Nature}\ }\textbf {\bibinfo {volume} {580}},\ \bibinfo {pages} {355} (\bibinfo {year} {2020})}\BibitemShut {NoStop}%
\end{thebibliography}%

\end{document}